\documentclass[twocolumn,pra,showpacs,superscriptaddress,amssymb,amsmath,amsmath,floatfix]{revtex4-2}

\usepackage{hyperref}
\usepackage{float}
\usepackage{amssymb}
\usepackage{bm}
\usepackage{orcidlink}

\usepackage{wrapfig}
\usepackage{eufrak}
\usepackage[utf8]{inputenc}
\usepackage{slashed}
\usepackage{latexsym}
\usepackage{amsfonts}
\usepackage{amssymb}

\begin{document}
\title{Coherent nonlinear Thomson scattering of Laguerre - Gauss beams on an electron sheet}

\author{Petru-Vlad Toma \orcidlink{0009-0002-9949-7266}}
\affiliation{Center for Advanced Laser Technologies,
    National Institute for Laser, Plasma and Radiation Physics, Magurele, Romania}
\affiliation{University of Bucharest, Department of Physics, Magurele, Romania}
\author{Andrei Cristian Opinca}
\affiliation{University of Bucharest, Department of Physics, Magurele, Romania}
\author{Virgil Baran \orcidlink{0009-0000-5585-9443}}
\affiliation{University of Bucharest, Department of Physics, Magurele, Romania}
\affiliation{Academy of Romanian Scientists, Strada Ilfov 3, Sector 5, 050044, Bucharest, Romania}
\author{Madalina Boca \orcidlink{0009-0001-5437-541X}}\email{madalina.boca@unibuc.ro}
\affiliation{University of Bucharest, Department of Physics, Magurele, Romania}
\date{\today}
\begin{abstract}

    We present a study of the scattering of a monochromatic helical laser beam, described by a Laguerre-Gauss solution of the Maxwell equations, on an electron sheet, initially at rest in the focal plane of the laser; the interaction is described in the framework of a local plane wave approximation. We calculate the scattered electromagnetic field observed in an arbitrary point at a large distance from the laser focal spot, by adding coherently the contributions of each electron in the electron sheet. Due to the interference effects, the radiation is emitted only into the forward direction, within a narrow cone, and it has a spatial structure that we analyze theoretically and numerically. For circularly polarized incident fields, the structure is also helical, with a helical index which  depends on the helical index of the incident radiation and the harmonic order.  These structures can be observed experimentally, as each harmonic order is emitted with a different frequency, and within cones of different opening angles. Our findings are in agreement with experimental results in the literature which demonstrate the generation of OAM carrying photons by radiation scattering on electrons.
\end{abstract}

\maketitle

{\it Introduction.} Starting from the experimental and theoretical results presented in  \cite{hems, taira}, demonstrating the generation of OAM-carrying photons in the laser electron interaction, we present here a study of the coherent non-linear Thomson scattering of Laguerre Gauss modes on  a statistical ensemble of electrons, modelled as an electron sheet.

Non-linear Thomson scattering is the scattering of the classical electromagnetic field on free electrons. Early studies such as  \cite{sarachik,esarey1, esarey2, salamin, salamin2,krafti, kraft}  presented the field scattered by a single electron calculated in the long distance approximation; a detailed review of the literature can be found, for example, in \cite{kam}. The angular momentum of the field emitted by a single particle was discussed in the papers by  Bordovitsyn et al \cite{borr}, Katoh et al \cite{Katoh},  Epp and  Guselnikova \cite{epp, epp2}. The topic of orbital angular momentum of the scattered light is especially important in the context of many applications of OAM carrying fields \cite{appoam}; Taira et al \cite{taira} studied the spatial structure of the gamma rays emitted in nonlinear inverse Thomson scattering of circularly polarized light on a single electron, and found vortices, with a helical phase structure $e^{i(N-1)\varPhi}$, where $N$ is the  harmonic order. Other works that studied the structure of the radiation emitted in the non-linear Thomson process are \cite{lan, kang, liu, wang}.

In the problem of scattering of electrons on structured fields, such as focused pulses or more complex, OAM carrying beams, the initial position of the particle determines the properties of the radiation emitted; several studies took into account not just the scattering on a single electron, but on a statistical ensemble of  electron distributed in the focal spot of the laser. Vais and Bychenkov  \cite{Vais} studied the incoherent  non-linear Thomson scattering of a tightly focused relativistically intense laser pulse by an ensemble of particles. Another study of incoherent scattering of radiation of a Laguerre-Gauss laser beam on electrons is presented in \cite{Pastor}.

In the  model of incoherent scattering, the energy radiated by each particle is simply added, without taking into account the interference effects of the electromagnetic field emitted from different points in the electron distribution. On the other hand, the model of coherent scattering, taking the interference into account was used in studies of Thomson scattering on plasma, (see  \cite{cp1} and the references therein).

The experimental observation of a structured field emitted coherently by an ensemble of electrons was reported as early as 2013 by Hemsing et al \cite{hems}; in their work, a Gaussian pulse was used, which was converted to an OAM mode in the interaction with a helically modulated electron-beam.

We analyze the  scattering of a Laguerre-Gauss laser beam on a statistical ensemble of electrons; in this case, the spatial structure of the incident laser beam realizes the modulation of the electron distribution. The temporal Fourier transform of the field emitted by each point of the sheet is calculated and then we take the {\it coherent} sum over the entire distribution for each harmonic. We show that the interference effects lead to a vortex field, as that predicted by Taira \cite{taira}, but with a more complex expression of the phase, which in our case, for the $N$-th harmonic generated by a mode of helical index $m_L$ leads to an electric field ${\bm E}\sim e^{i(Nm_L\pm(N-1))\varPhi}$; the supplemental term $Nm_L\varPhi$ in the phase is a pure interference effect. Our numerical results are similar to those observed  by Hemsing et al \cite{hems}. To our knowledge this study is the first to consider the coherent Thomson scattering of structured light on an ensemble of electrons; our results  could be also of importance for various applications of the interaction of helical light with matter \cite{s1,s2,s3}.

The analytical formulas are written in the international system of units, and for the numerical results we adopted the atomic system of units.

    {\it The local plane wave approximation}. We consider the coherent scattering of a Laguerre - Gauss mode of the  monochromatic electromagnetic field of frequency $\omega_L$ on a  system of electrons initially at rest, distributed uniformly in a disk of radius $\varrho_m$, equal to at least the beam radius, placed in the focal plane of the laser. The electromagnetic  field   propagating along the $Oz$ direction of the reference frame can be written in the paraxial approximation in terms of an approximate solution of the Helmholtz equation, named Laguerre Gauss solution $u_{p_Lm_L}({\bm r})$; its expression  in  cylindrical coordinates $(\varrho_0,\varPhi_0,z)$ is $u_{p_Lm_L}({\bm r}) = v_{p_Lm_L}(\varrho_0,z)e^{im_L\varPhi_0}$   the product between a function $v$ dependent on $\varrho_0$ and $z$ and the exponential $e^{im_L\varPhi_0}$. The detailed expressions of  $v_{p_L,m_L}$ and  of the electromagnetic field are discussed  in the supplemental material \cite{supp}, Sect. I. A.
We shall study the scattering of a monochromatic  LG beam (assumed to be adiabatically introduced in the distant past) on a statistical ensemble of electrons, initially at rest, positioned in the focal plane of the laser. The laser wavelength that we consider is $\lambda=800$ nm, corresponding to $\omega_L=0.057$ au, and the beam waist $w_0$ will be taken $w_0=75 \lambda$, well within the validity of the paraxial approximation (see e.g. \cite{siegman}, Cap. 7). For this choice the variation of the field in the transversal plane is very slow and it can be considered as constant along distances of the order of the wavelength $\lambda$ in the $Oxy$ plane.

The motion of an electron in a plane wave monochromatic field of wavelength $\lambda$, and dimensionless intensity parameter $\xi_0=\frac{|e_0|E_0}{\omega m_ec}$ consists in an oscillation  in the polarization plane, of amplitude $r_\perp\sim \frac{c\xi_0}{\omega_L}\sim\xi_0\lambda$, and a displacement along the propagation direction $ z\sim\frac{c\xi_0^2}{2\omega_L}\sim\xi_0^2\lambda$. From this estimation it follows that an electron interacting with  a structured field, such as a helical beam, will see only a monochromatic plane wave field whose amplitude depends on its initial position in the focal plane if the variation of the field intensity in the focal plane is slow and the intensity $\xi_0\lessapprox1$. For the case of a helical beam, this condition is satisfied  if the beam waist parameter $w_0$ is large with respect to $\lambda$; in our work, we consider $w_0=75\lambda$.

Then, if the initial cylindrical  coordinates of the electron in the focal plane are $\varPhi_0$, $\varrho_0$ the electromagnetic field felt by the electron can be approximated as
\begin{align}
     & \widetilde{\bm E}({\bm r},t) = {\cal E}_0(\varrho_0)\Re\left\{{\boldsymbol{\epsilon}}e^{-i(\omega_L t-kz- m_L\varPhi_0)}\right\},\nonumber \\
     & \widetilde{\bm B}({\bm r},t) = \frac1c{\bm e}_z\times\widetilde{\bm E}({\bm r},t),\label{ebapprox}
\end{align}
where ${\cal E}_0(\varrho_0)$ is the local amplitude of the field ${\cal E}_0(\varrho_0)= E_0 v_{p_Lm_L}(\varrho_0,z=0)$ and ${\boldsymbol{\epsilon}}$ is the polarization vector, chosen as ${\boldsymbol{\epsilon}}={\bm e}_x\zeta_x+i{\bm e}_y\zeta_y$. We can define a local intensity parameter $\xi(\varrho_0)=\xi_0 v(\varrho_0,z=0)$; in our model each electron feels a plane wave monochromatic field whose magnitude ${\cal E}_0$ and initial phase $ m_L\varPhi_0$ are remnants of the incident field structure. We name this approach the local plane wave approximation (LPWA) and we can test its validity by comparing the solution of the exact equations of motion with those for a particle in the external field (\ref{ebapprox}).
A detailed analysis  presented in the supplemental material \cite{supp} shows that for $\xi_0\lessapprox1$ and $w_0\ge70\lambda$, LPWA can be used instead of the exact treatment.

    {\it Coherent Thomson scattering in LPWA}. The problem of the non-linear Thomson scattering  of a monochromatic plane wave on a charged particle is well studied in the literature  \cite{eberly,salamin}; we review here briefly the main results.

We will consider an electron of charge $e_0$ having the initial position ${\bm r}(t_{in})={\boldsymbol{\mathfrak{R}}}_0$ in the focal plane $Oxy$, of cylindrical coordinates ${\mathfrak R}_0\equiv(\varrho_0,\varPhi_0,z=0)$, moving along the trajectory ${\bm r}(t)$ in the field (\ref{ebapprox}). We define the electron trajectory  relative to the initial position of the particle ${\boldsymbol{\mathfrak R}}_0$, ${\bm r}_0(t)={\bm r}(t)-{\boldsymbol{\mathfrak R}}_0$, and the relative observation point ${\bm x}_0={\bm x}-{\boldsymbol{\mathfrak R}}_0$,  whose unity vector is ${\bm n}_0=\frac{{\bm x}_0}{|{\bm x}_0|}$ of polar angles $\theta_0$, $\phi_0$.

In the LPWA the electron is at the initial moment in the plane $Oxy$, having the cylindrical coordinates  $(\varrho=\varrho_0,\varphi=\varPhi_0,z=0)$ and moves under the influence of a local plane wave field (\ref{ebapprox}), of intensity dependent on the radial coordinate $\xi(\varrho_0)$, and initial phase dependent on the initial azimuthal coordinate $\varPhi_0$ of the electron. Using the analytical expression of the trajectory and of the velocity of the particle  we obtain, after a straightforward  calculation \cite{salamin} the Fourier transform of the emitted fields.  The frequency spectrum of the emitted radiation is discrete, consisting of equidistant lines
\begin{align}
    \omega_N=N\frac{\omega_L}{\alpha},\qquad \alpha = 1+\frac{\xi^2(\varrho_0)}4(1-\cos\theta_{0})\label{omegan-1}
\end{align}
the Fourier transform of the  field being expressed as an infinite series of the form
\begin{align}
     & {\bm E}({\bm x},\omega;\varrho_0,\varPhi_0)=\sum\limits_{N=-\infty}^{\infty}\tilde{\bm E}_N({\bm x};\varrho_0,\varPhi_0)\delta(\omega-\omega_N)\nonumber \\
     & {\bm B}({\bm x},\omega;\varrho_0,\varPhi_0)={{\bm n}_0}\times{\bm E}({\bm x},\omega;\varrho_0,\varPhi_0)\label{EBncirc}
\end{align}
where the coefficients $\tilde{\bm E}_N$ can be written in terms of generalized Bessel functions; for the details of calculation and their explicit form see the supplemental material \cite{supp}, Sect. II. C; in the previous equation we indicated explicitly the dependence on the initial coordinates of the electron $\varrho_0$, $\varPhi_0$.  The calculation  is done in the long distance approximation, which requires $|{\bm x}_0|\gg|{\bm r}_0|$, with no restriction on ${\boldsymbol{\mathfrak{R}}}_0$. We can easily check that $\tilde{\bm E}_{-N}({\bm x})=({\tilde{\bm  E}_{N}}({\bm x}))^*$ as it should be, since the electric field is real  ${\bm E}({\bm x},t)\in{\mathbb{R}}^3$, so in the following we can consider only the positive values of the harmonic index, $N\ge1$.

In the second part of our calculation, we consider the coherent emission of radiation by a collection of electrons distributed in the focal plane of the laser; this distribution is modelled as a continuous thin electron sheet, characterized by the uniform surface charge density $\sigma_0$; for simplicity, in the numerical calculation, we choose $\sigma_0=1$ au. Summation over the electrons will be then replaced by integration over all the values of ${\boldsymbol{\mathfrak R}}_0\in Oxy$ plane. We will present results for  the field measured on a screen, orthogonal to the laser propagation direction, at a large distance  $z=Z\gg w_0$ from the focal plane.

The temporal Fourier transform of the total field measured in a point ${\bm x} = (x,y,Z)$ on the screen will be
\begin{align}
  {\bm E}_{tot}({\bm x},\omega) &= \sum\limits_{N=-\infty}^{\infty} \int\limits_{{\boldsymbol{\mathfrak R}}_0} d^2a\tilde{\bm E}_N({\bm x};\varrho_0,\varPhi_0)\delta(\omega-\omega_N)\nonumber\\
  &\equiv\sum\limits_{N=-\infty}^{\infty}\tilde{\bm E}_{N,tot}({\bm x},\omega)\label{int-En}
\end{align}
In the previous equation $\omega_N$, defined in (\ref{omegan-1}) depends on the integration variable ${\boldsymbol{\mathfrak R}}_0$ through $\theta_0$  and the argument $\varrho_0$ of $\xi$. However,   due to the interference effects, at large $Z$ the radiation is confined inside of a cone of opening angle of the order of magnitude $\sim \frac{\lambda}{w_0}$ which means that scattering angles appearing in the  frequency formula are very small  $\theta_0\sim10^{-2}$. Taking into account also the fact the $\xi(\varrho_0)\le1$ we can see that when the integration variable ${\boldsymbol{\mathfrak R}}_0$ changes in the focal plane of the laser, the values of $\omega_N$  corresponding to a fixed $N$ have only very small variations
\begin{align}
    \omega_N(\varrho_0,\Phi_0) =N\omega_L(1+\varepsilon(\varrho_0,\Phi_0)),\qquad \varepsilon(\varrho_0,\Phi_0)\ll1.
\end{align}
Then we define the $N$th Fourier component of the total  field as an integral over a small interval $\Delta\omega\ll \omega_L$
\begin{align}
    {\bm E}_{N,tot}({\bm x})\equiv & \int\limits_{N\omega_L-\Delta/2}^{N\omega_L+\Delta/2} d\omega \tilde{\bm E}_{N,tot}({\bm x},\omega)\nonumber            \\
                                   & = \int\limits_0^{\varrho_m}d\varrho_0\varrho_0\int\limits_0^{2\pi}d\varPhi_0 \tilde{\bm E}_N({\bm x};\varrho_0,\varPhi_0)\label{intPhi0}    \nonumber \\
    {\bm B}_{N,tot}({\bm x})\equiv & \frac1{c}\int\limits_0^{\varrho_m}d\varrho_0\varrho_0\int\limits_0^{2\pi}d\varPhi_0 ({\bm n}_0\times\tilde{\bm E}_N({\bm x};\varrho_0,\varPhi_0))
\end{align}

If the incident field is circularly polarized $\zeta_x=\pm\zeta_y$ the calculation presented in detail in \cite{supp}, Sect. II. C shows that the field recorded in a point of cylindrical coordinated $\varrho$, $\varPhi$ on the  screen at the distance $Z\gg w_0$ from the focal spot can be written as
\begin{align}
    \tilde{\boldsymbol{E}}_{N, t o t}(\boldsymbol{x})=e^{i\left(m_L N +\epsilon_L(N-1) \right)\varPhi} \int\limits_0^{\rho_m} d \varrho_0 \varrho_0 \boldsymbol{h}\left(\varrho, Z ; \varrho_0\right)\label{phiex}
\end{align}
where $\epsilon_L$ is the handedness of the circular polarization $\epsilon_L=\frac{\zeta_x}{\zeta_y}=\pm1$;
one can see that the field  has a spatial structure of helical shape  with a phase index
\begin{align}
    q = m_L N +\epsilon_L(N-1)
\end{align}
dependent on the harmonic index and on the helical index of the incident beam. This structure should be observable experimentally, since the frequency of each harmonic is distinct.

    {\it Results and discussion}. In this section we present numerical results for the scattered field for different parameters of the laser beam, in all cases we shall take the dimensionless intensity parameter $\xi_0 =0.1$, the beam waist parameter $w_0=75\lambda$, the laser wavelength $\lambda=800$ nm, which corresponds to a frequency $\omega_L=0.057$ au and the helicity $\epsilon_L=1$.

We start by analyzing the Poynting vector of the emitted radiation.
\begin{figure}[h]
    \includegraphics[scale=0.15]{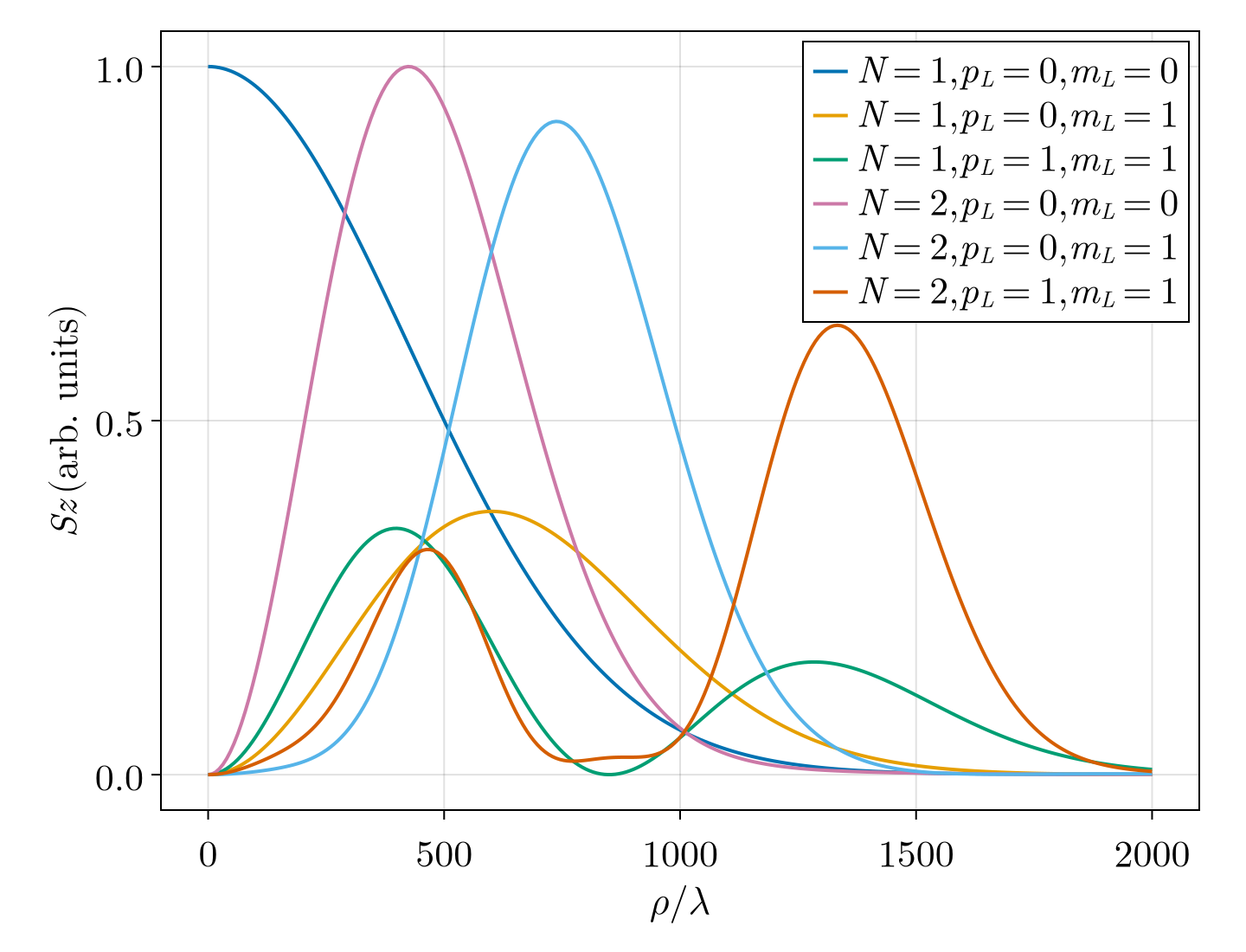}
    \caption{The $Oz$ component of the Poynting vector ${\bm S}$ measured in the plane $Z=2\times 10^5\lambda$ ($\sim$16cm) for $N=1,2$, and for different mode indices $p_L$, $m_L$ (as written in the legend). For each $N$, the values of $S_z$ were scaled by the factor $s=\max(S_z)(p_L=m_L=0)$ (the maximum of $S_z$ for the same $N$ and $p_L=m_L=0$). The scaling factors are $s=0.1$ au ($N=1$), and $s=1.7\times10^{-9}$ au ($N=2$).  .
        \label{fig1main}}
\end{figure}
In Fig \ref{fig1main} we presented the component $S_z$ measured on a screen  at the distance $Z=2\times10^5\lambda$ from the focal plane, for three combinations of values for the indices $p_L, m_L$: $(0,0)$, $(0,1)$, and $(1,1)$, for the first and the second harmonic. In all cases the the shape of $S_z$ distribution has cylindrical symmetry in the screen plane, so it is enough to represent it as a function of the cylindrical coordinate $\varrho$.  As the distance $Z$ is much larger than the laser wavelength, due to the interference effects between the electromagnetic field emitted from different points in the focal plane, the radiation is emitted into a cone of very small opening angle (similar to the Fraunhofer diffraction). The opening angle can be estimated from the ratio of the radial extension of the distribution measured on the screen to the value of $Z$. We obtain in all cases $\Theta_0\sim\frac1{100}$,  in agreement with the ratio of the laser wavelength to the beam diameter.  The interference pattern consists of concentrical rings whose position and size depend on $N,p_L, m_L$; the only case when a central maximum is present is for the fundamental harmonic and $p_L=m_L=0$.

We can see very large differences between the numerical values of the fundamental and the second harmonic (see the values of the scaling factors $s$ defined in the figure caption); they are due to two factors: one the one hand,  the electric field depends on the the "local scattering angle"  $\theta_0$ as ${\bm E}_n\sim\theta_0^{N-1}$  and  as the scattering takes place mostly in the forward direction, $\theta_0$ will be of the order of $\Theta_0\approx10^{-2}$. On the other hand, the emitted field $E_N$ depends also on the dimensionless parameter $\xi_0$ as $E_N\sim\xi_0^N$ which introduces an additional factor of $10^{2}$ between $S_z(N=1)$ and $S_z(N=2)$

Besides the cylindrical symmetry, the Poynting vector obeys also a scaling law  with respect to $\varrho$ and $Z$. We have checked that for $Z\ge10^5\lambda$ the relation
\begin{align}
    Z^2 S_z(\varrho, Z) = \text{fct}(\frac{\varrho}{Z})\label{scalingZ}
\end{align}
is obeyed with very good accuracy, which means that at large distance from the focal plane the shape of the energy distribution of the emitted radiation  propagates undistorted inside a cone of given opening angle.

However, as we will see next,  the electric and  magnetic components of the field have more complex properties. The following two figures present the electric field and its phase, for the fundamental Gaussian mode of the incident field ($p_L=m_L=0$) and for a Laguerre - Gauss incident beam ($m_L\ne0$).
\begin{figure*}[h]
    \includegraphics[scale=0.1]{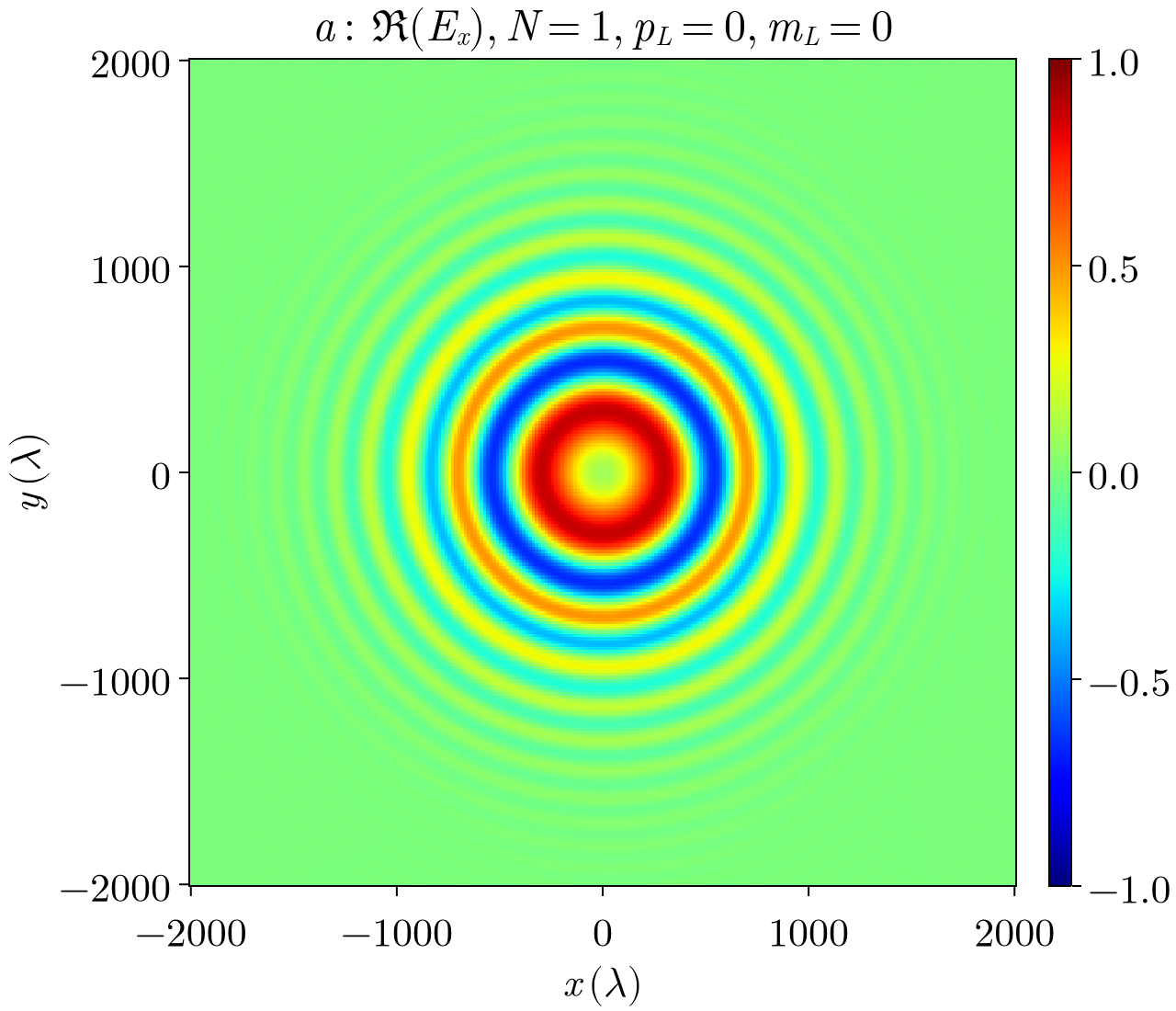}\includegraphics[scale=0.1]{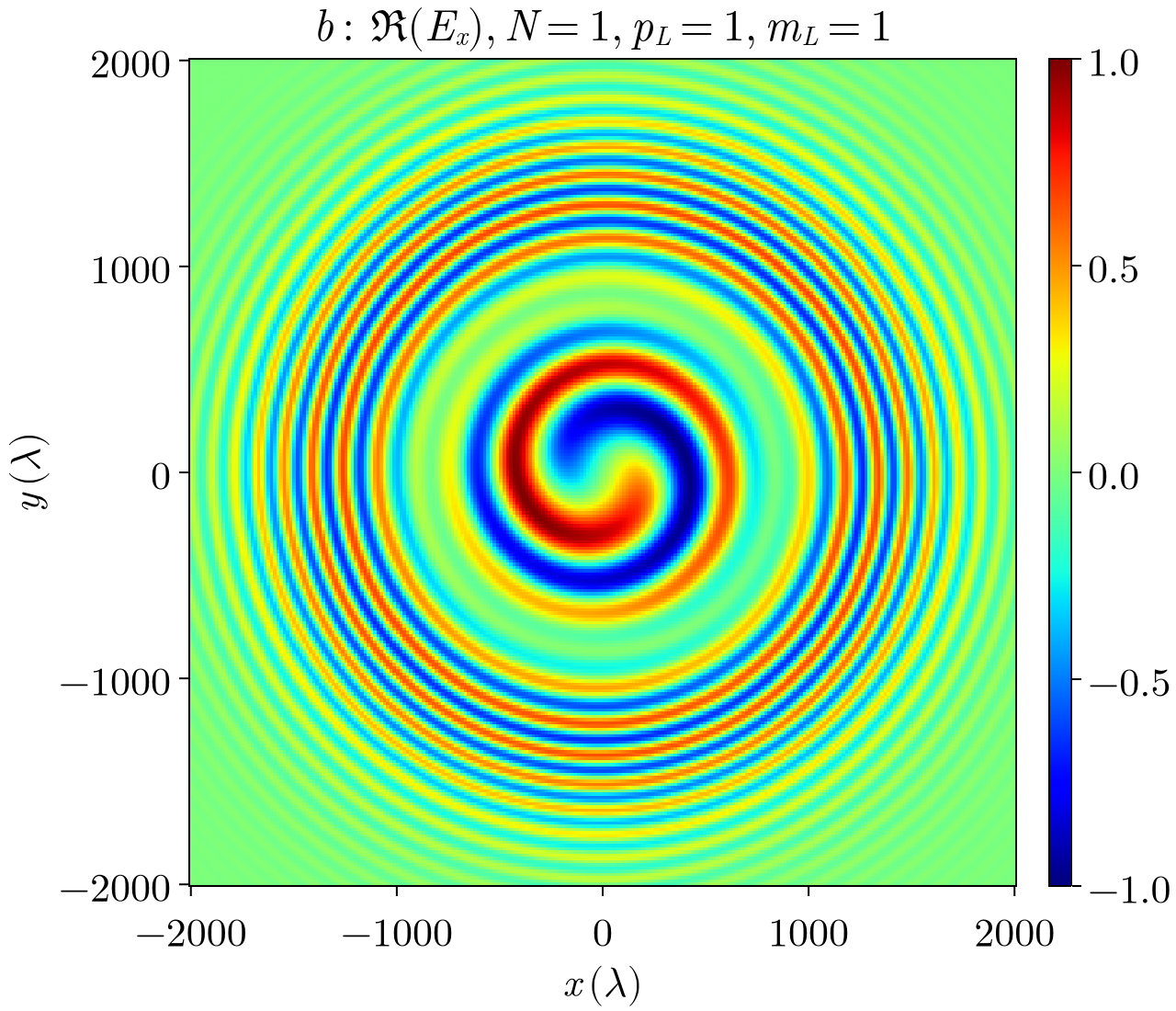}\includegraphics[scale=0.1]{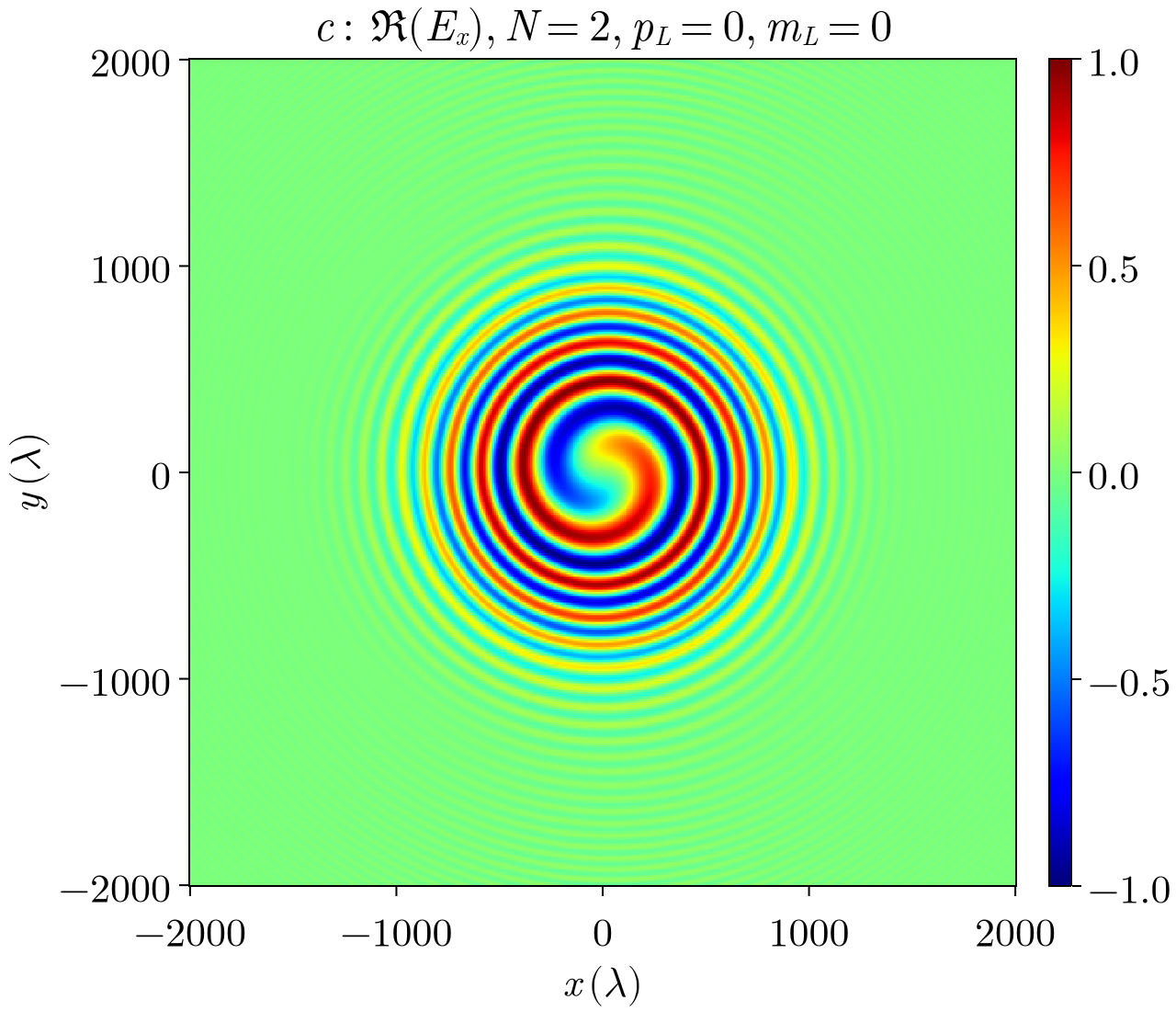}

    \includegraphics[scale=0.1]{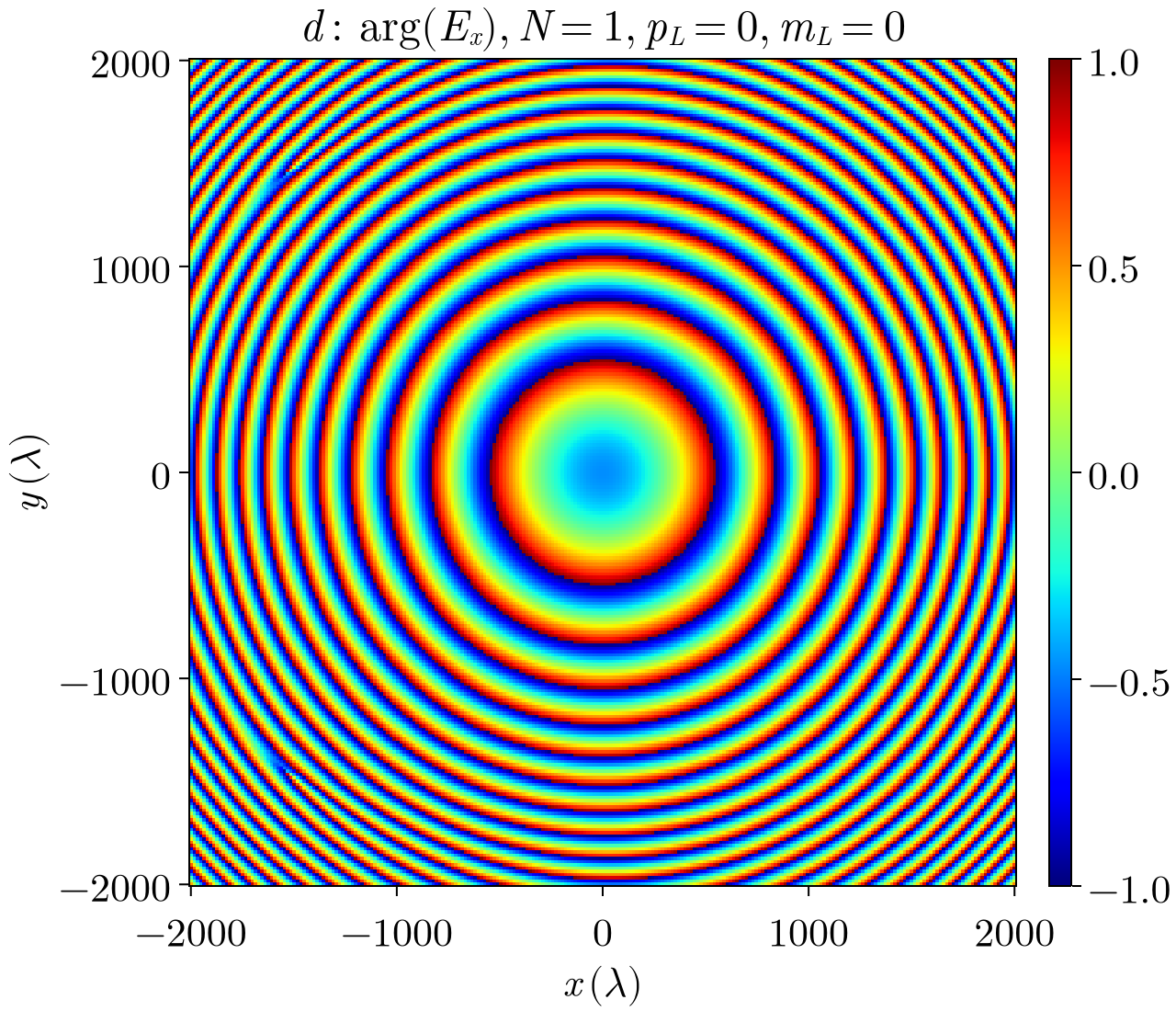}\includegraphics[scale=0.1]{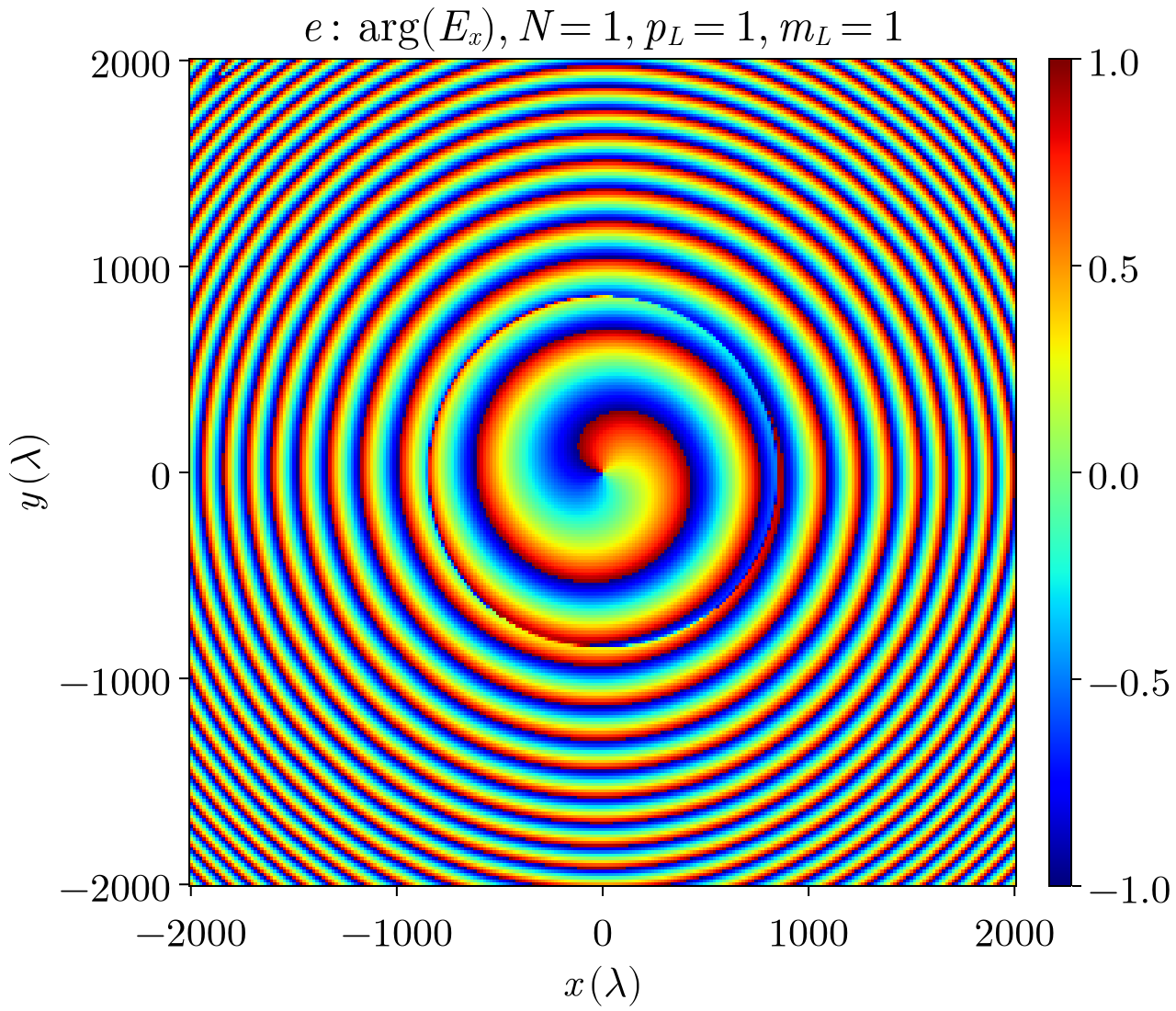}\includegraphics[scale=0.1]{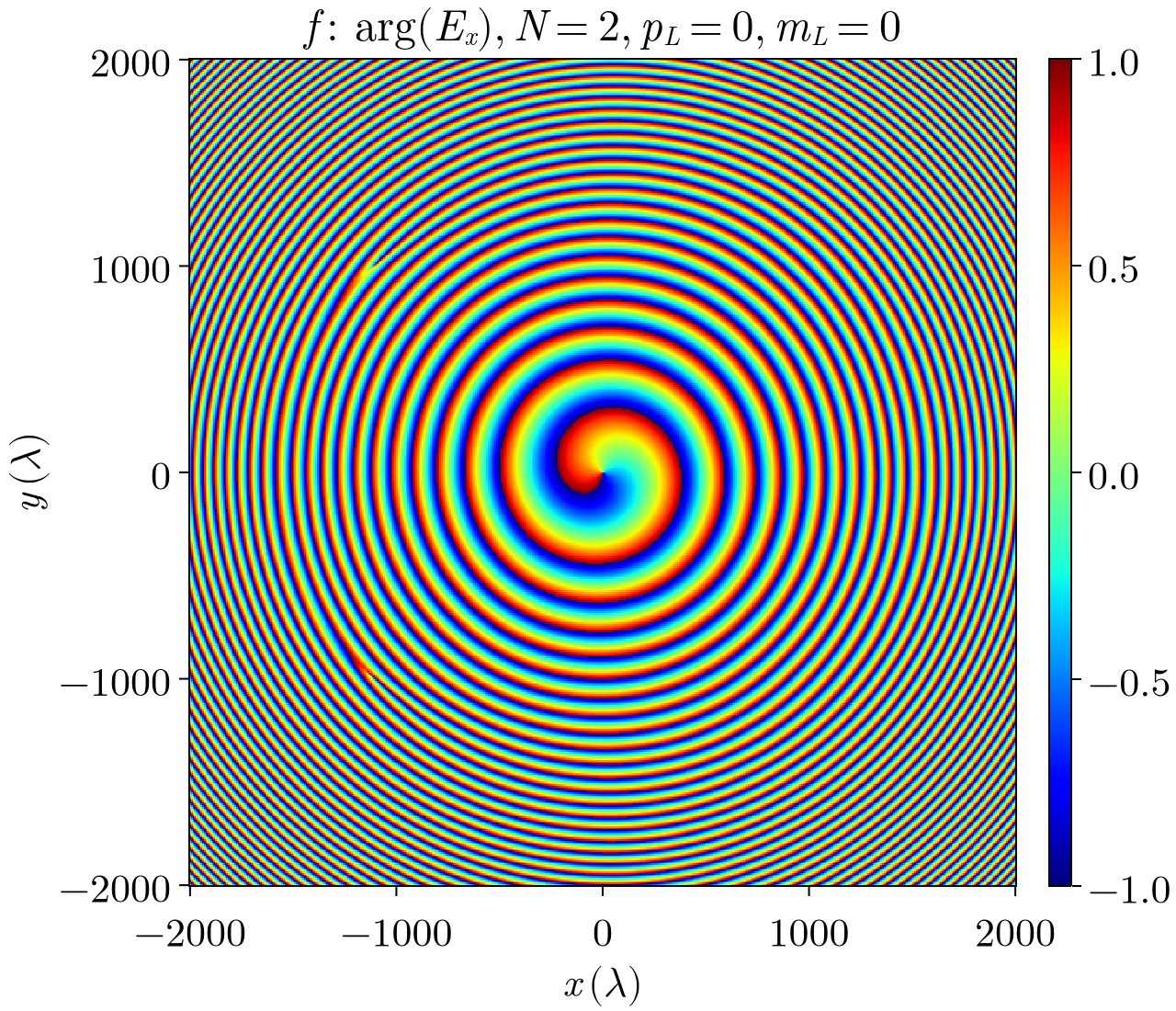}

    \caption{The $Ox$ component of the scattered field measured on a screen at $Z=2\times10^5\lambda$: the real part of the electric field scaled to $(-1,1)$ (the upper row), and its phase in units of $\pi$ (the lower row). The laser field is circularly polarized with $\epsilon_L=1$ and $\xi_0=0.1$. (a)  $\Re(E_{x})$ for $N=1$, $p_L=m_L=0$, (scaling factor $0.052$ au), (b)  $\Re(E_{x})$ for $N=1$ $p_L=m_L=1$ (scaling factor $s=$0.039 au); (c) $\Re(E_{x})$ for $N=2$, $m_L=p_L=0$ (scaling factor $s=8.9\times10^{-6}$ au); (d), (e), (f) the phase of $E_{x}$ in units of $\pi$ for  the same parameters as (a), respectively (b) and (c). \label{fig3amain}}
\end{figure*}

In Fig. \ref{fig3amain} we present the $Ox$ component of the emitted electric field for $p_L=m_L=0$,  $N=1$  ($a$, $d$) and $N=2$ ($c$, $f$) and for $p_L=m_L=1$, $N=1$ ($b$, $e$). In all cases the distance $Z$ is $2\times10^5\lambda$. The upper row shows the real part of $E_x$, and the lower row, its phase.  For  $N=1$  the emitted field is essentially identical to the incident one: for $m_L=0$ there is no azimuthal dependence of the electric field and for $m=1$, it has a helical structure $E_x\sim e^{i\varPhi}$. A similar azimuthal structure of the emitted field is formed also for $m_L=0$  $N=2$ ($b$ and $e$). In all cases the dependence of $\varPhi$ is in agreement with Eq. (\ref{phiex}); we thus obtained the same shape of the helical wavefront in two different scenarios: the fundamental of the helical mode $m_L=1$, or the second harmonic of the fundamental Gaussian mode. It is also in good agreement with the experimental result presented by Hemsing {\it et al} \cite{hems}; in their work, the field was obtained by scattering of a Gaussian mode on a helically modulated electron beam, while in out case the modulation is done by the incident field itself.

The radial dependence of the electric field is different in the three cases, (see also the energy distribution  for the same parameters in the previous figure) and the numerical values are much lower for the second harmonic than the fundamental.

\begin{figure*}[h]
    \includegraphics[scale=0.1]{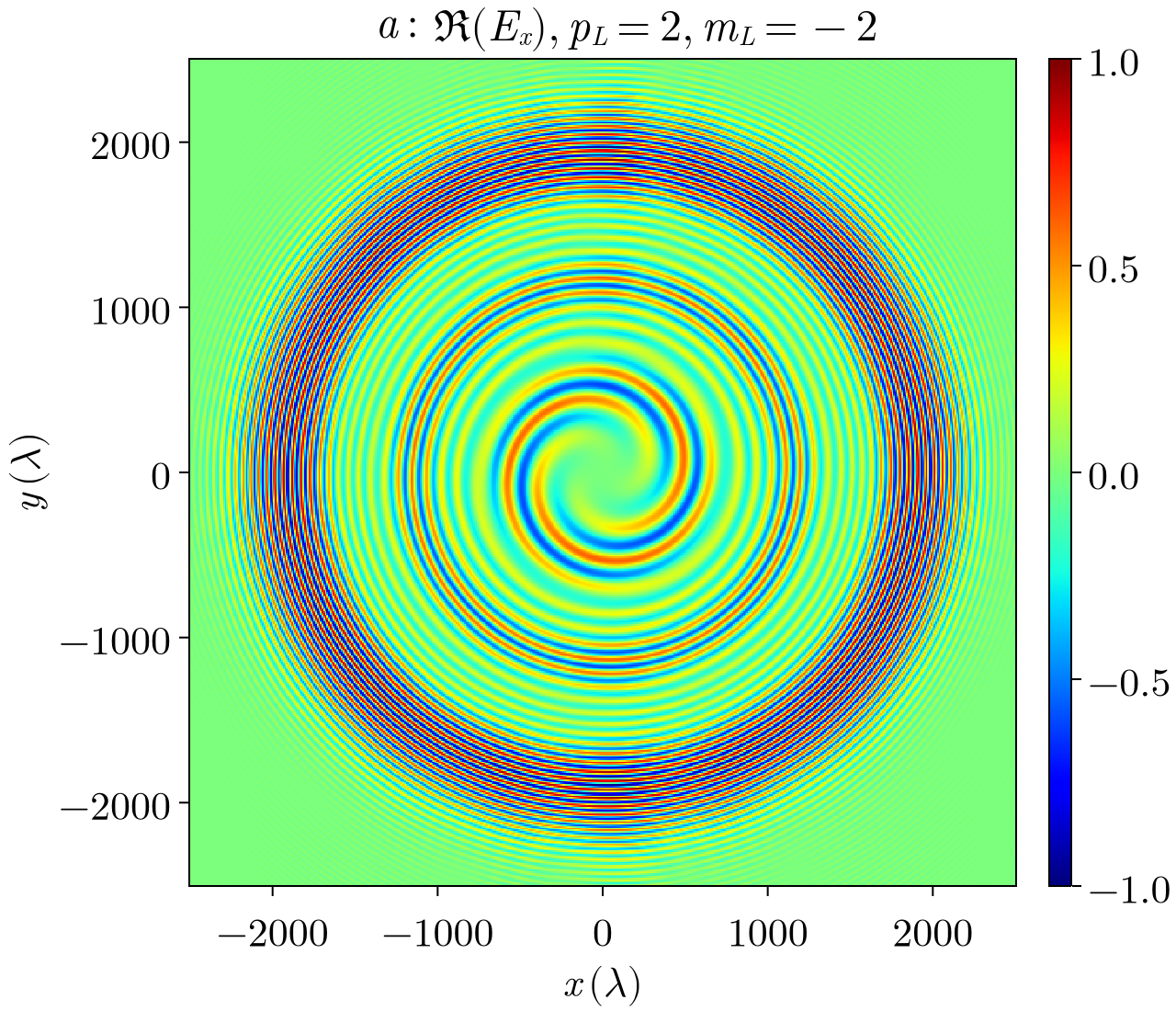}\includegraphics[scale=0.1]{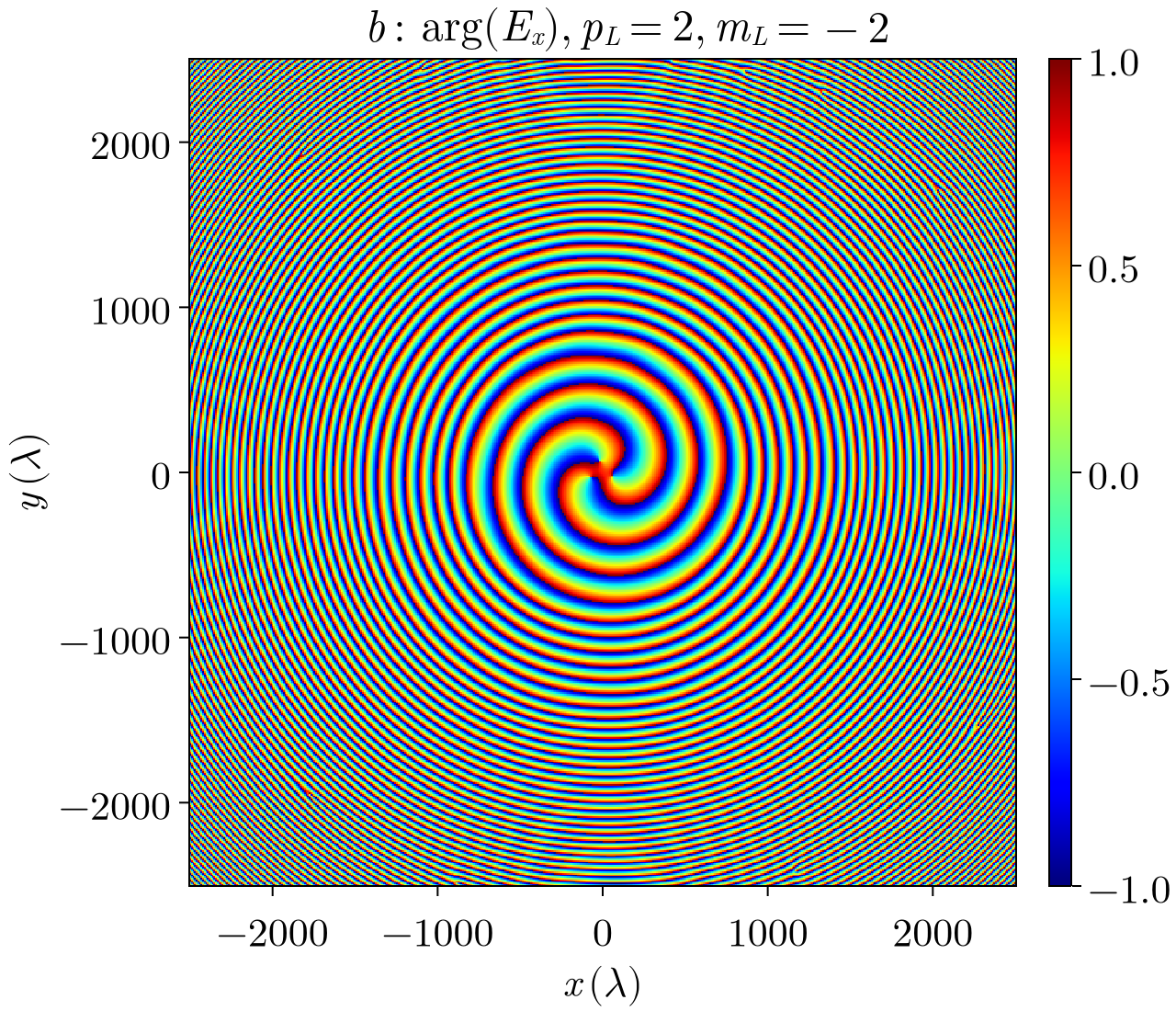}\includegraphics[scale=0.1]{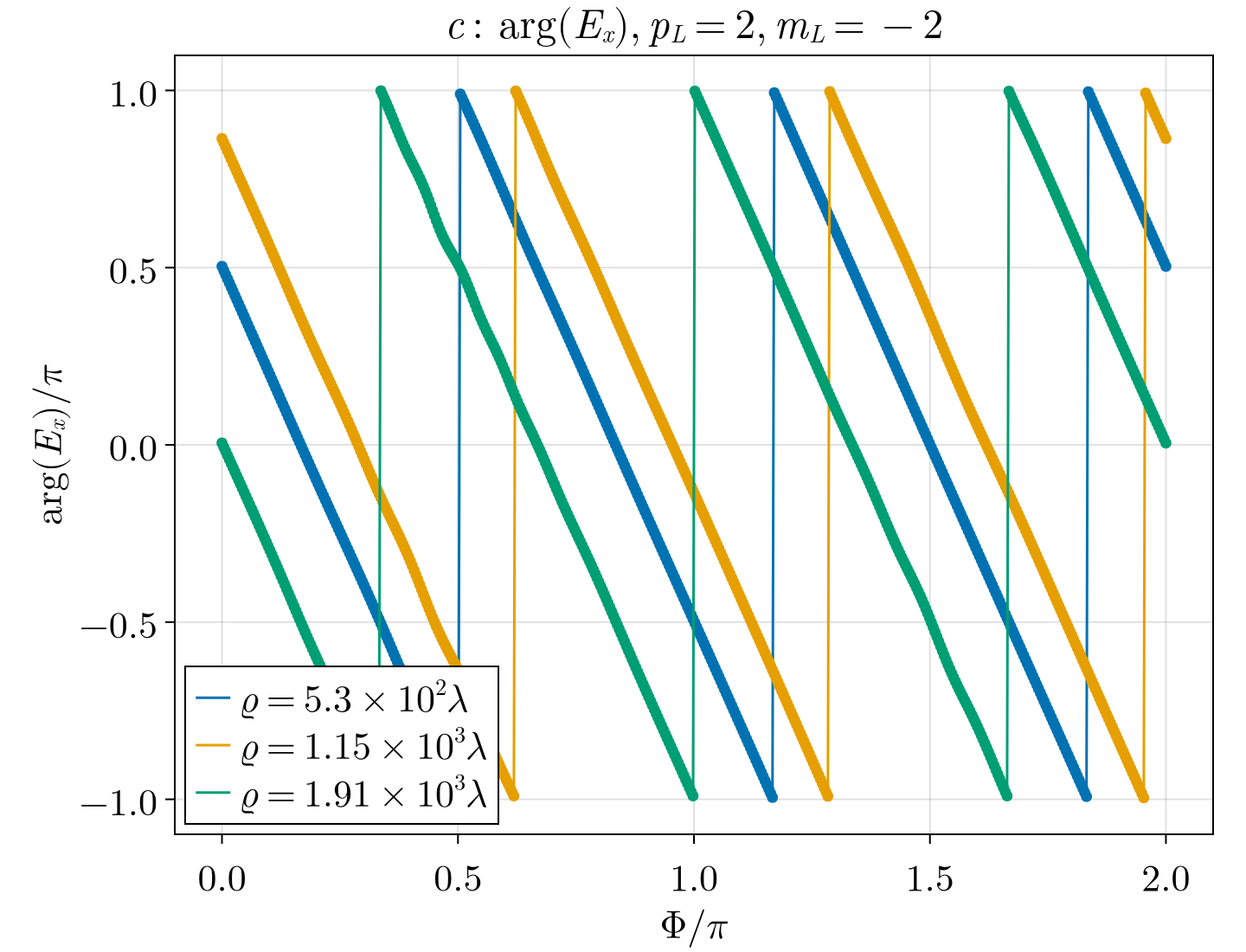}

    \includegraphics[scale=0.1]{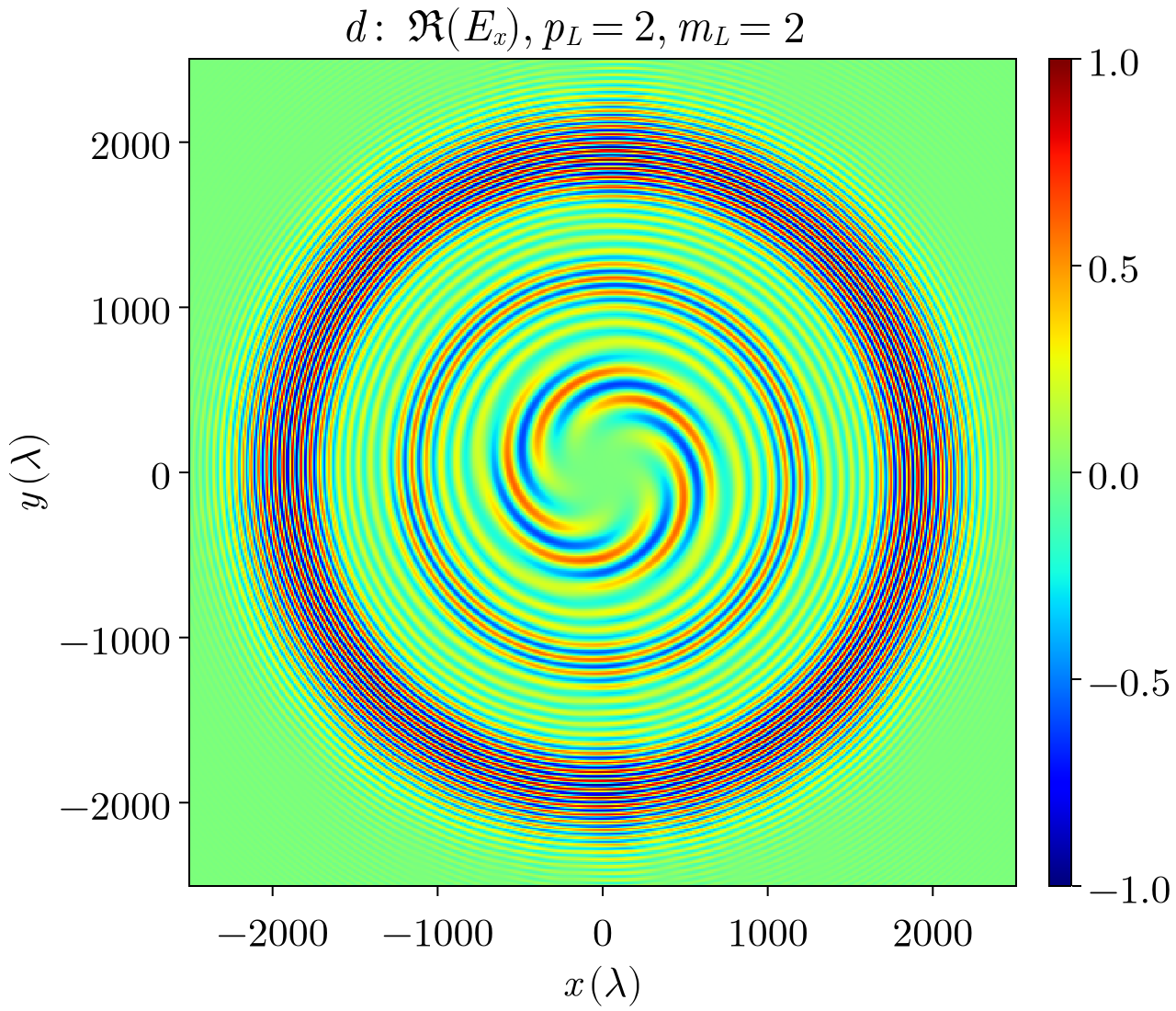}\includegraphics[scale=0.1]{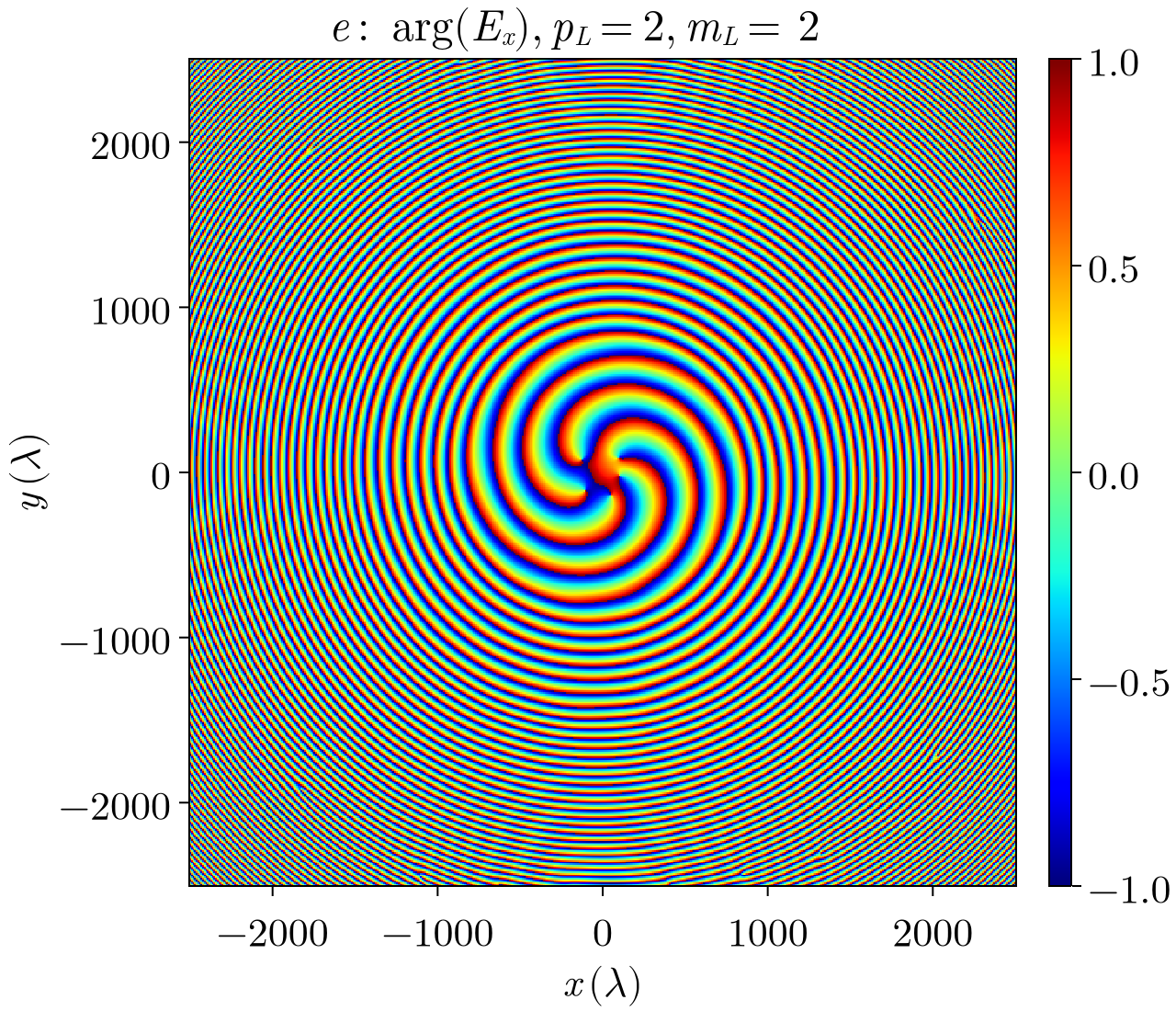}\includegraphics[scale=0.1]{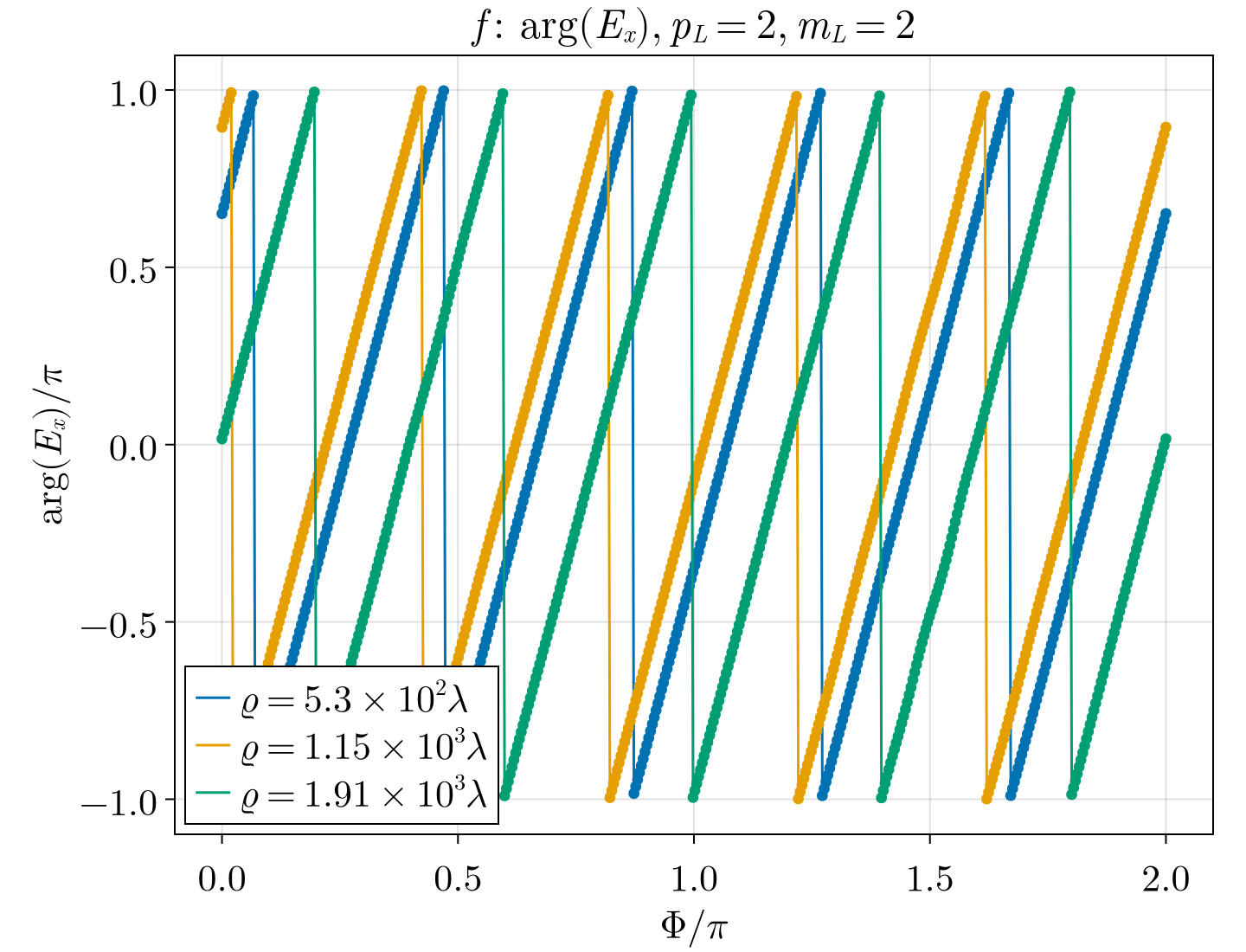}

    \caption{The scattered field for $Z=2\times 10^5\lambda$, $N=2$, $p_L=2$.  $(a)$, $(d)$:  the real part of $E_{x}$  for $m_L=2$, respectivelly $m_L=-2$ (scaling factor $1.0\times10^{-9}$ au), $(b)$, $(e)$: the phase of $E_{x}$ in units of $\pi$, $(c)$, $(f)$: the phase of $E_{x}$ as a function of the azimuthal angle $\varPhi$ in the observation plane for three fixed values $\varrho$ written in  the legend, in units of $\pi$. \label{fig4main}}
\end{figure*}

The results are similar for the case $N=2$, $p_L=2$, $m_L=\pm2$ presented in Fig. \ref{fig4main} showing the real part of the  component $E_x$ of the electric field (panels (a) and (d)) and the phase of the $E_x$ (panels (b) and (e)), at the same distance $Z=2\times10^5\lambda$. The Poynting vector has a structure of concentric rings similar to those in Fig. \ref{fig1main} and is not represented.
In the graphics of the phase of $E_x$ there are two distinct regions: an inner region of radius $\varrho_1\sim400\lambda$ in which the phase changes slowly with $\varrho$, and the outer region in which the phase changes very fast with $\varrho$, giving rise to a spiral-like figure, in agreement with the behavior predicted in the supplemental material \cite{supp}, and with the experimental observation presented in \cite{hems}.
Panels (c) and (f) contain graphics of the phase of $E_x$ as a function of the angle $\varPhi$ in the observation plane for three values of $\varrho$, chosen to correspond to the radii of the concentric rings in which the radiation is emitted $\varrho = 530\lambda$, $1150\lambda$ and $1910\lambda$. In all cases we can see the very good agreement with the behavior predicted by Eq. (\ref{phiex}); even in the outer region, where  $\text{arg}(E_x)$  changes very fast with $\varrho$, for fixed $\varrho$ the phase depends linearly on $\varPhi$, with the coefficient
\begin{align}
    q = Nm_L+(N-1)\epsilon_L\label{defq}=\left\{\begin{array}{ll}-3,&m_L=-2\\5,&m_L=2\end{array}\right.
\end{align}
We have also checked that in all cases     $\text{arg}E_x=\frac{\pi}2-\text{arg}E_y$, 
which confirms that the emitted radiation is circularly polarized.

{\it Conclusion}. In conclusion, we have demonstrated that the coherent scattering of a monochromatic field described by a  Laguerre-Gauss mode  by a 2D continuos electron distribution  will generate  structured helical light. Our approach is based on a local plane wave approximation whose validity is estimated theoretically and tested numerically (see the supplemental material \cite{supp}); the  model is valid for incident fields having the dimensionless intensity parameter up to $\xi_0\lessapprox1$ and the beam waist $w_0\ge75\lambda$.
Our numerical calculations show that due to the interference effects the emitted field is concentrated into the forward direction, similar to the Fraunhofer diffraction, if the incident field is  circularly polarized, each harmonic has a helical spatial structure, with a given helical index $q$ defined in (\ref{phiex}); we can expect it to carry angular momentum proportional to the new helical index

A further test of our results will be to calculate the angular momentum of the scattered radiation in order to check the correlation with the spatial structure of the electric field described in the present paper.

    {\it Acknowledgments}.    P.-V. Toma acknowledges the financial support from the     Romanian Ministry of Education consisting in a doctoral     scholarship received through the Doctoral School of Physics
of the University of Bucharest. This paper was supported     by Council for Doctoral Studies (C.S.U.D.), University of Bucharest. P.-V. Toma was also funded from the Romanian Ministry of Education and Research under the Romanian National Nucleus Program LAPLAS VI, contract no.
30N/2023.

\bibliography{main}
\end{document}


\title{Supplemental material for \\  "Coherent Thomson scattering of Laguerre - Gauss beams on an electron sheet"}

\author{Petru-Vlad Toma \orcidlink{0009-0002-9949-7266}}
\affiliation{Center for Advanced Laser Technologies,
     National Institute for Laser, Plasma and Radiation Physics, Magurele, Romania}
\affiliation{University of Bucharest, Department of Physics, Magurele, Romania}
\author{Andrei Cristian Opinca}
\affiliation{University of Bucharest, Department of Physics, Magurele, Romania}
\author{Virgil Baran \orcidlink{0009-0000-5585-9443}}
\affiliation{University of Bucharest, Department of Physics, Magurele, Romania}
\affiliation{Academy of Romanian Scientists, Strada Ilfov 3, Sector 5, 050044, Bucharest, Romania}
\author{Madalina Boca \orcidlink{0009-0001-5437-541X}}\email{madalina.boca@unibuc.ro}
\affiliation{University of Bucharest, Department of Physics, Magurele, Romania}
\date{\today}

\maketitle

\tableofcontents

\section{The electromagnetic field and the local plane wave approximation}

\subsection{The Laguerre-Gauss modes of the electromagnetic field}\label{s1a}
In the present work we consider the coherent scattering of a Laguerre - Gauss mode of the  monochromatic electromagnetic field of frequency $\omega_L$ on a  system of electrons initially at rest, distributed uniformly in a disk of radius $\varrho_m$, at least equal to the beam radius, placed in the focal plane of the laser. The incident electric  field  of the beam propagating along the $Oz$ direction of the reference frame can be written in the paraxial approximation as \cite{10.1063/5.0150590}
\begin{align}
      & E_{x}({\bm r},t) = E_0\Re\left\{\zeta_{x} u_{p_Lm_L}({\bm r})f(\varphi)\right\},\quad E_{y}({\bm r},t) = E_0\Re\left\{i\zeta_{y} u_{p_Lm_L}({\bm r})f(\varphi)\right\},\quad f(\varphi)=e^{-i\varphi },\quad \varphi = \omega_L t-k_Lz                                      \nonumber \\
      & E_z({\bm r},t) = E_0\Re\left\{-\frac1{ik}\left[\zeta_x\frac{\partial u_{p_Lm_L}({\bm r})}{\partial x}+i\zeta_y\frac{\partial u_{p_Lm_L}({\bm r})}{\partial y}\right]f(\varphi)\right\}\label{eexact}
\end{align}
In the expressions above the components in the plane orthogonal on the propagation direction, $E_{x,y}$ are written in terms of the Laguerre-Gauss solution of the Helmholtz equation in the paraxial approximation, $u_{p_Lm_L}({\bm r})e^{ik_Lz}$ with $k_L=\frac{\omega_L}c$, the polarization being fixed by the two real constants $\zeta_x$, $\zeta_y$ chosen such that $\zeta_x^2+\zeta_y^2=1$; for example, $\zeta_x=\pm \zeta_y=\frac1{\sqrt{2}}$ corresponds to circular polarization and $\zeta_x=1$, $\zeta_y=0$ describe the linear polarization along $Ox$.  The amplitude of the electric field is conveniently expressed in terms of a dimensionless parameter $\xi_0$
\begin{align}
     E_0 = \frac{\omega_L{ m_e c \xi_0}}{|e_0|}\label{defxi}
\end{align}
with $m_e$ and $e_0$ the electron mass and charge, and $c$ the velocity of light in vacuum. The component of ${\bm E}$ along the propagation direction, $E_z$ is calculated  form the condition that ${\boldsymbol{\nabla}}\cdot{\bm E} =0$, using, in the paraxial approximation ${\partial_zE_z\approx ik_L E_z}$. Similarly, the magnetic field is calculated from the Maxwell equations
\begin{align}
     \partial_t {\bm B} =-{\boldsymbol{\nabla}}\times{\bm E} ,\qquad {\boldsymbol{{\nabla}}}\cdot{\bm B}=0,
\end{align}
using $\partial _t{\bm B} =-i\omega_L{\bm B}$ and again the paraxial approximation, with the result
\begin{align}
      & B_x({\bm r},t) = \frac{E_0}c\Re\left\{-i\zeta_y u_{p_Lm_L}({\bm r})f(\varphi)\right\},\quad  B_y({\bm r},t) = \frac{E_0}c\Re\left\{\zeta_x u_{p_Lm_L}({\bm r})f(\varphi)\right\}      \nonumber               \\
      & B_z({\bm r},t) = \frac{E_0}c\Re\left\{-\frac1{ik}\left[-i\zeta_y\frac{\partial u_{p_Lm_L}({\bm r})}{\partial x}+\zeta_x\frac{\partial u_{p_Lm_L}({\bm r})}{\partial y}\right]f(\varphi)\right\}\label{bexact}
\end{align}
The explicit expression of $u_{p_Lm_L}({\bm r})$ is, in cylindrical coordinates $\varrho_0,\varPhi_0,z$,
\begin{align}
     u_{p_Lm_L}(\varrho_0, \varPhi_0, z) \equiv & v_{p_Lm_L}(\varrho_0,  z)e^{im_L\Phi_0}=\nonumber                                                                                                          \\
     u_{p_Lm_L}(\varrho_0, \varPhi_0, z) =      & N_{p_L|m_L|}\frac{w_0}{w(z)}\left(\frac{\sqrt{2}}{w(z)}\right)^{|m_L|}e^{-i(2p_L + |m_L| +1)\psi_G(z)}e^{-\frac{k}2\frac{\varrho_0^2}{z_R+i z}}\,\nonumber \\
                                                & \times{}_1F_1(-p_L, |m_L|+1, \frac{2\varrho_0^2}{w(z)^2}) \varrho_0^{|m_L|} e^{im_L\varPhi_0},
     \qquad p_L\in{\mathbb{N}},\qquad m_L\in{\mathbb{Z}}\label{def-upm}
\end{align}
where    $w(z)=w_0\sqrt{1+\frac{z^2}{z_R^2}}$, $w_0$ being the beam waist, $z_R$ is the Rayleigh radius, $z_R = \frac{kw_0^2}{2}$, and $\psi_G$ the Gouy phase $\Psi_g = \arctan\frac{z}{z_R}$.
The normalization constant is
\begin{align}
     N_{p_Lm_L} = \frac{\sqrt{2}}{|m_L|!}\sqrt{\frac{(p_L+|m_L|)!}{p_L!}}
\end{align}
chosen such that the "nominal intensity" of the beam  ${\cal I}_0$ (the total power ${\cal P}$ of the beam divided by $\pi w_0^2$ ) to be ${\cal I}_0=\frac{\epsilon_0 c E_0^2}2$ \cite{don}.

\begin{figure}[H]
     \includegraphics[scale=0.08]{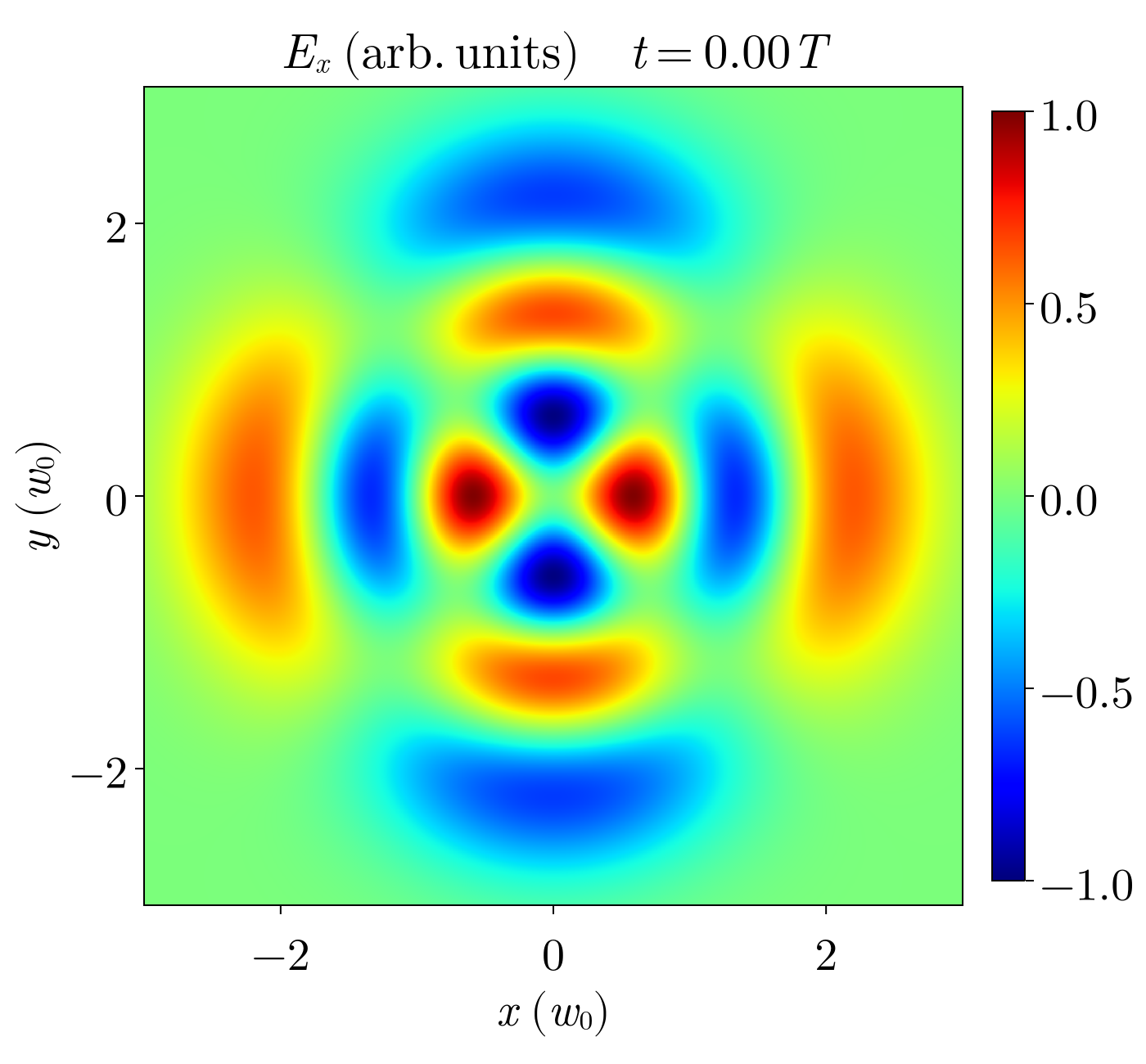} \includegraphics[scale=0.08]{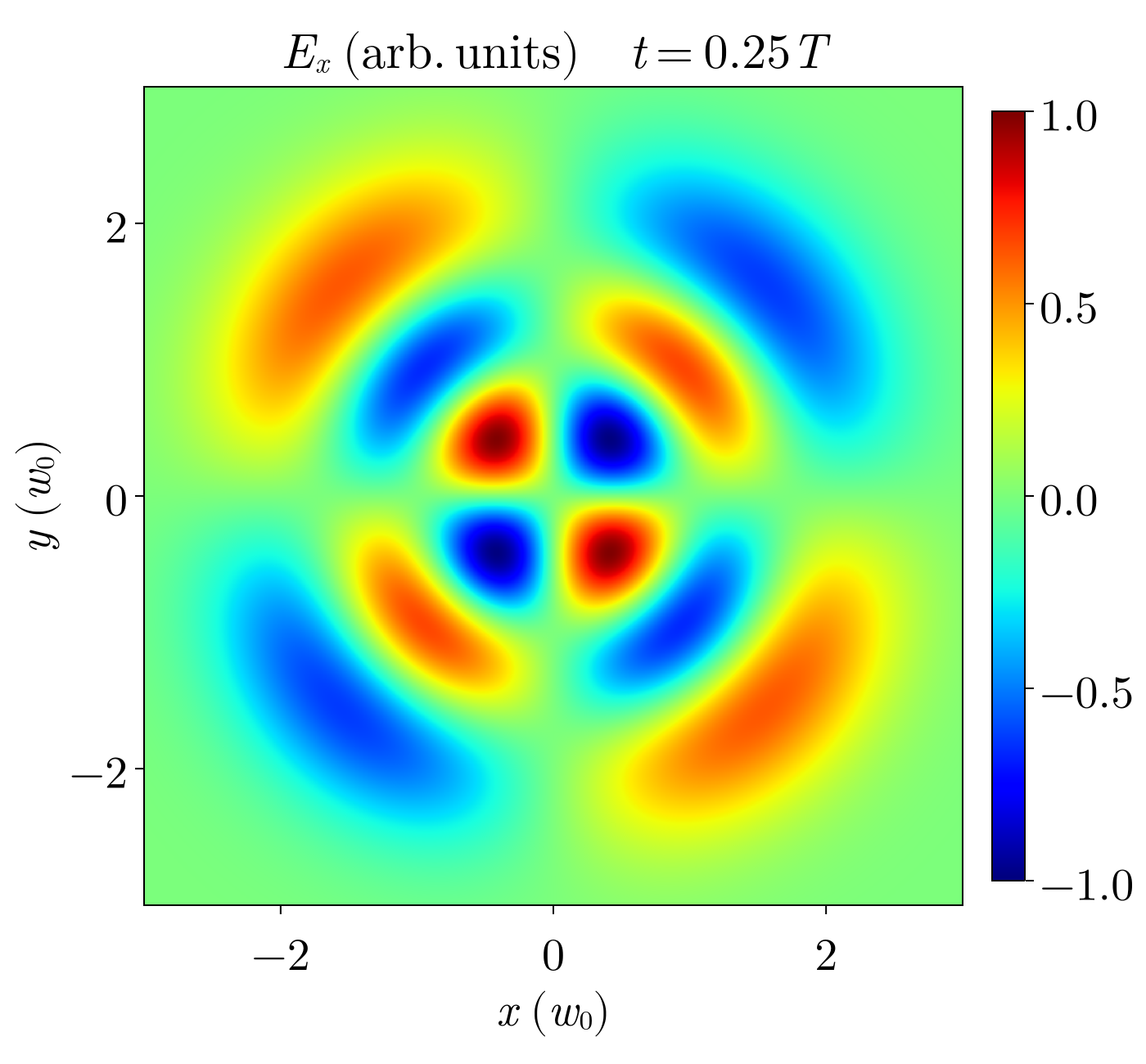} \includegraphics[scale=0.08]{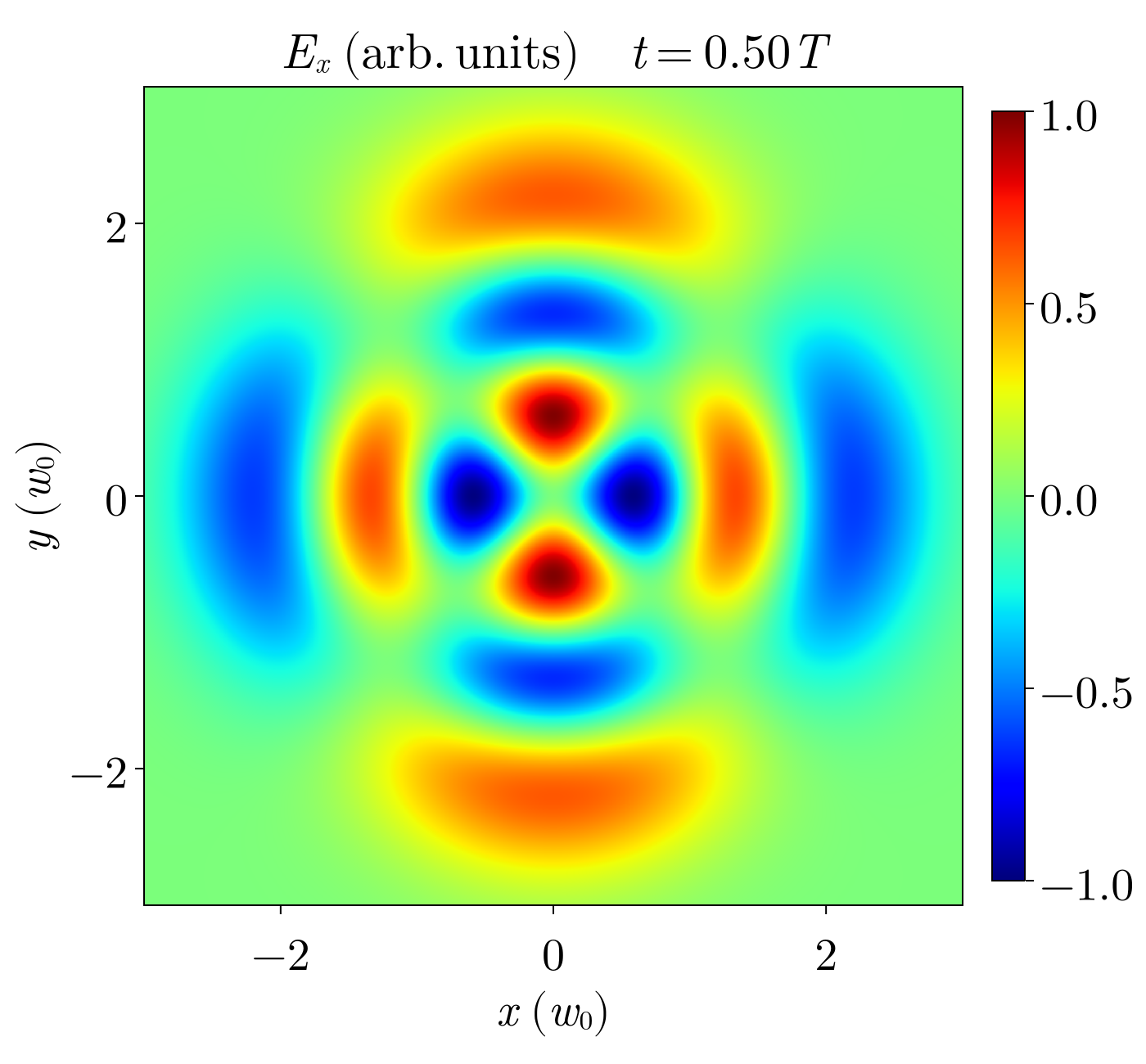} \includegraphics[scale=0.08]{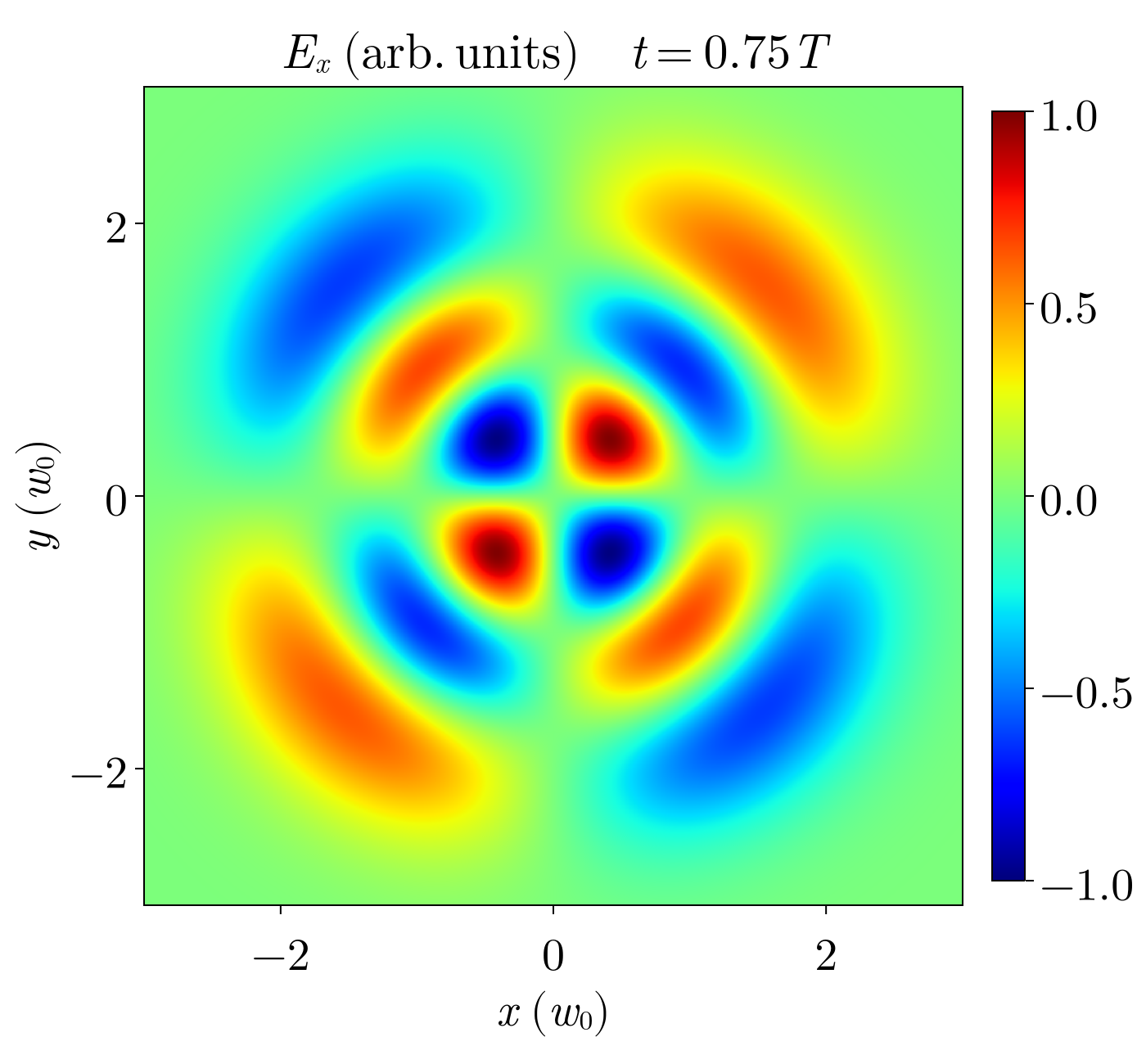}
     \caption{The $E_x$ component of the electromagnetic field represented in the focal plane $z=0$ at $t=0$, $t=0.25T$, $t=0.5T$, $t=0.75T$.\label{plot-field}}
\end{figure}

In Fig \ref{plot-field} below we present  the component $E_x$ of the field,  in the focal plane of the laser ($z=0$), at the moments $t=0$, $t=T/4$, $t=T/2$, and $t=3T/4$, $T$ being the period of the field $T=2\pi/\omega_L$. The parameters of the LG mode are $p_L=m_L=2$, and the polarization is circular $\zeta_x=\zeta_y=\frac1{\sqrt{2}}$. The magnitude of the field, here in arbitrary units, is represented by color, according to the color scale  next to each graph, and the units on the $x$, $y$ axes are $w_0$. We can see the apparent "rotation" of the spatial structure  due to the combination of the spatial and temporal phase of the field
\begin{align}
     {\bm E}\sim e^{i m_L\varPhi_0-i\omega_L t},\qquad \varPhi_0=\arctan(y/x).
\end{align}

\subsection{The local plane wave approximation (LPWA)}\label{sectlpwa}

We shall study the scattering of a monochromatic  LG beam (assumed to be adiabatically coupled in the very distant past) on a statistical ensemble of electrons, initially at rest, positioned in the focal plane of the laser. The laser wavelength that we will consider is $\lambda=800$ nm, corresponding to $\omega_L=0.057$ au and the beam waist $w_0$ will be taken $w_0=75 \lambda$, well within the validity of the paraxial approximation (see e.g. \cite{siegman}, Cap. 7). We notice that for this choice the variation of the field in the transversal plane is very slow and the  field can be considered as constant along distances of the order of the wavelength $\lambda$ in the $Oxy$ plane.

We will show here that for beam waists much larger that $\lambda$ and for the dimensionless intensity parameter $\xi_0\lessapprox1$ we can use an approximation based on the assumption that each electron, interacting with the laser beam feels the electromagnetic field as a plane wave of amplitude and initial phase determined by the electron initial position.

Considering the expression of the trajectory of a charged particle of mass $m_e$ and electric charge $e_0$, initially at rest in a plane wave field , described in the Coulomb gauge by the vector potential ${\bm A}(\phi)$ (see, for example  \cite{Boca2011ThomsonAC}, Eq. (A1))
\begin{align}  z(t)= & \frac{c}{2\omega_L}
               \int_{\varphi_0}^\varphi d \chi \frac{e_0^2
                    \bm{A}^2(\chi)}{(m_ec)^2}  ,\qquad  \bm{r}_{\perp}(t)=  \frac{c}{\omega_L}
               \int_{\varphi_0}^\varphi d \chi \frac{e_0 \bm{A}(\chi)}{m_ec},
               \quad \varphi=\omega_L t-k{z} .
\end{align}
it follows that the amplitude of the motion of the electron in a monochromatic field is of the order of magnitude of
\begin{align}
     r_\perp\sim \frac{c\xi_0}{\omega_L}\sim\xi_0\lambda,\qquad z\sim\frac{c\xi_0^2}{2\omega_L}\sim\xi_0^2\lambda\label{estimate}
\end{align}
where $\xi_0$ is the dimensionless intensity parameter defined in Eq. (\ref{defxi}).
The analytical expression of the electron  momentum is
\begin{align}
      & \bm{p}_{\perp}(\phi)=-e\left(\bm{A}(\phi)-\bm{A}_0+\frac{\bm{p}_{0 \perp}}{e}\right), \nonumber                                                                                                                                                                               \\
      & p_z(\phi)=\frac{e^2 {\boldsymbol{\mathcal A}}^2(\phi)}{2 \left(p_0^0-p_0^3\right)}+p_{0z},\qquad {\boldsymbol{\mathcal A}}^2(\phi)=\left(\mathbf{A}(\phi)-\mathbf{A}_0\right)^2-\frac{2 \mathbf{p}_{0 \perp}}{e} \cdot\left(\mathbf{A}(\phi)-\mathbf{A}_0\right) ,\label{pan}
\end{align}
with ${\bm p}_0$ and ${\bm A}_0$  the initial values of the momentum and, respectively, vector potential.

From the estimation (\ref{estimate}) it follows that an electron in a helical beam of waist $w_0\gg\lambda$ and $\xi_0 \lessapprox1$ will see only a monochromatic plane wave field of amplitude given by its initial position in the focal plane; if the initial cylindrical  coordinates of the electron in the focal plane are $\varPhi_0$, $\varrho_0$ then the electromagnetic field felt by the electron can be approximated as
\begin{align}
     \widetilde{\bm E}({\bm r},t) = {\cal E}_0(\varrho_0)\Re\left\{({\bm e}_x\zeta_x+i{\bm e}_y\zeta_y)e^{-i(\omega_L t-kz- m_L\varPhi_0)}\right\},\qquad \widetilde{\bm B}({\bm r},t) = \frac1c{\bm e}_z\times\widetilde{\bm E}({\bm r},t)\label{ebapprox}
\end{align}
and
\begin{align}
     {\cal E}_0(\varrho_0)= \frac{m_ec\omega_L \xi(\varrho_0) }{ |e_0|},\qquad \xi(\varrho_0)= \xi_0 N_{p_Lm_L}\left(\frac{\sqrt{2}\varrho_0}{w_0}\right)^{|m_L|}e^{-\frac{\varrho_0^2}{w_0^2}}\, _1F_1(-p_L, |m_L|+1, \frac{2\varrho_0^2}{w_0^2})\label{defxirho}
\end{align}
i.e. each electron feels a plane wave monochromatic field whose magnitude ${\cal E}_0$ and initial phase $ m_L\varPhi_0$ are remnants of the initial field structure. The function $\xi(\varrho_0)$ defined in Eq. (\ref{defxirho}) is a local,  position dependent intensity parameter of the plane waves. We shall use the name {\it local plane wave approximation (LPWA)} for this approach and in the following we shall discuss its validity.

\begin{figure}[H]\begin{center}
     \includegraphics[scale=0.15]{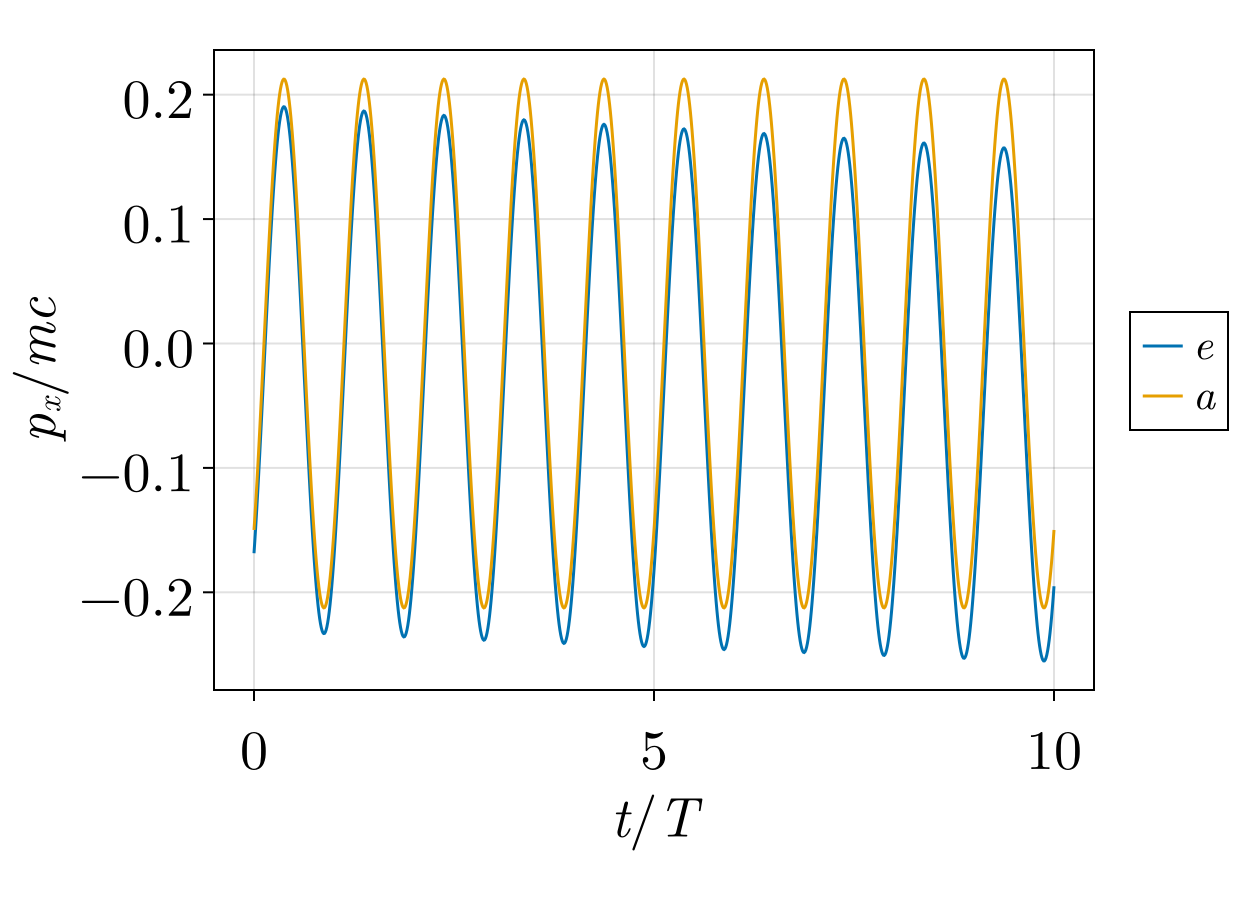} \includegraphics[scale=0.15]{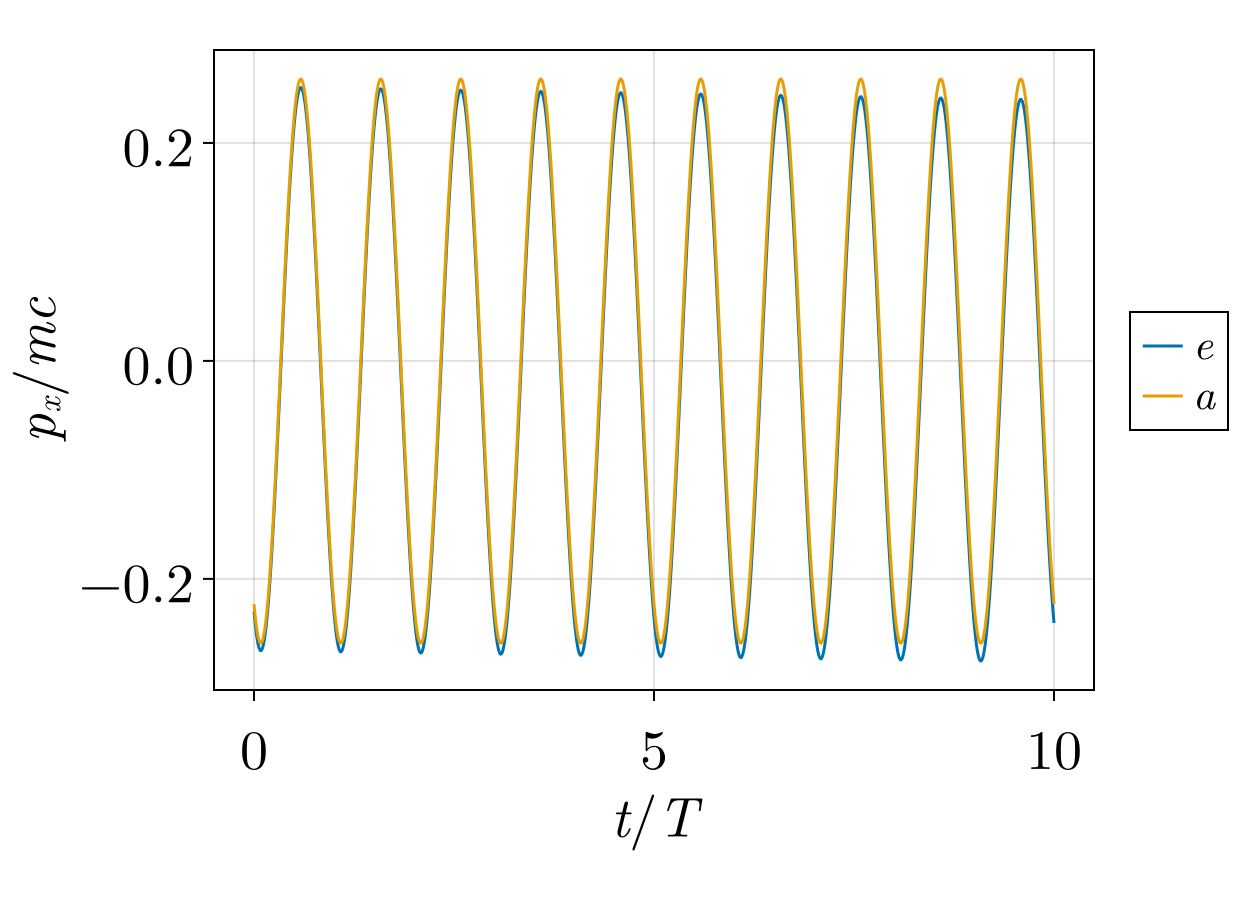}\end{center}
     \caption{Comparison between the exact results and the LPWA for the $Ox$ component of the momentum of an electron interacting with a LG laser beam having $\xi_0=1$, $p_L=2$, $m_L=2$, $w_0=75\lambda$, $\omega_L=0,057$  au. Left: the initial position of the electron on the $z=0$ plane is $(x_0,y_0) = (70\lambda, -32\lambda)$; right: $(x_0,y_0) = (160\lambda, -90\lambda)$ \label{Fig1}}
\end{figure}

We have integrated numerically the relativistic equations of motion for an electron in the presence of the electromagnetic field described exactly and in LPWA and we have compared the corresponding  trajectories and momenta.  In both cases (exact and LPWA) the particle-field coupling  was introduced using a smooth envelope of Gaussian shape
\begin{align}
      & ({\bm E},\,{\bm B})\rightarrow  (f(\varphi){\bm E},\,f(\varphi){\bm B}),\qquad  f(\varphi)=\left\{\begin{array}{ll}e^{-\frac{\varphi^2}{\tau^2}},&\varphi<0\\1,&\varphi>0\end{array}\right.\qquad \text{with}\quad  \tau = 10\pi.
\end{align}
and the electron was initially at rest in the plane $z=0$; the parameters of the field were $\xi_0=1$, $w_0=75\lambda$, $p_L=2$, $m_L=2$, circular polarization.
We present results for time $t\in(0,10T)$ (after the onset of the stationary regime). In the left panel of Fig. \ref{Fig1}  below we present the comparison for the variation in time of the component along the $Ox$ axis, $p_x$, of a particle with the initial position $(x_0,y_0) = (70\lambda, -32\lambda)$, which shows differences of about 10\% between the exact and the approximated results. However,  the accuracy of the approximation depends strongly on the initial position of the particle. For example, for a particle initially at $(x_0,y_0) = (160\lambda, -90\lambda)$, the exact and the approximate results are practically  indistinguishable, as one can see in the right panel of Fig \ref{Fig1}.

Since the validity of the approximation depends strongly on the initial position of the particle in the focal plane, we considered next a statistical ensemble of 10000 particles distributed randomly inside a disk of radius $R_0=3.5w_0$ in the focal plane of the laser and we compared the exact and approximate results for each particle. We used the same parameters as before, $p_L=m_L=2$, $w_0=75\lambda$, $\omega_L=0.057$ a.u. and we consider two values of the dimensionless intensity parameter $\xi_0=0.5$ and $\xi_0=1$.

We compare the time evolution of the electron  momenta for the two calculations: the exact one one and within LPWA. We represent the electrons as dots in the plane $Oxy$, each dot having the same coordinates as those of the initial position of the corresponding particle. The numerical values of a cartesian component of the momenta are represented by the color of the dot; such that the variation of the momentum of an electron is seen as the change in the color of the dot in the $Oxy$ plane.

\begin{figure}[H]
     \includegraphics[scale=0.08]{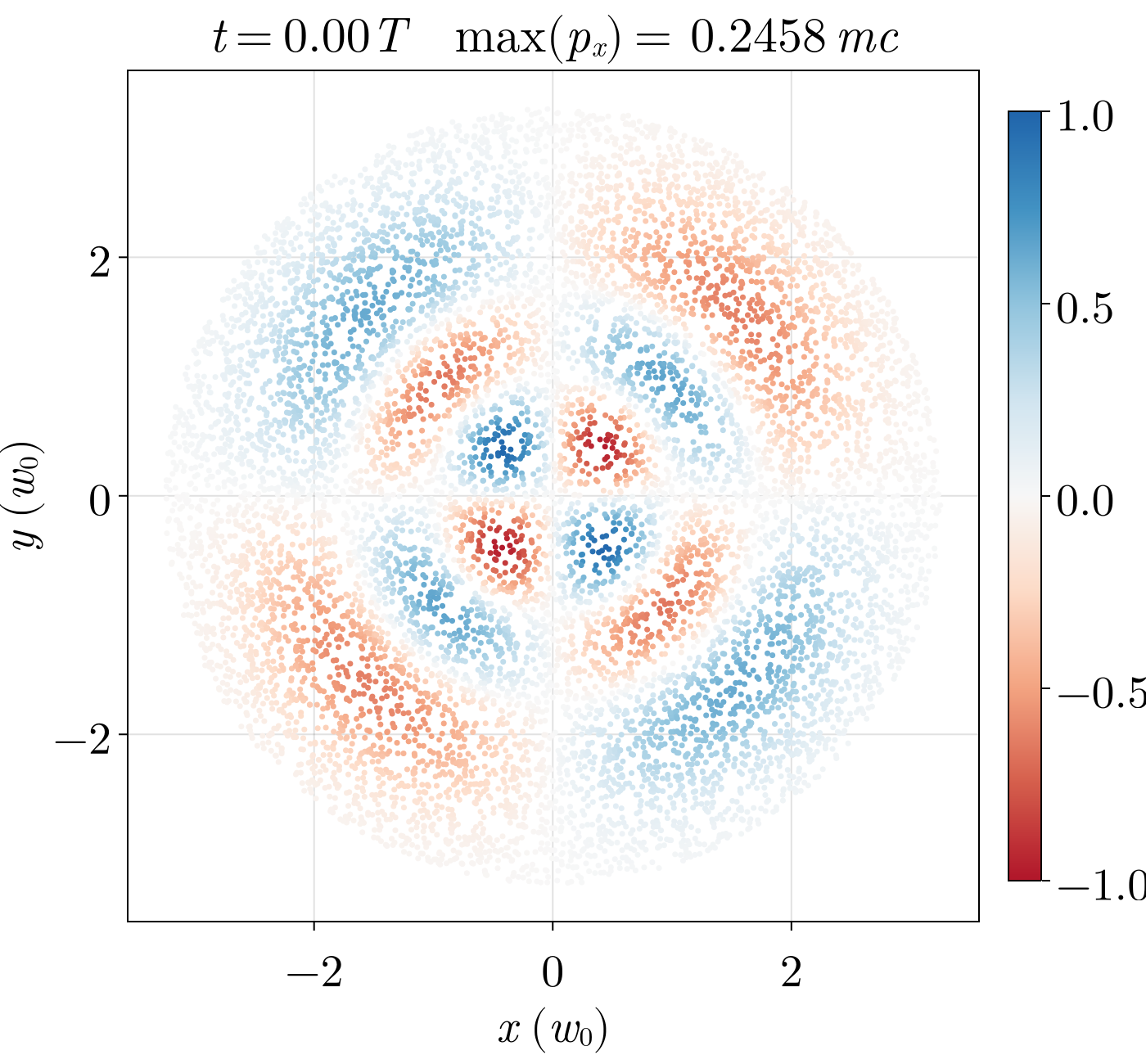}\includegraphics[scale=0.08]{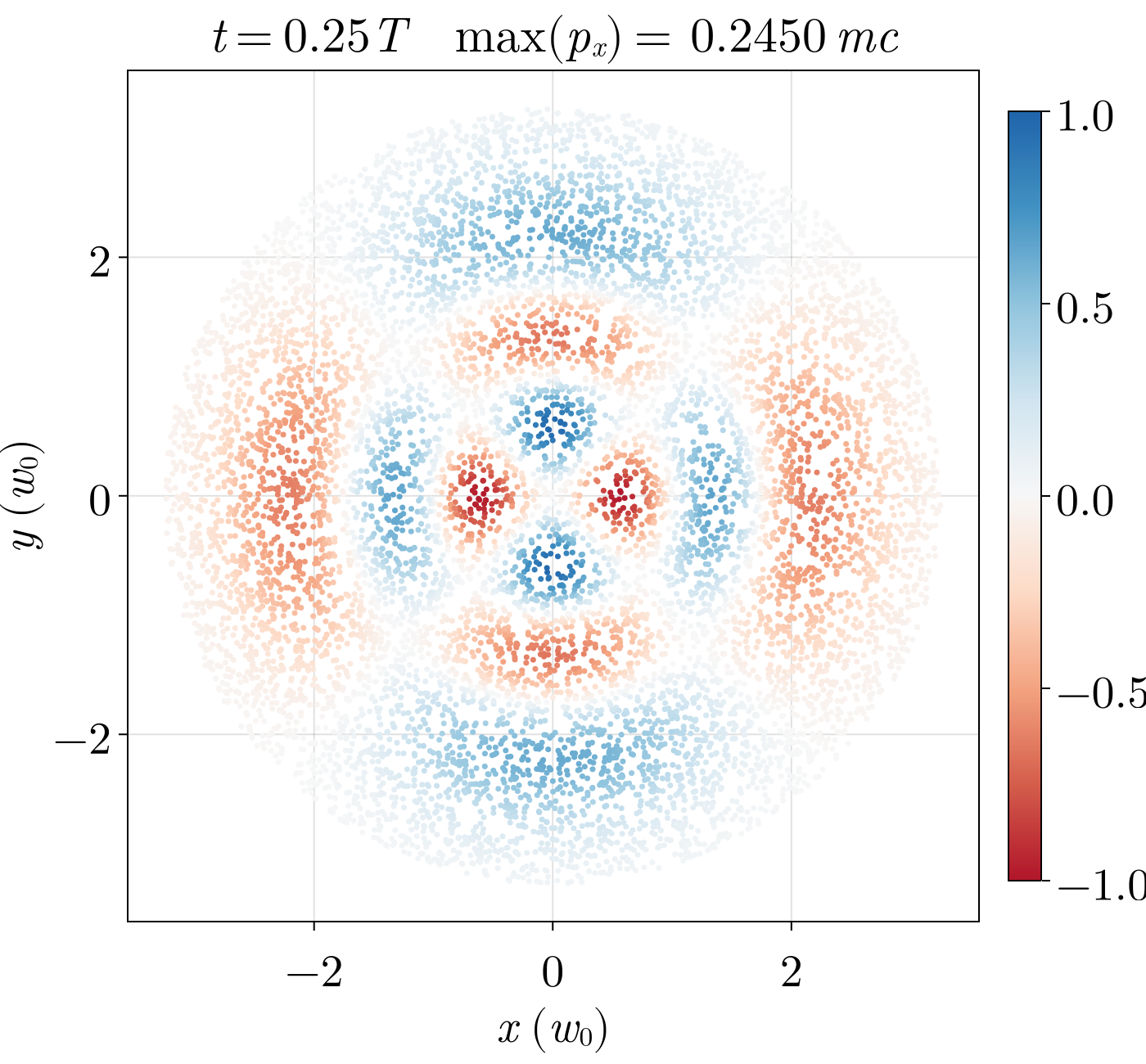}\includegraphics[scale=0.08]{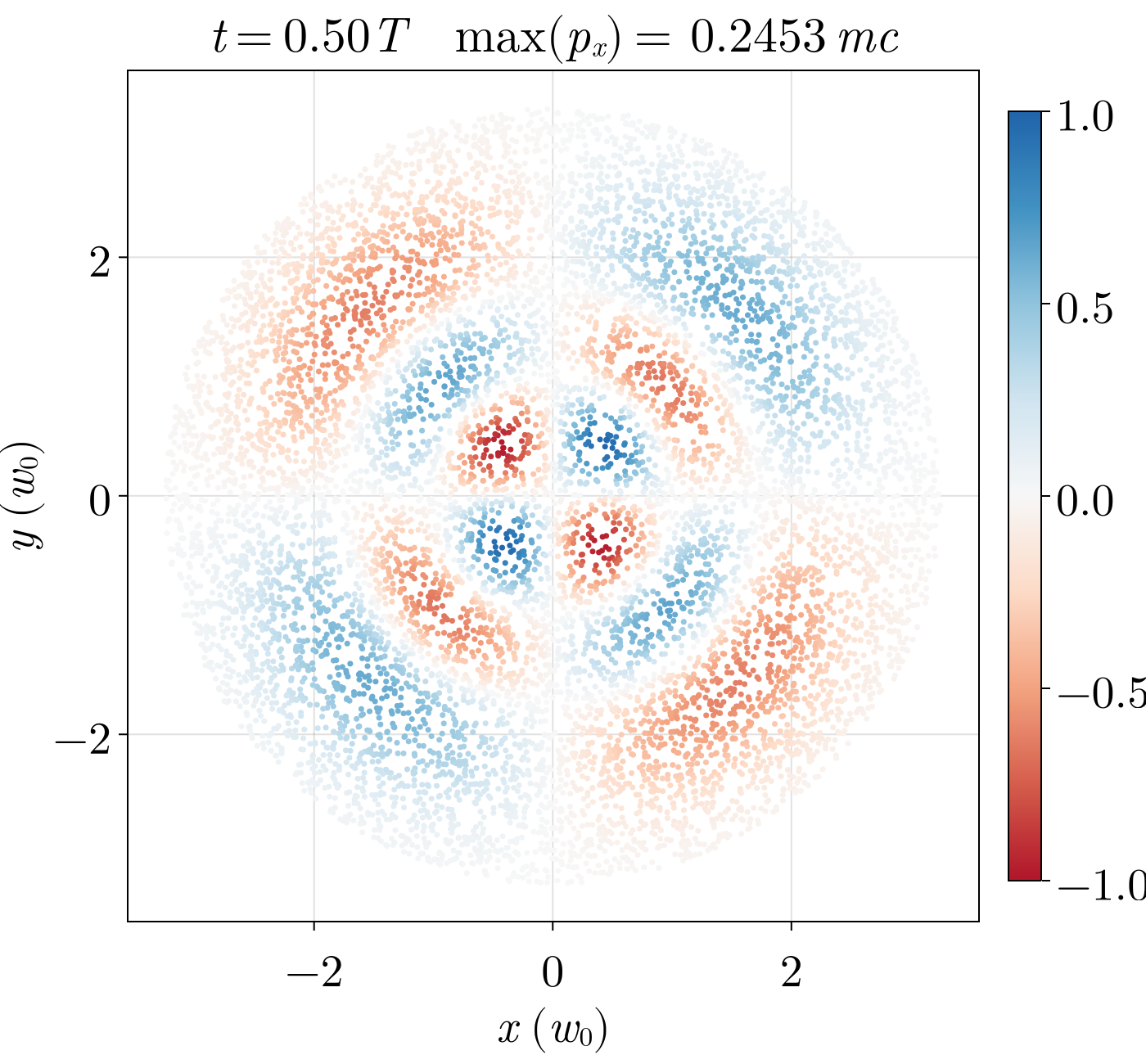}\includegraphics[scale=0.08]{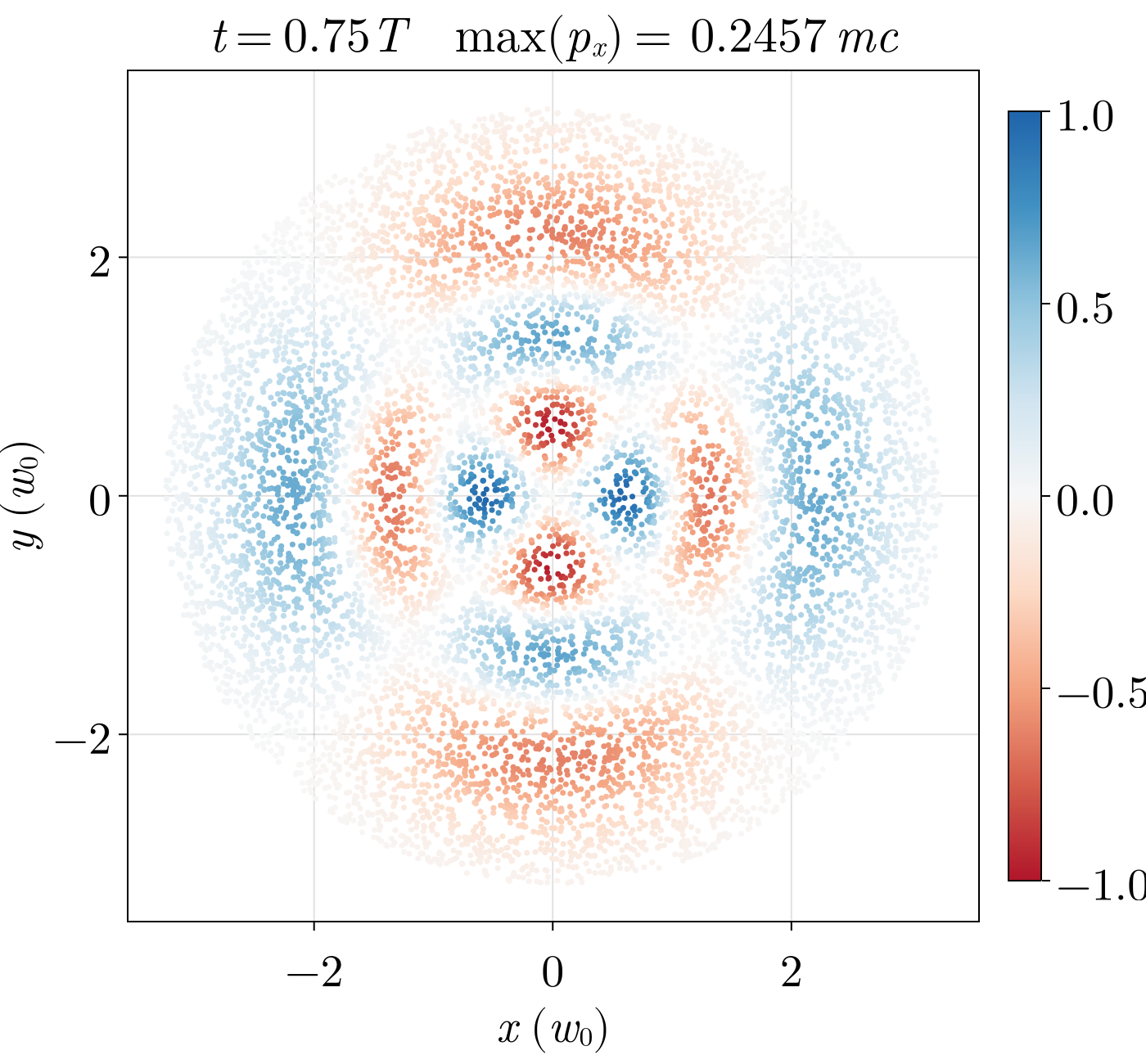}

     \includegraphics[scale=0.08]{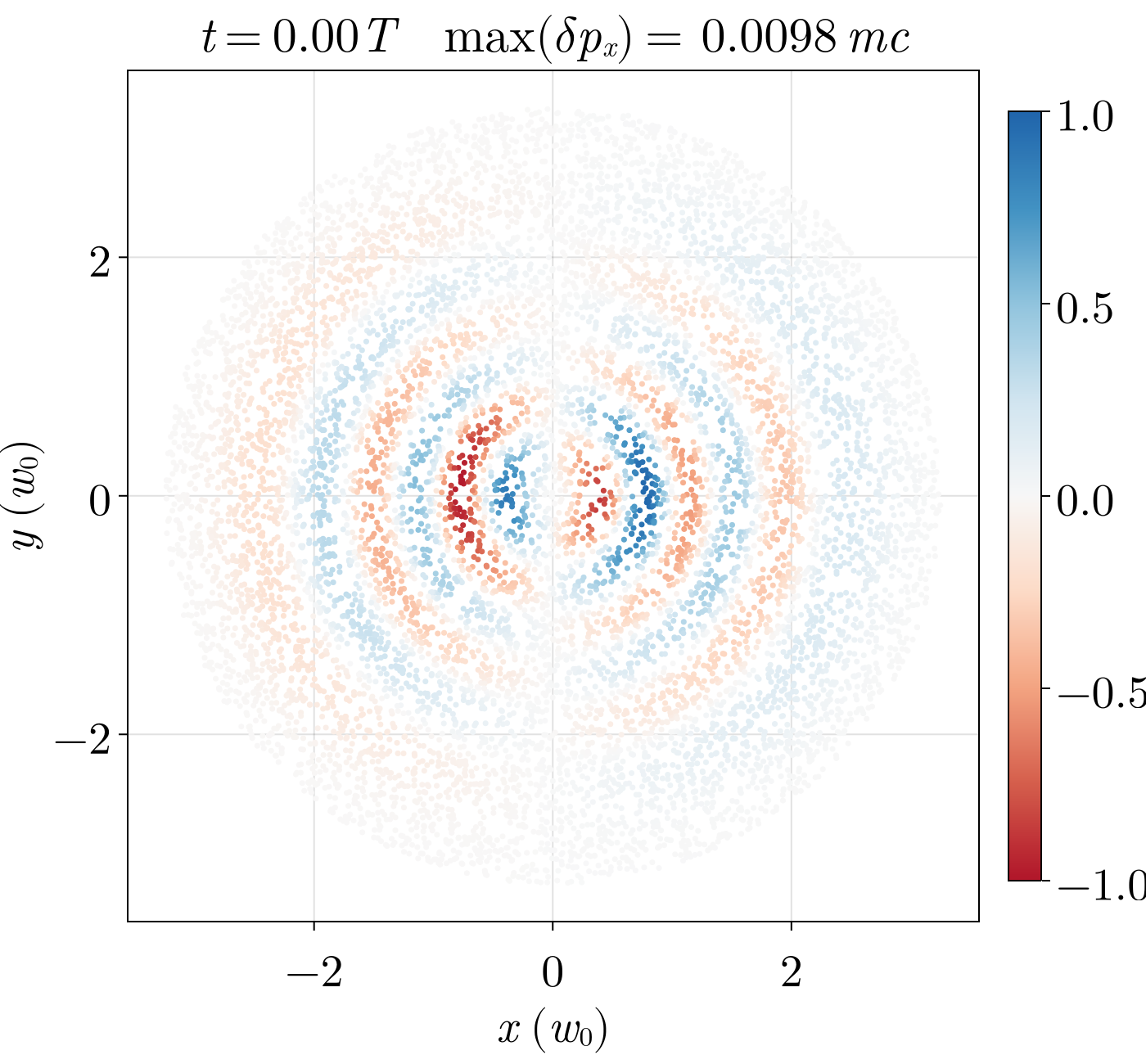}\includegraphics[scale=0.08]{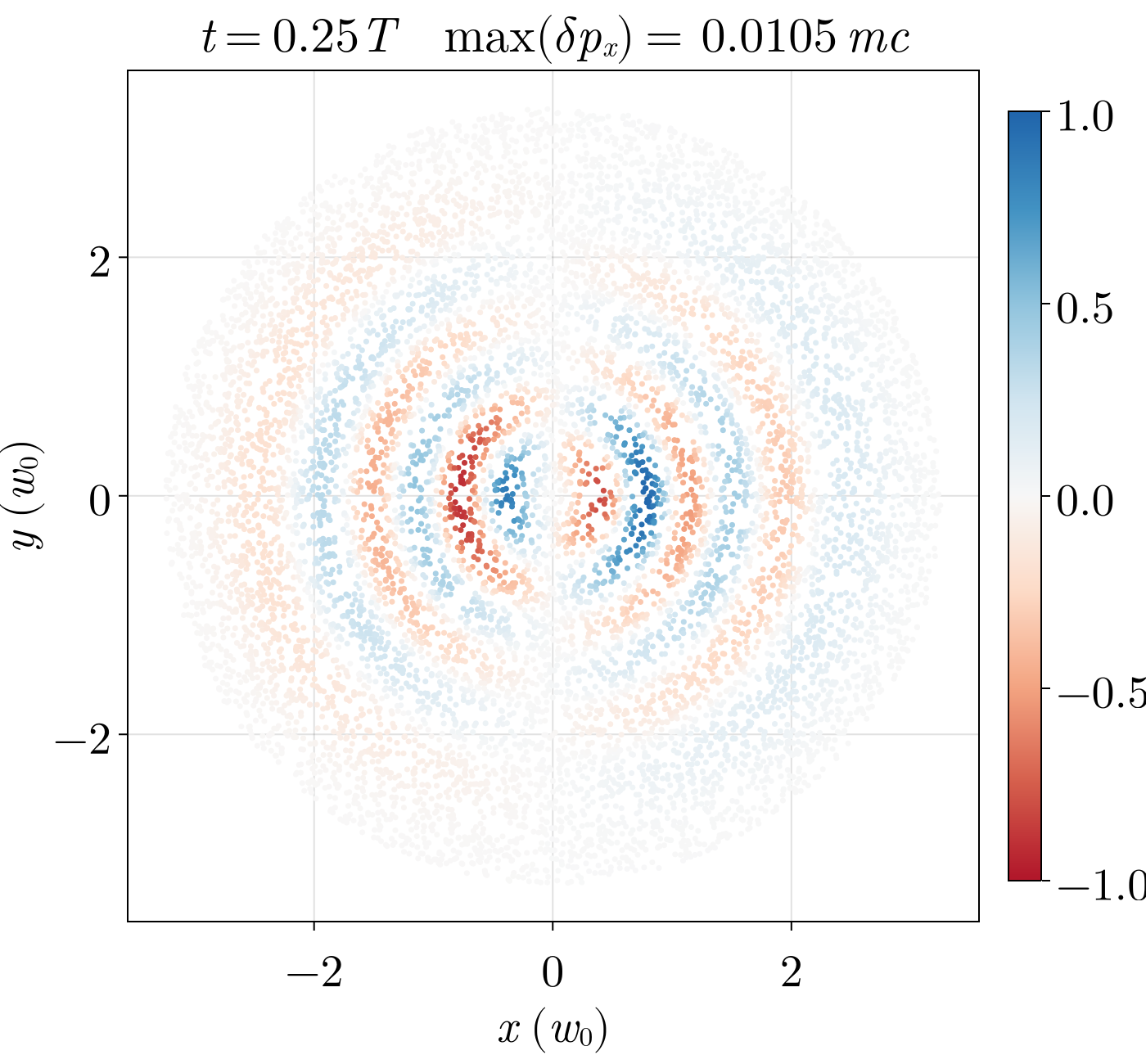}\includegraphics[scale=0.08]{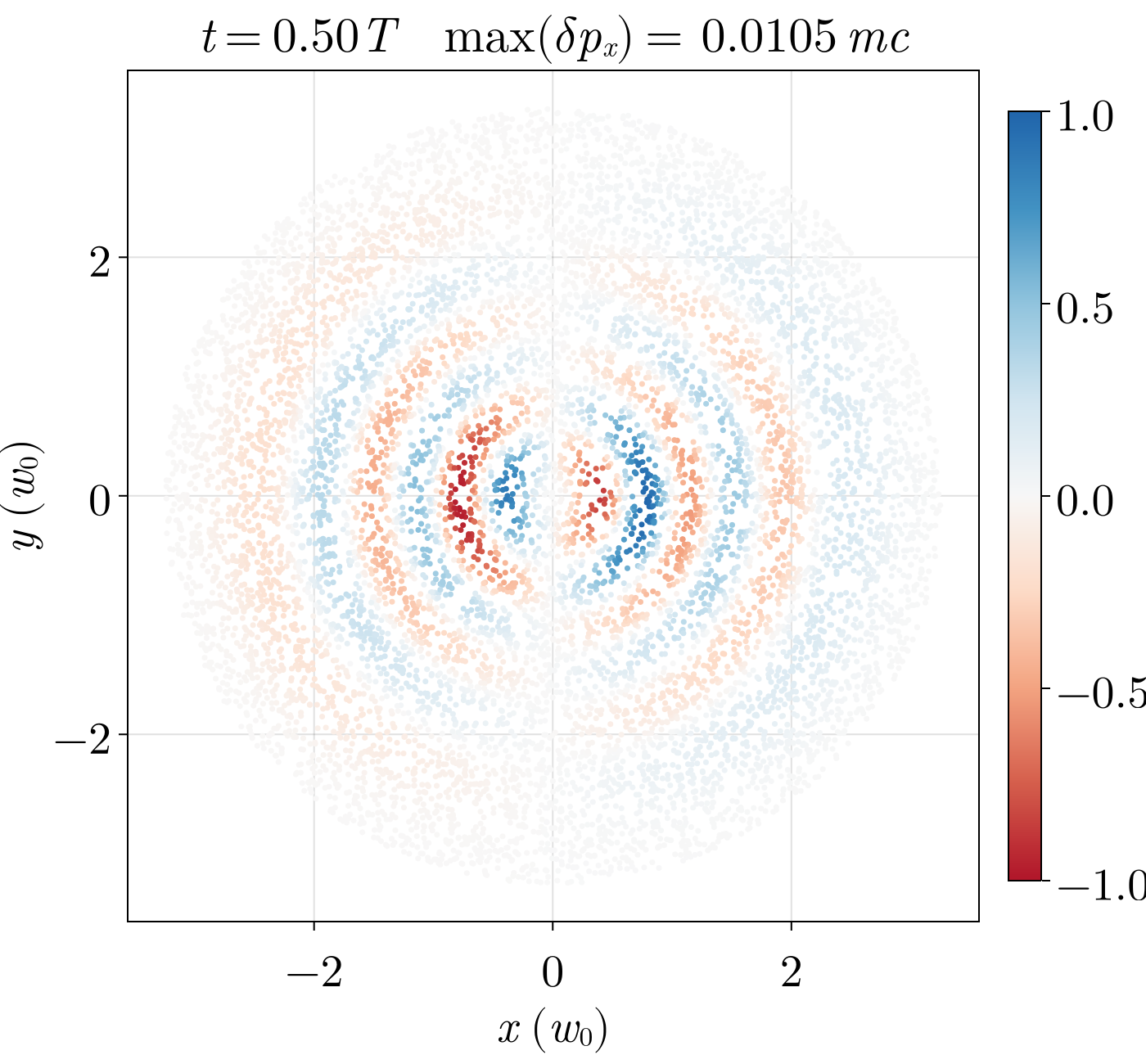}\includegraphics[scale=0.08]{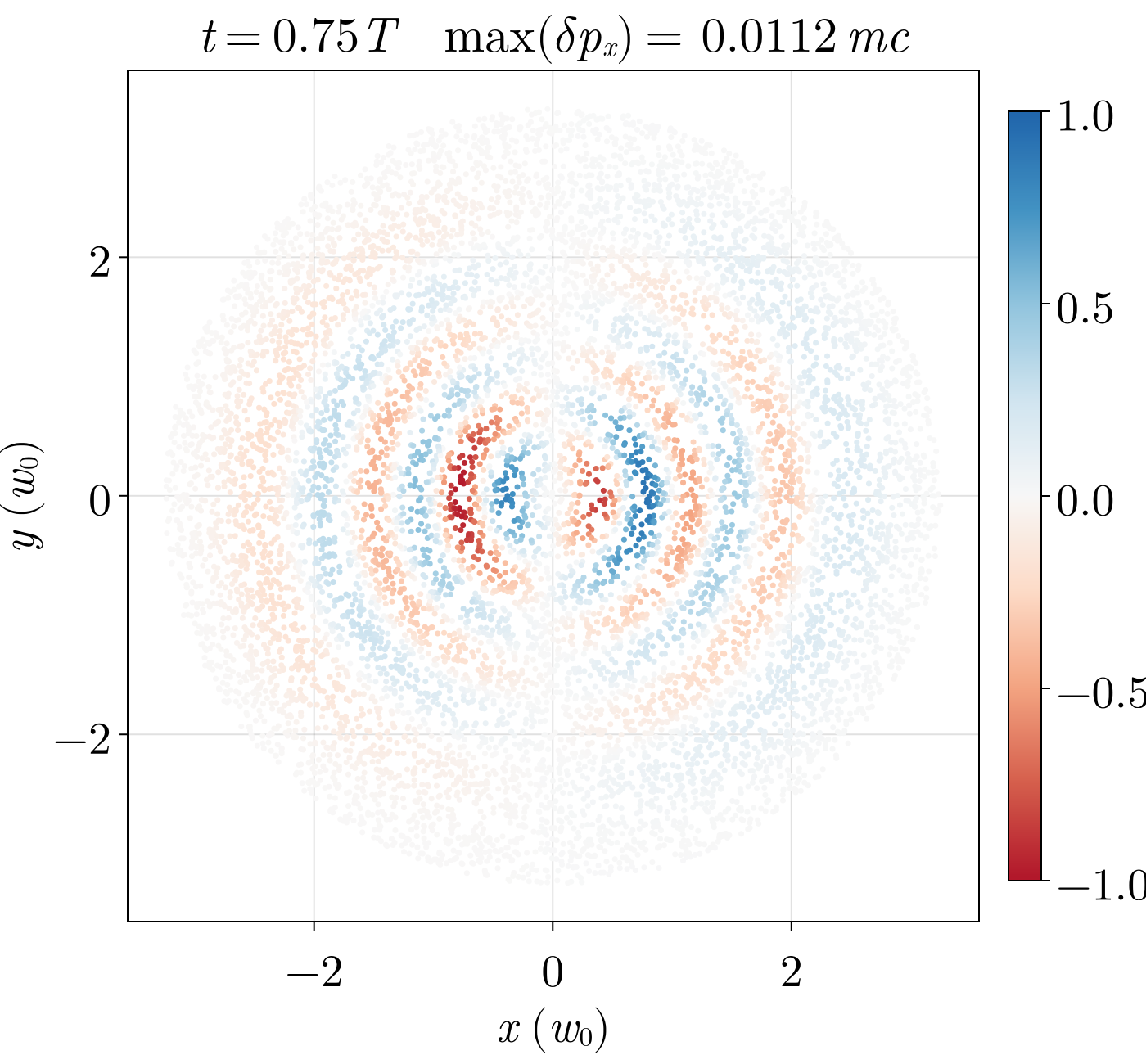}

     \vspace*{1cm}

     \includegraphics[scale=0.08]{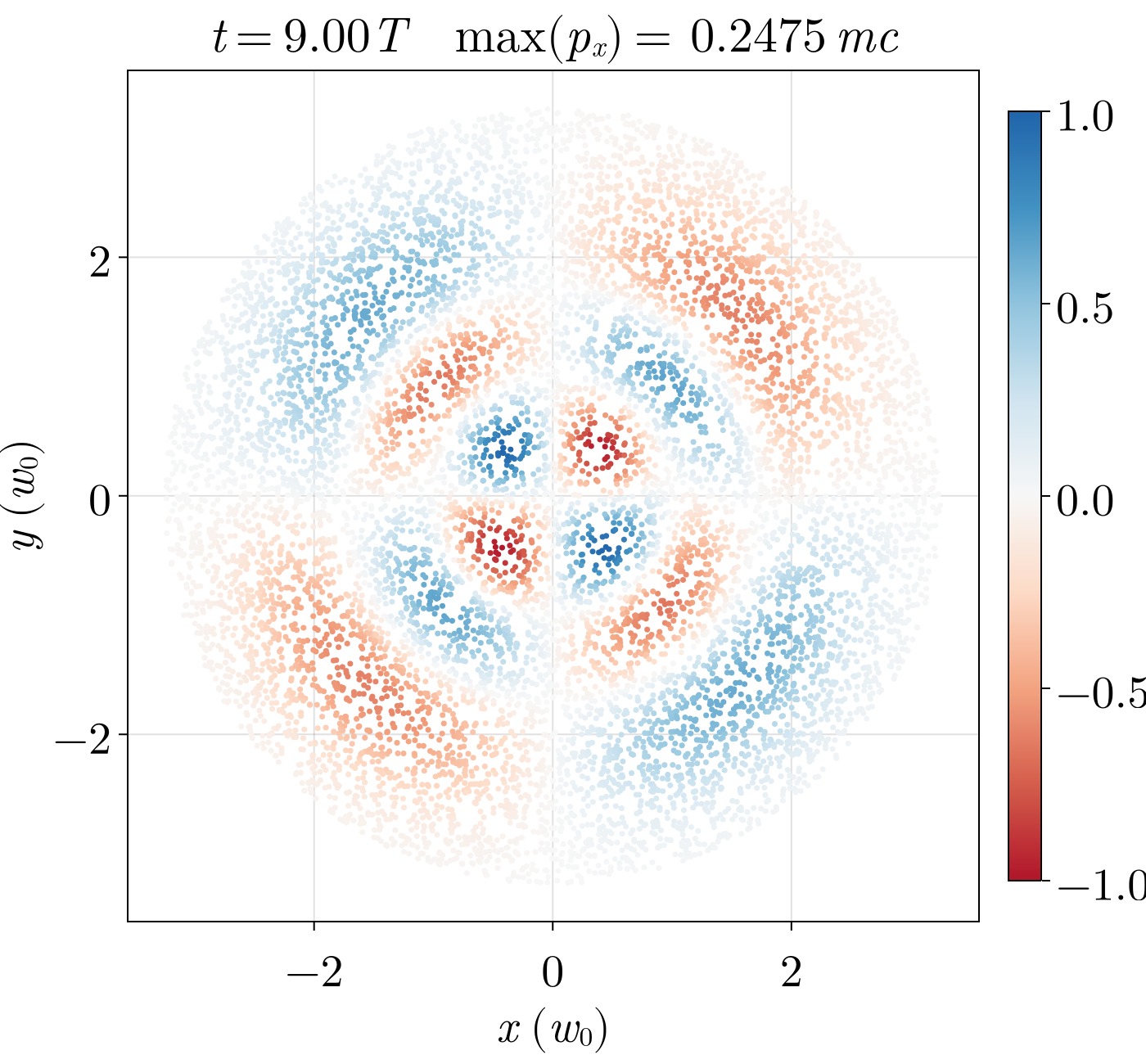}\includegraphics[scale=0.08]{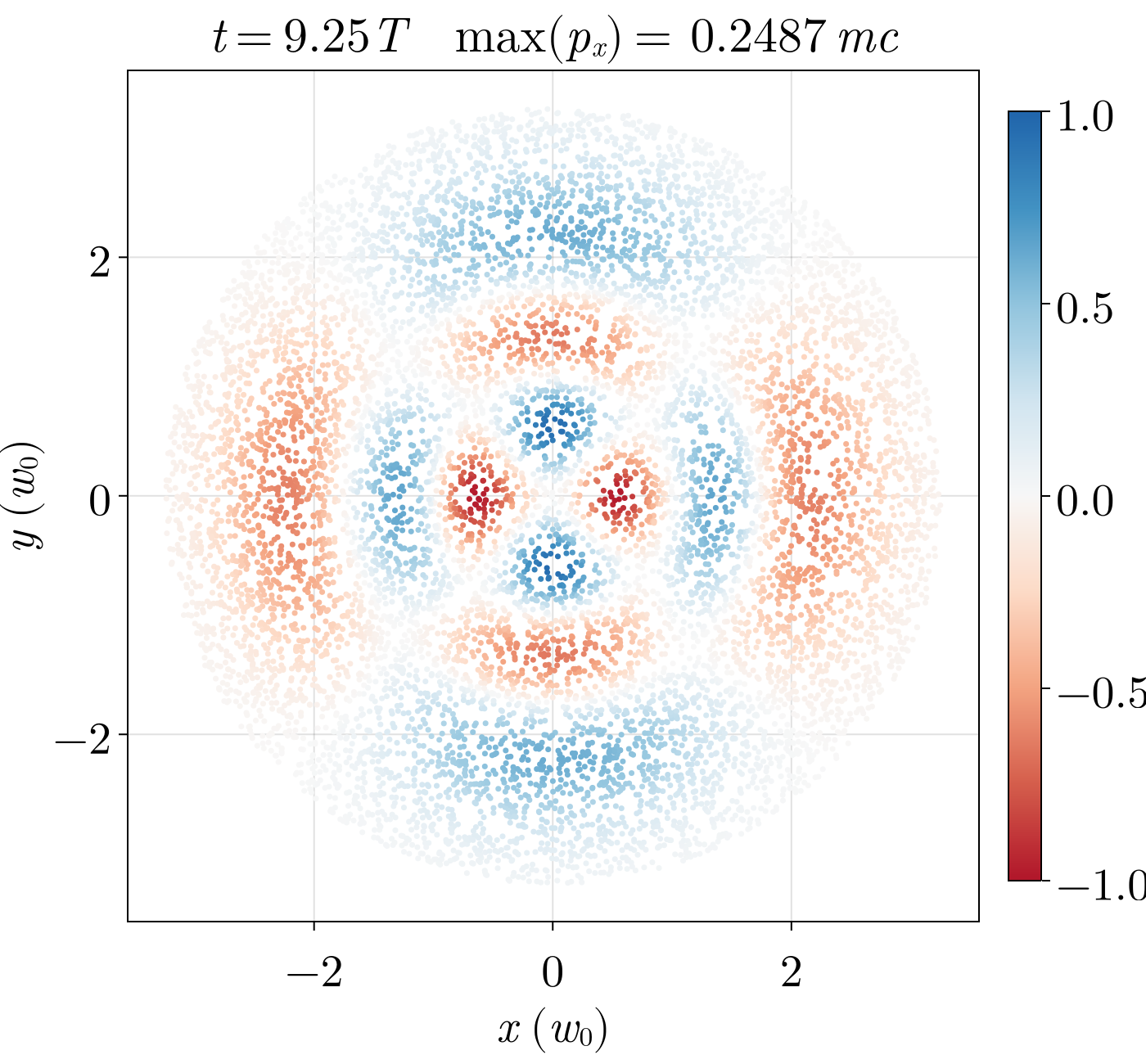}\includegraphics[scale=0.08]{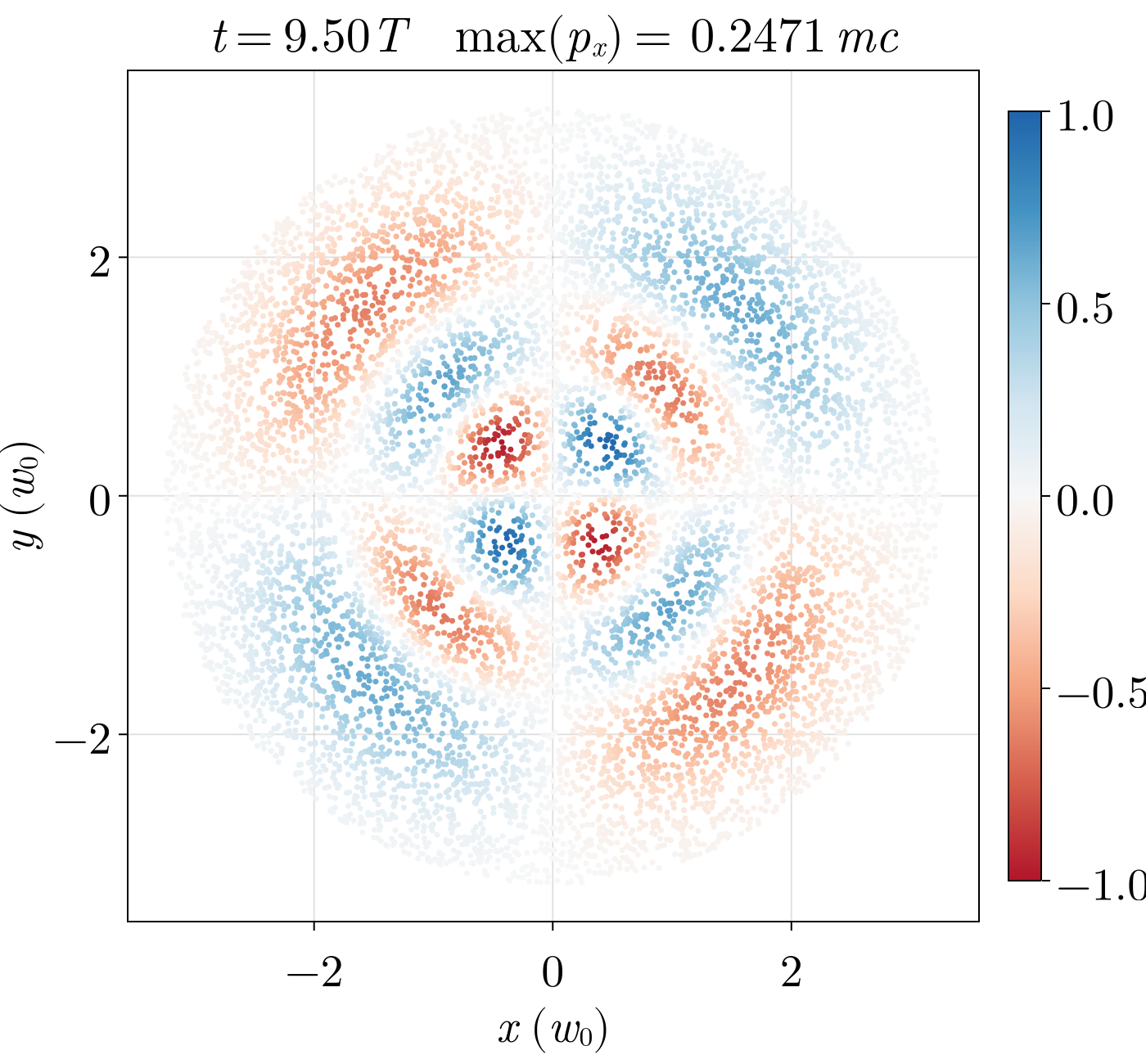}\includegraphics[scale=0.08]{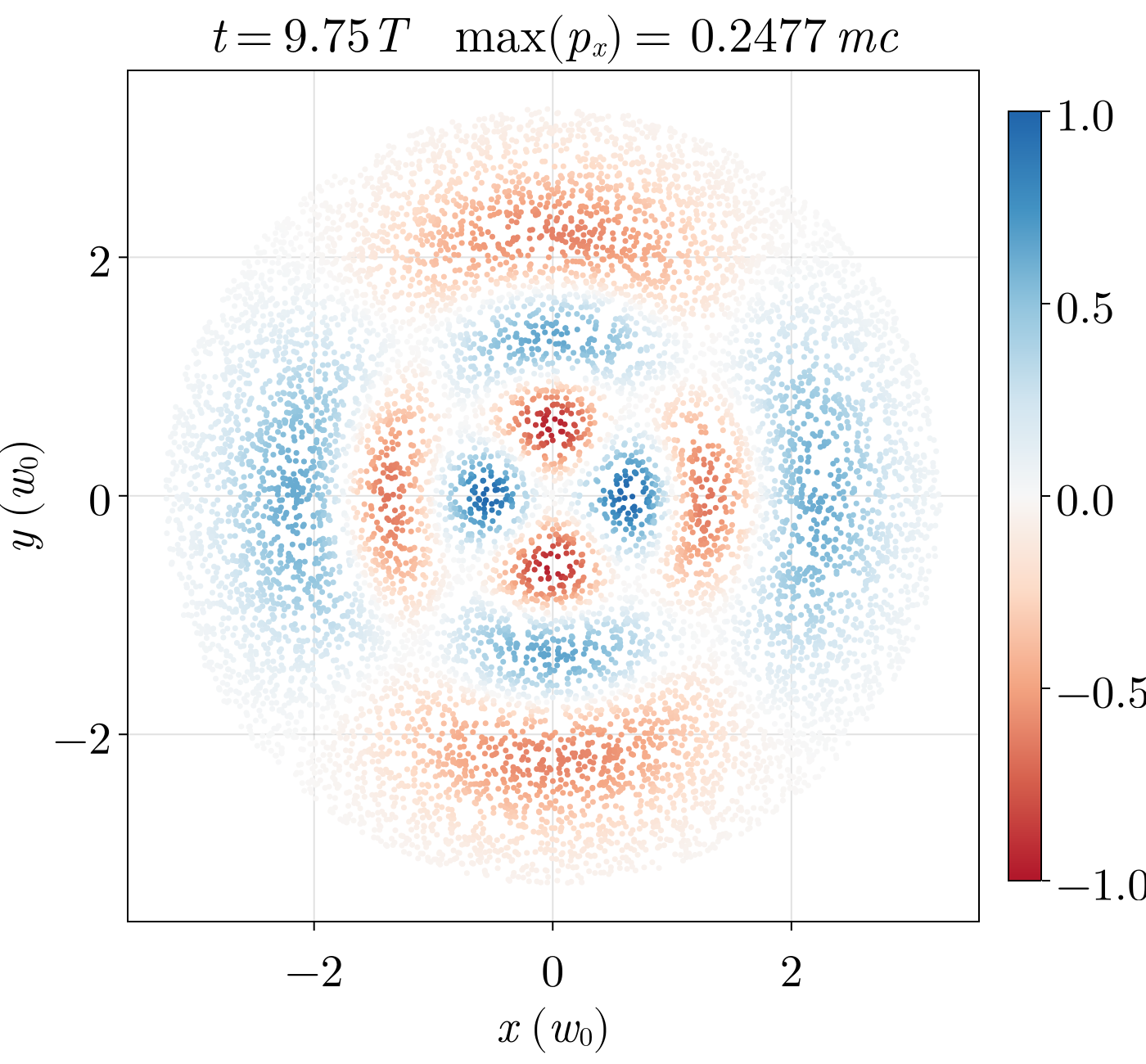}

     \includegraphics[scale=0.08]{supp-fig/f3/dpx/fig_0001.png}\includegraphics[scale=0.08]{supp-fig/f3/dpx/fig_0006.png}\includegraphics[scale=0.08]{supp-fig/f3/dpx/fig_0011.png}\includegraphics[scale=0.08]{supp-fig/f3/dpx/fig_0016.png}

     \caption{Snapshots of $p_x^{\text{exact}}$ (upper rows) and $\delta p_x=p_x^{\text{exact}}-p_x^{\text{LPWA}}$ for each particle in a statistical ensemble of electrons in interaction with a LG beam with $p_L=2$, $m_L=2$, $\xi_0=0.5$ for different times $t\in(0,10T)$ written on each graphic; the maximum value of  $|p_x|$ and respectively $|\delta p_x|$  is written at every moment in the title of each graphic. \label{fig-px}}
\end{figure}

In Fig \ref{fig-px} we show in the upper rows 8 snapshots of the time evolution of the component $p_x$ for $\xi_0=0.5$, calculated using the exact equations of motion; the times $t$ are written on top of each graph. The values of $p_x$ are scaled to the interval $(-1,1)$, as shown by the colorbar, and the scaling factor (which is in this case the maximum of $|p_x|$ over the entire electron collection) is written at every moment also in the title of the graphic. We can see the spatial structure, which is due to the fact that the "initial phase" of each electron momentum is given by the particle initial position in the plane, and also the spatial structure has an apparent rotation motion, due to the fact that the value of $p_x$ for each electron oscillates in time, as can be seen, e.g. in  Eq. (\ref{pan}). The first 4 snapshots are taken for $t\in(0,T)$ and the last four are for $t\in(9T, 10T)$. Below each graphic with the distribution of $p_x$ we represent a second graphic which contains in the same representation  the difference $\delta p_x=p_x^{\text{exact}}-p_x^{\text{LPWA}}$. As before the values are scaled to the interval $(-1,1)$ and the scaling factor and the time is indicated on top of each graphic. The ratio of the maximum of $|\delta p_x|$ to the maximum of $|p_x|$ (the ratio of the two scaling factors) is an indication of the accuracy of the LPWA approximation.

\begin{figure}[H]\begin{center}
     \includegraphics[scale=0.08]{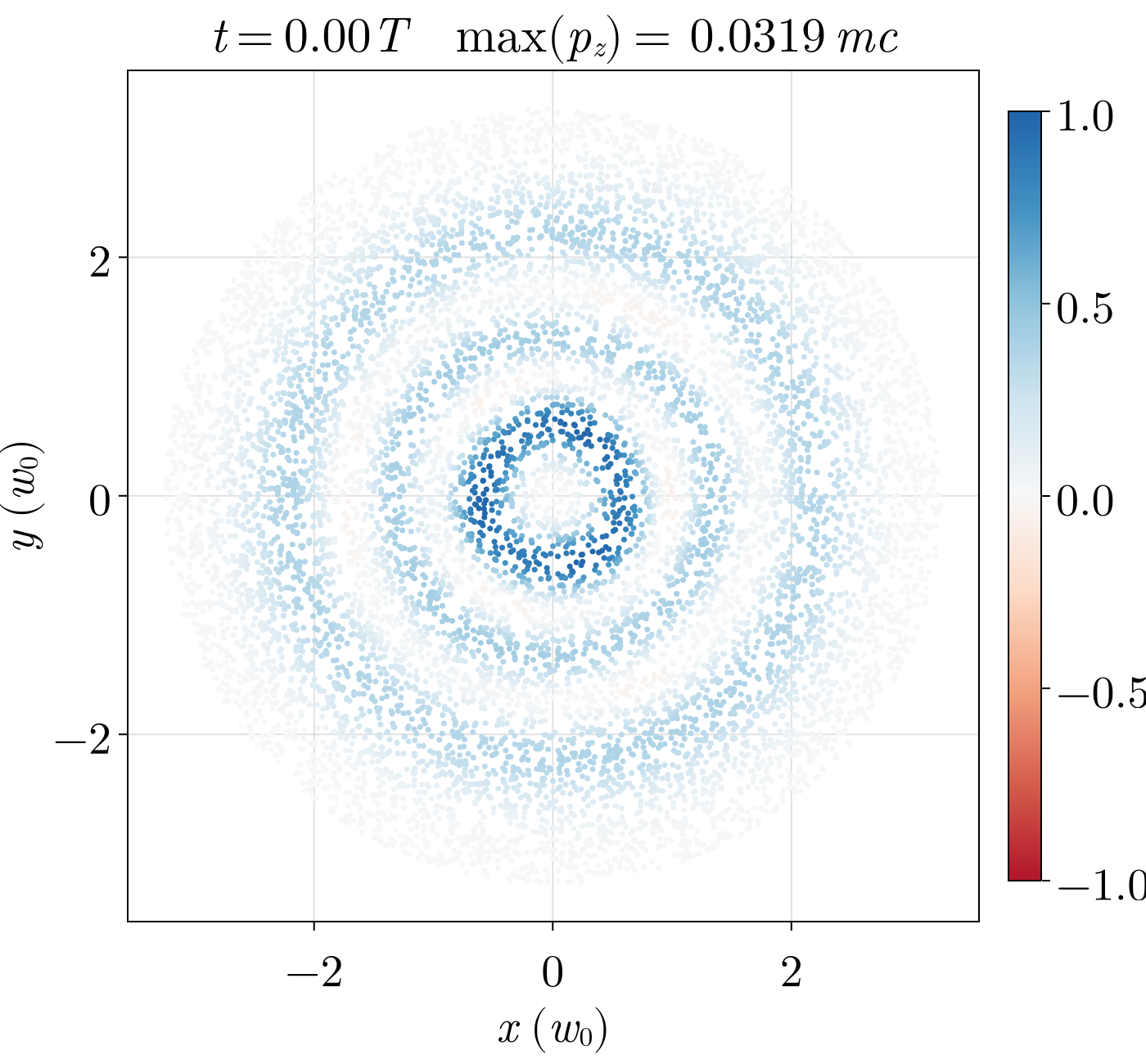}\includegraphics[scale=0.08]{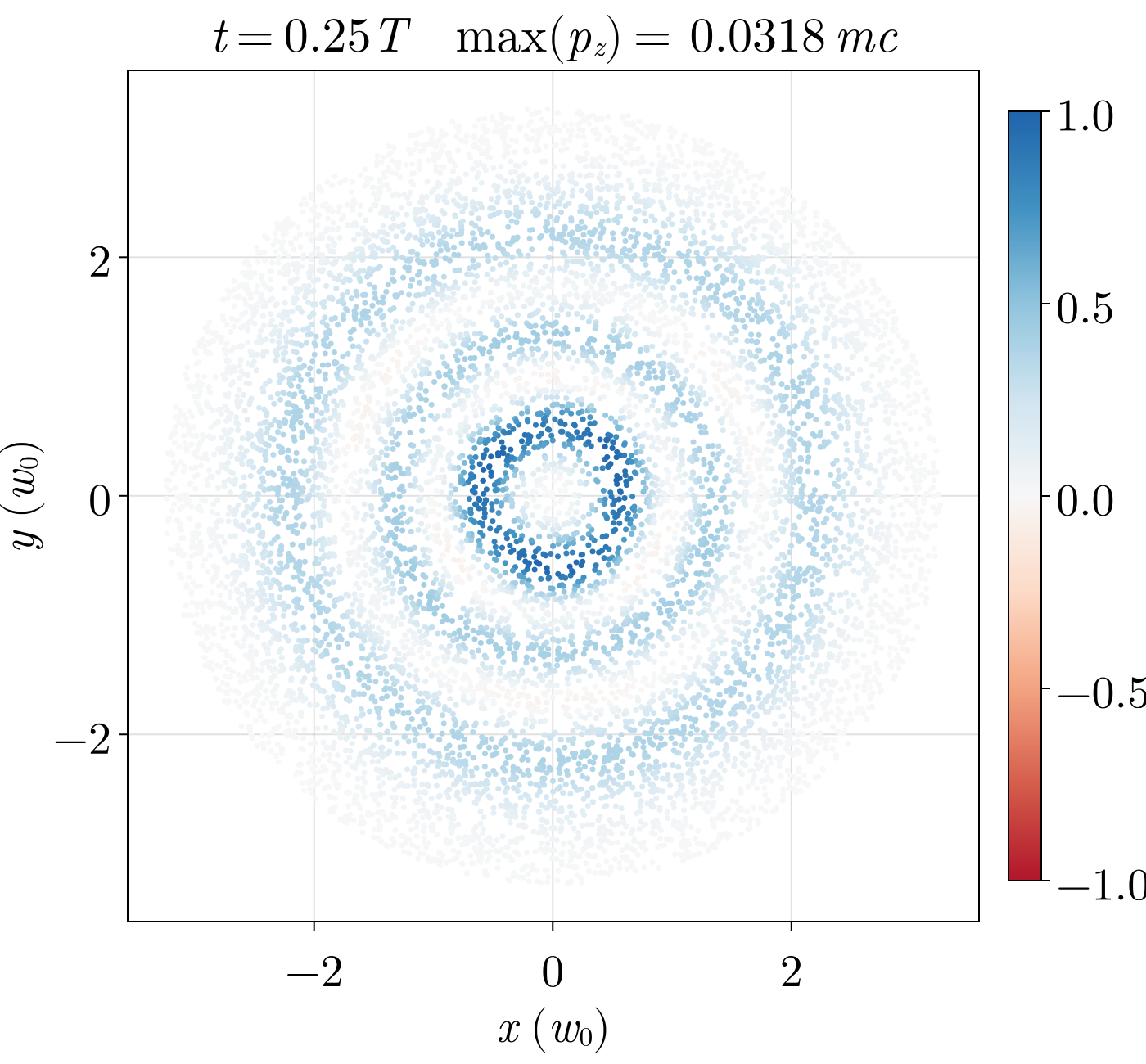}\includegraphics[scale=0.08]{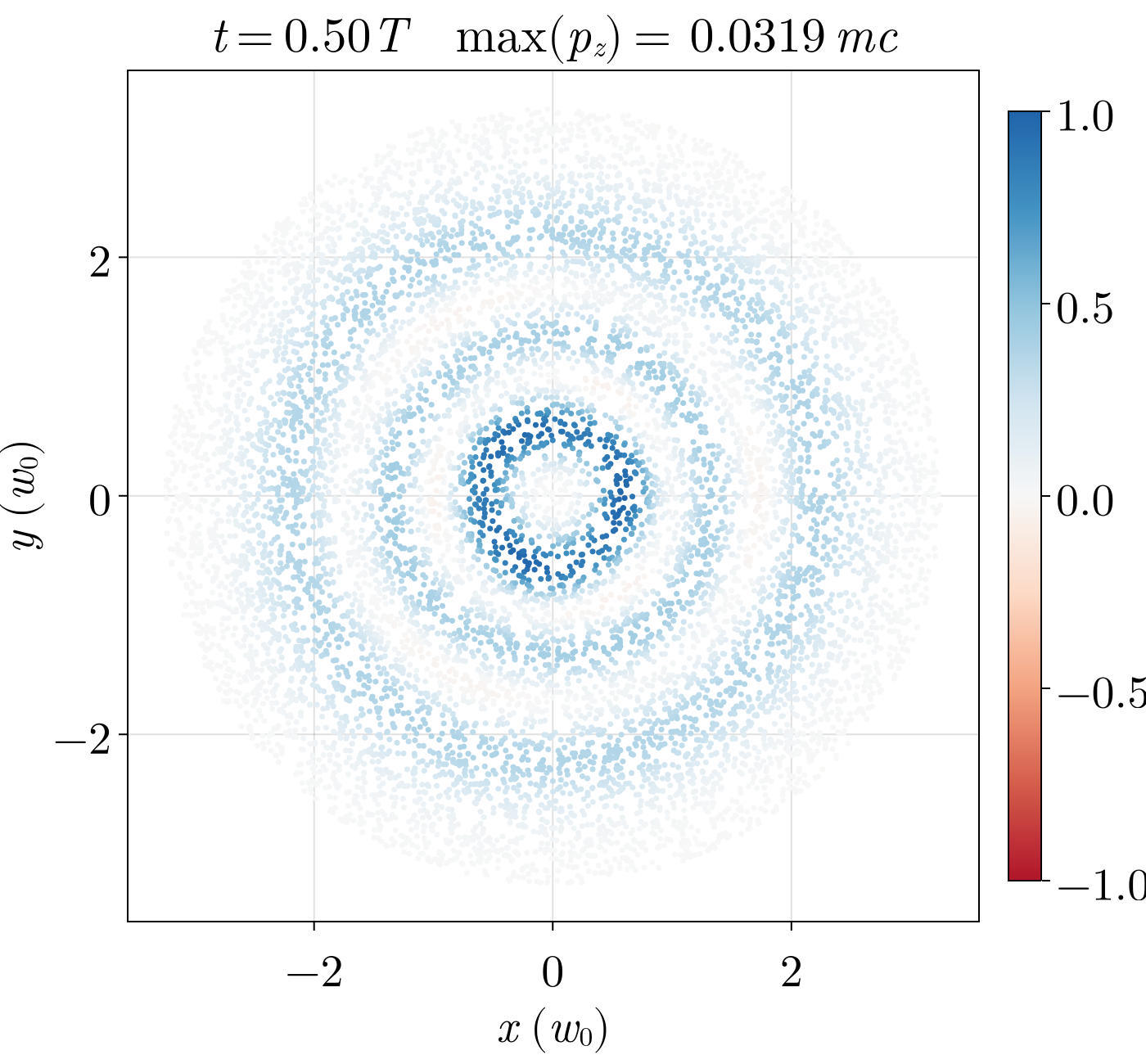}\includegraphics[scale=0.08]{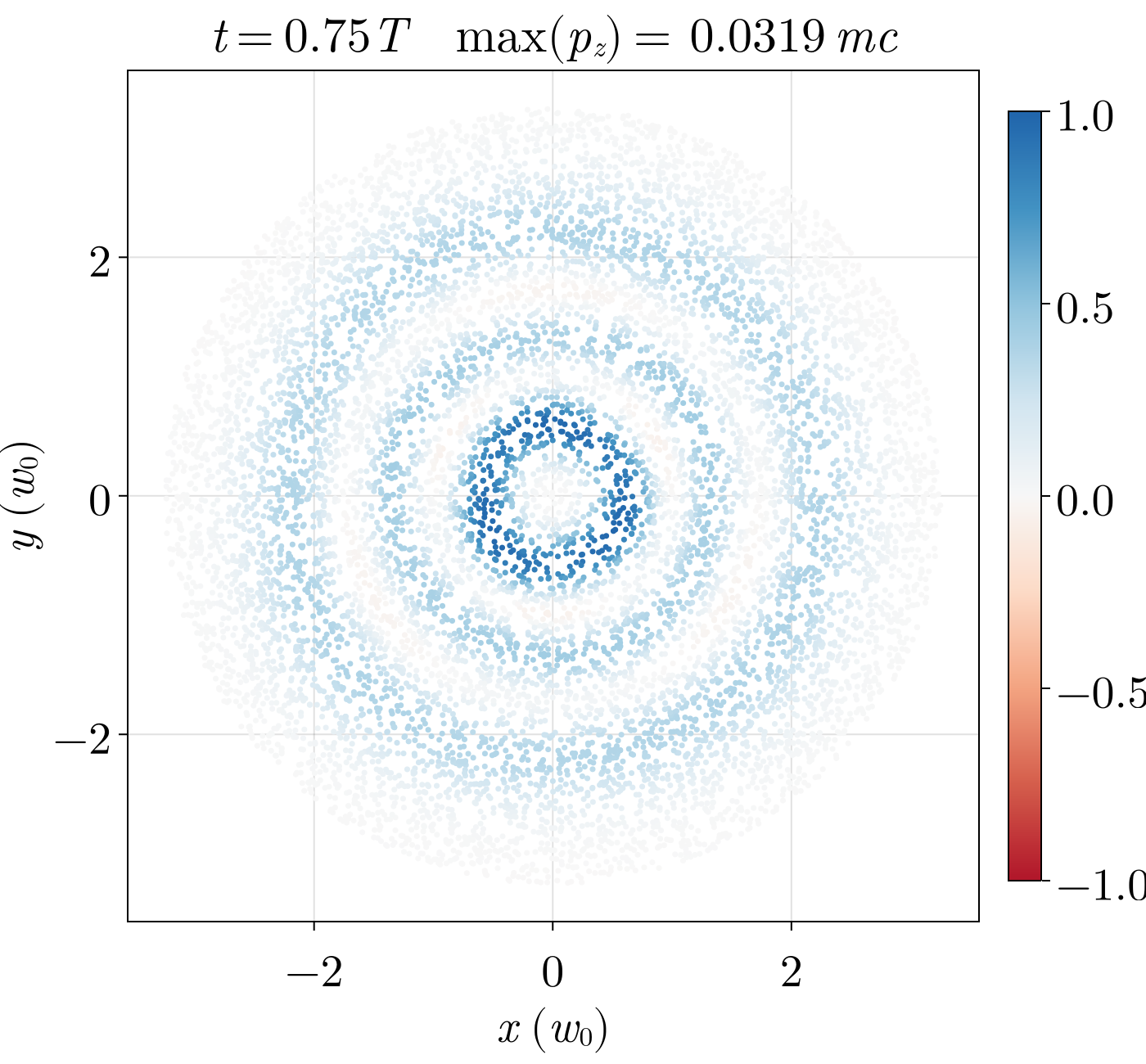}

     \includegraphics[scale=0.08]{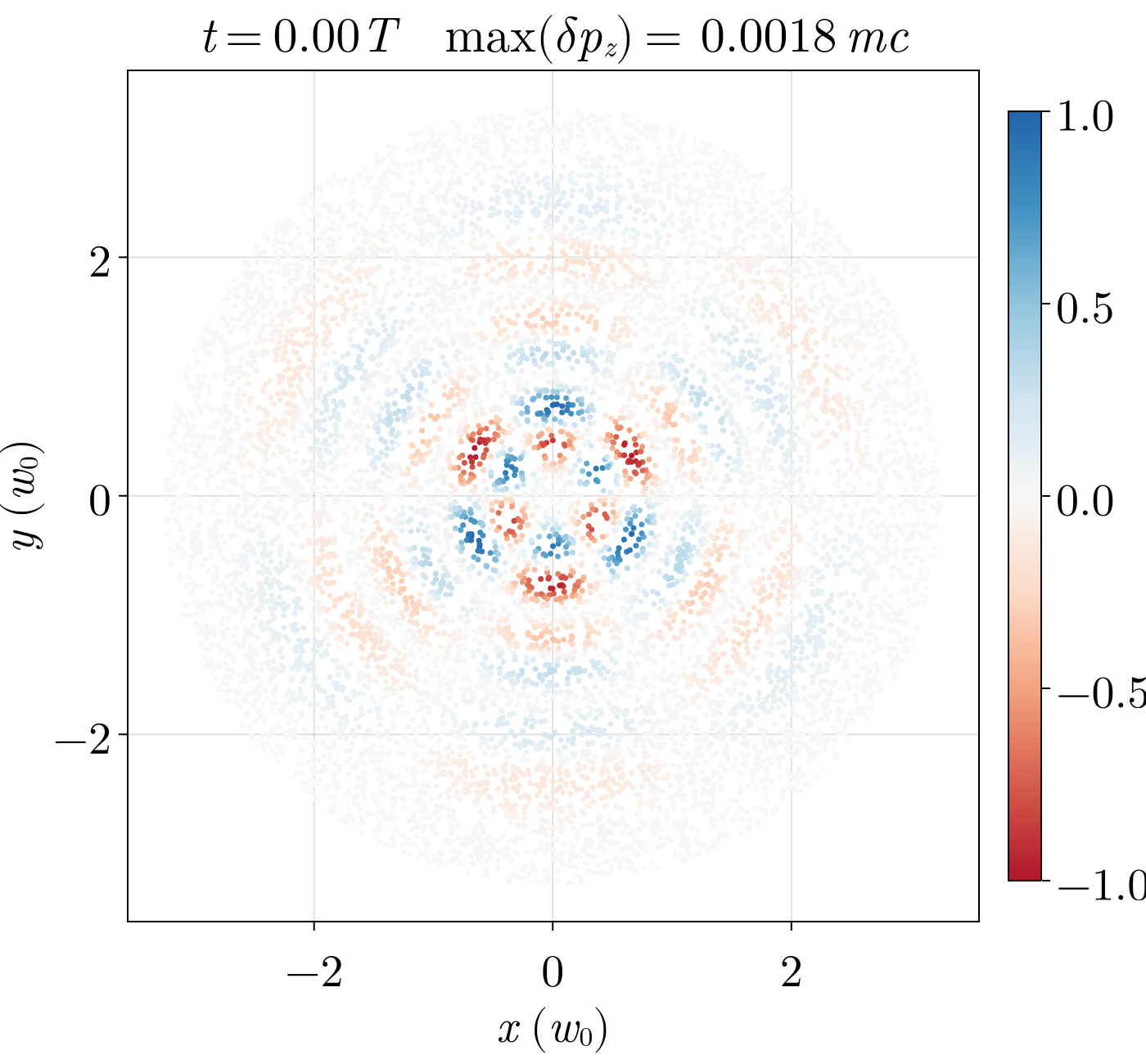}\includegraphics[scale=0.08]{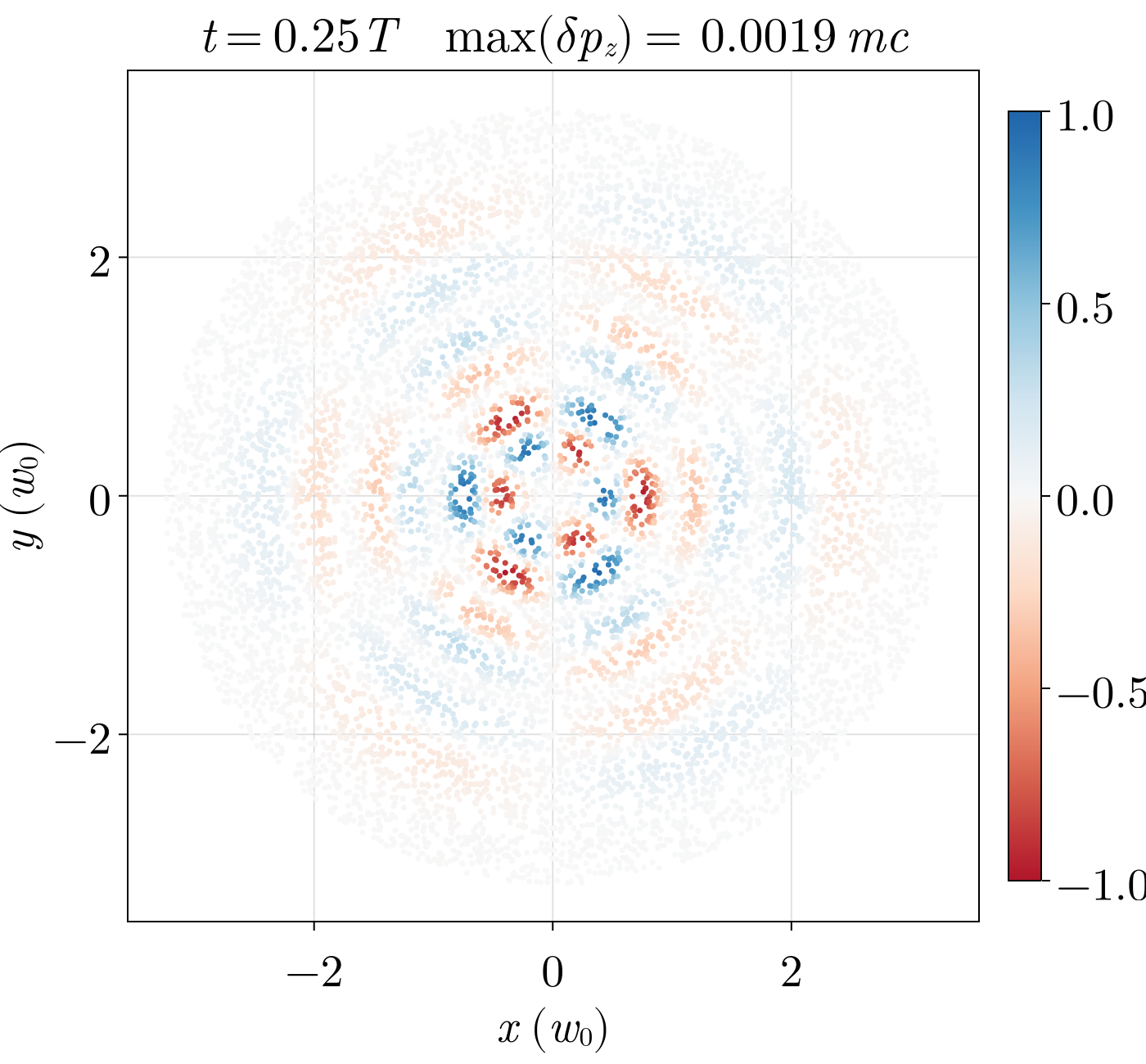}\includegraphics[scale=0.08]{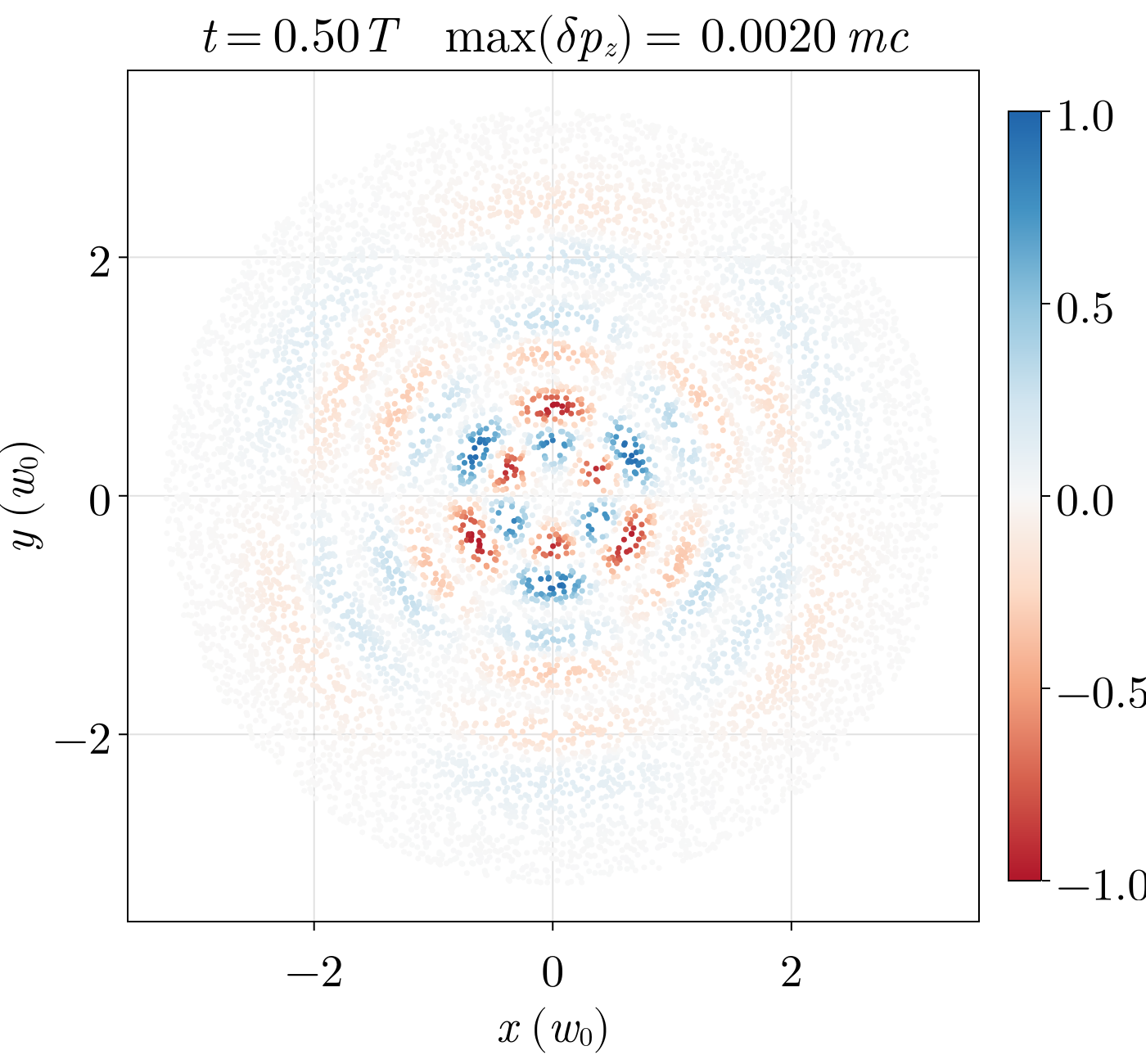}\includegraphics[scale=0.08]{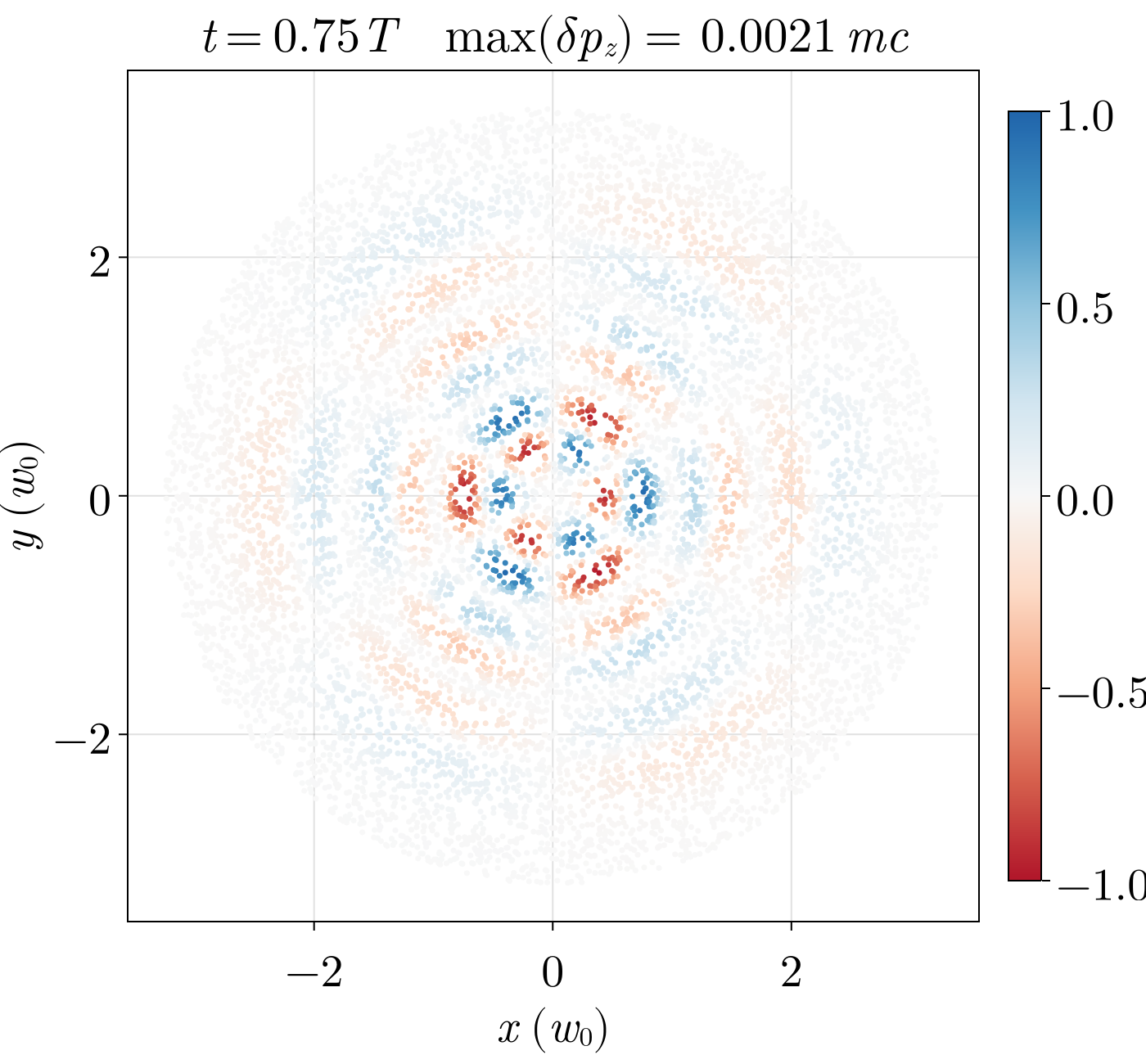}

     \vspace*{1cm}

     \includegraphics[scale=0.08]{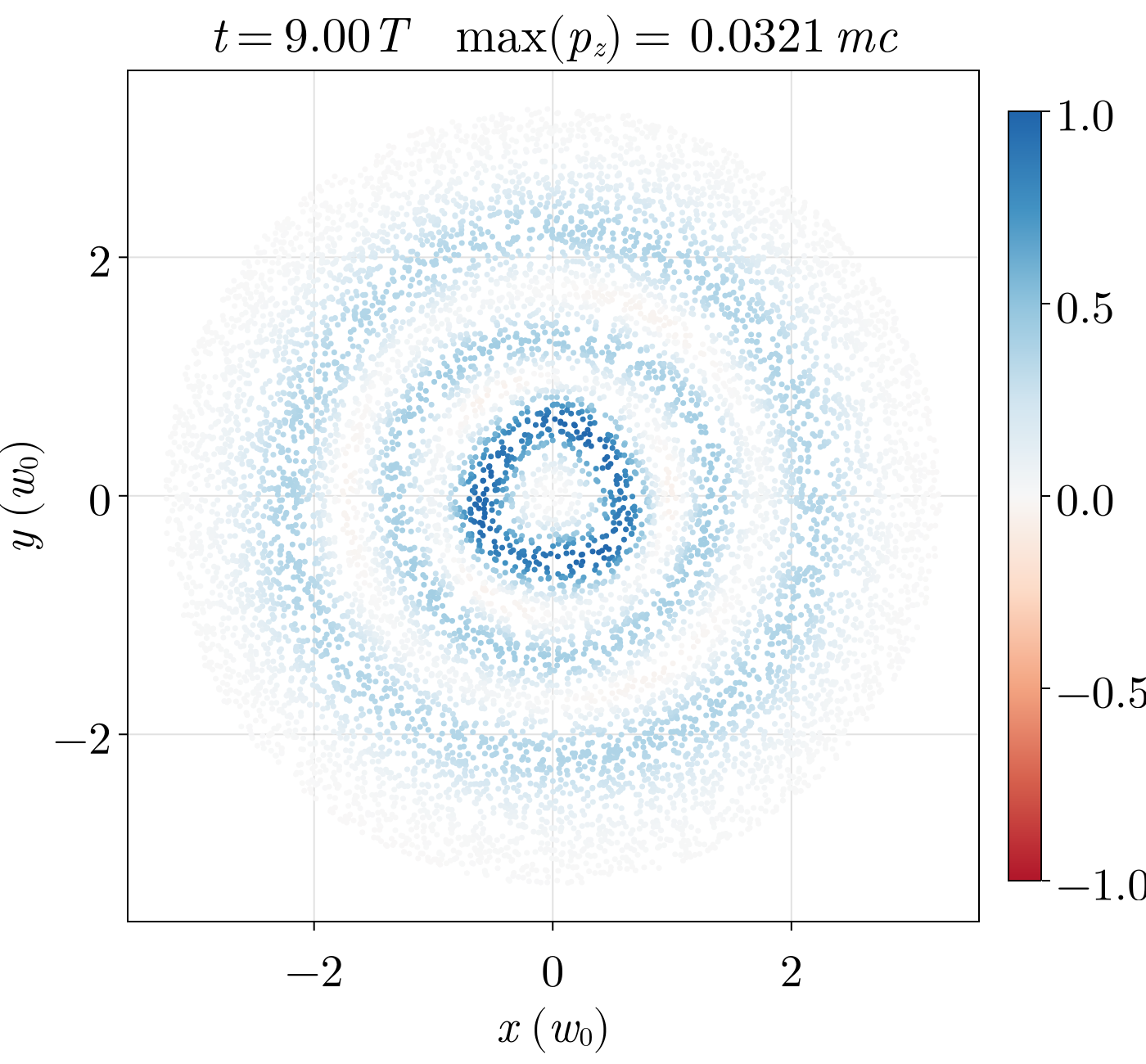}\includegraphics[scale=0.08]{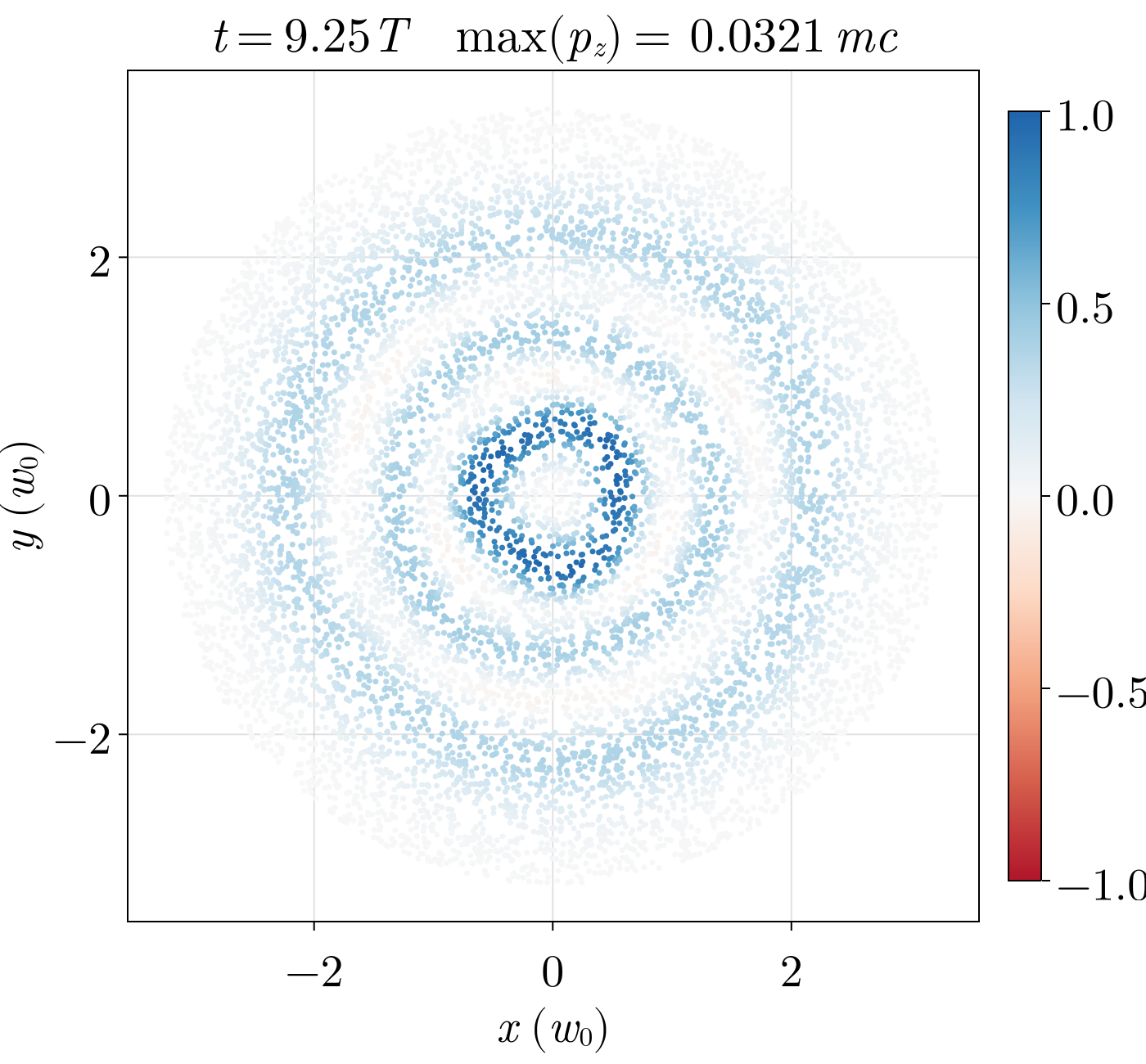}\includegraphics[scale=0.08]{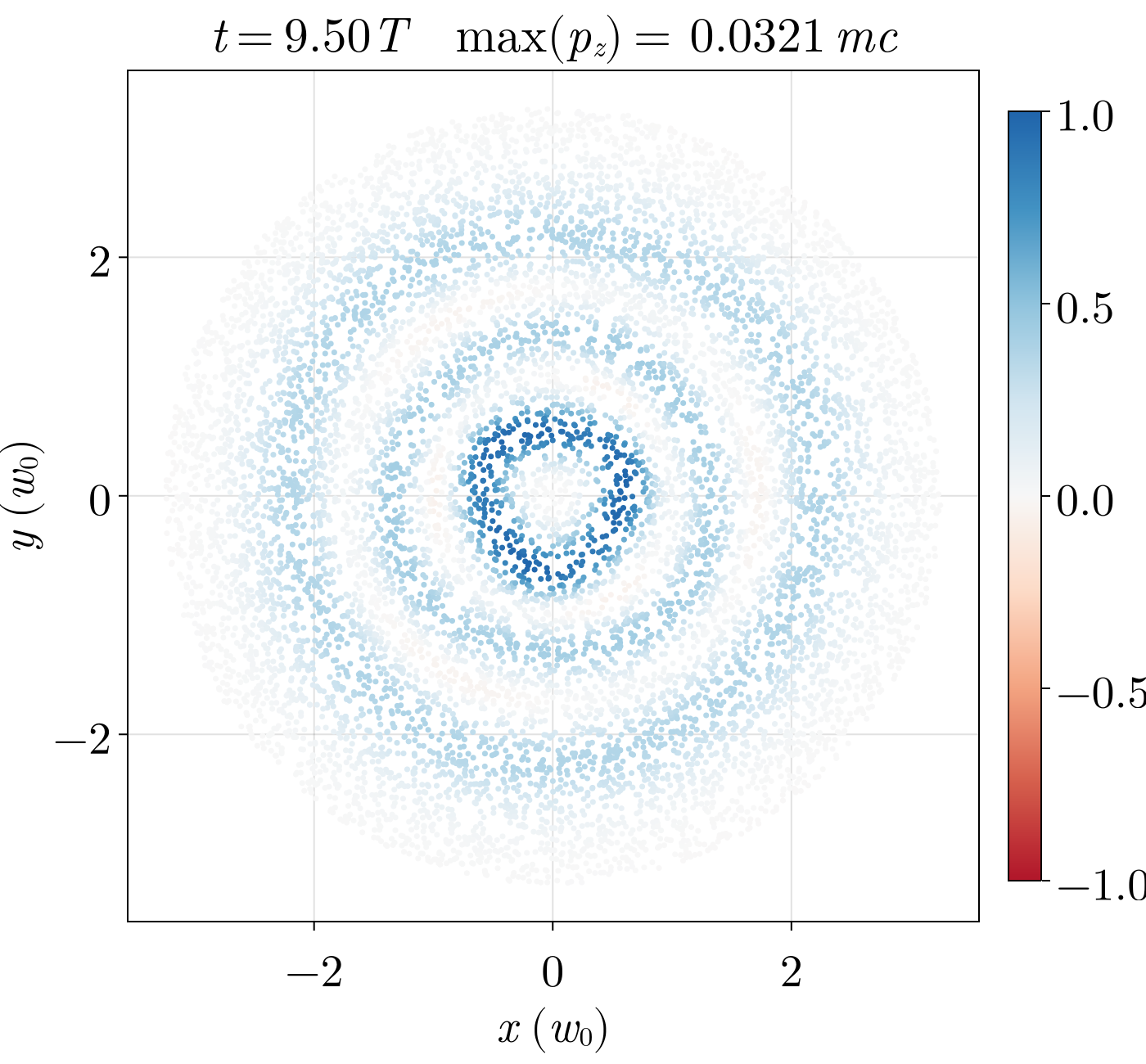}\includegraphics[scale=0.08]{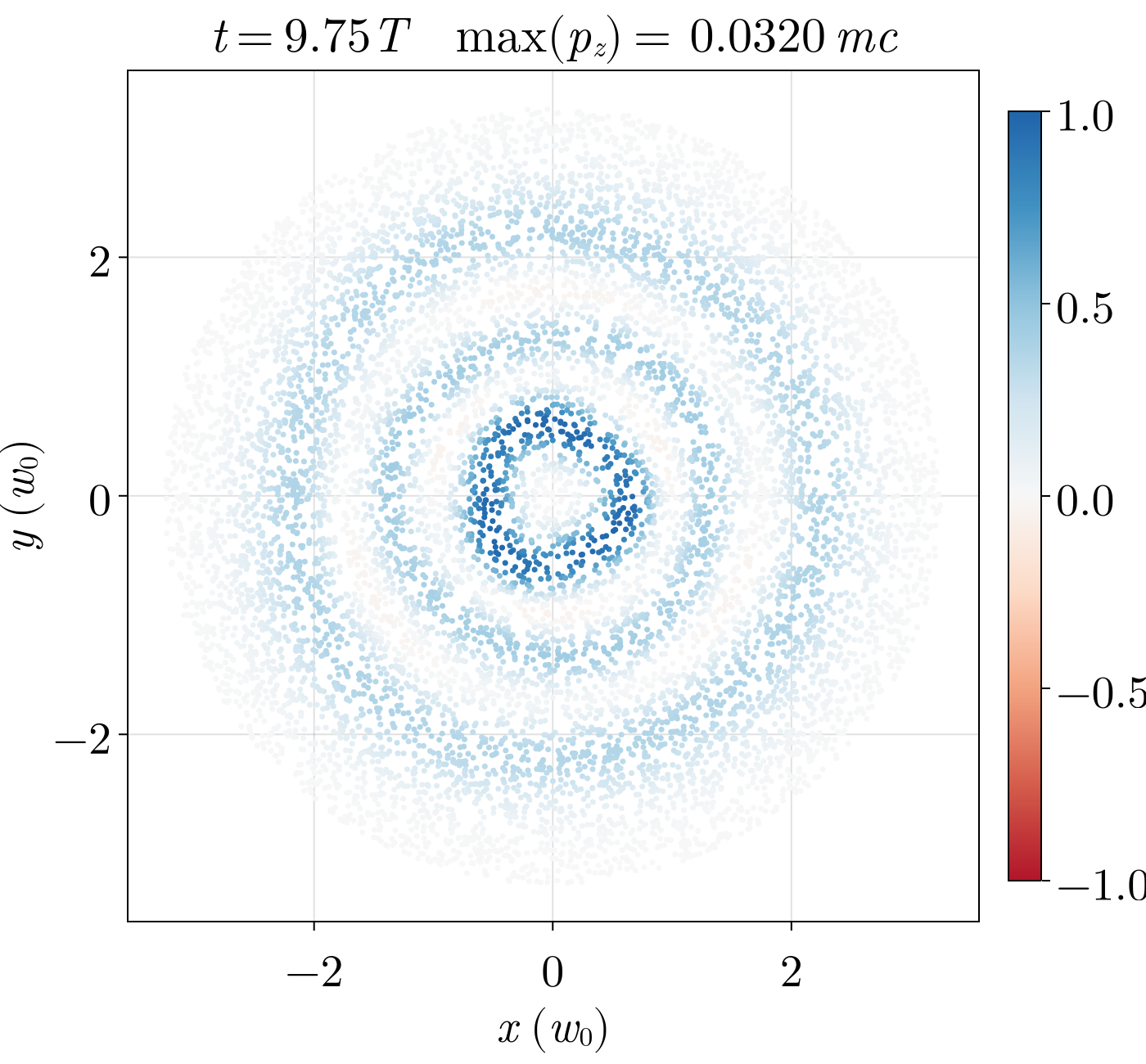}

     \includegraphics[scale=0.08]{supp-fig/f4/dpz/fig_0001.png}\includegraphics[scale=0.08]{supp-fig/f4/dpz/fig_0006.png}\includegraphics[scale=0.08]{supp-fig/f4/dpz/fig_0011.png}\includegraphics[scale=0.08]{supp-fig/f4/dpz/fig_0016.png}\end{center}

     \caption{Snapshots of $p_z^{\text{exact}}$ (upper rows) and $\delta p_z=p_z^{\text{exact}}-p_z^{\text{LPWA}}$ for each particle in a statistical ensemble of electrons in interaction with a LG beam with $p_L=2$, $m_L=2$, $\xi_0=0.5$ for different times $t\in(0,10T)$ written on each graphic; the maximum value of  $|p_z|$ and respectively $|\delta p_z|$  is written at every moment in the title of each graphic. \label{fig-pz}}
\end{figure}

We can see that for each particle $Ox$ component of  the momentum oscillates with the frequency $\omega_L$ of the laser, and the amplitude is essentially constant, dependent only on the initial position of  the electron in the plane.
By comparing the maximum values written in the title of each figure we can see that the error   is at most  4\% at $t=0$  (this is the error accumulated during the turn-on of the field along the envelope $f(\phi)$) and increases linearly in time up to about 7.5\% at $t=10T$. The comparison of the $Oy$ components of the momentum is not presented here since, given  the circular polarization of the field, $p_x$ and $p_y$ behave identically; in Fig. \ref{fig-pz} we present the analysis of the $Oz$ component of the momentum.

In this case the spatial distribution of $p_z$ is almost constant; this property can be explained using the analytical expression of the momentum in a plane wave field (\ref{pan}) with the initial condition $p_0^3=0$, $p_0^0=mc$, ${\bm A}_0=0$. We can see that if the field is circularly polarized $p_z$ is constant, proportional to the square of the field amplitude, so the concentric circles of different amplitudes in Fig. \ref{fig-pz} are in fact a representation of the  transversal intensity profile of the beam. Also in this case, from the graphics  of $\delta p_z$  we  see that the relative error changes between 6\% at $t=0$ and 10\% at $t=10T$. Finally,  by comparing the representations of $p_x$ and $p_z$ it follows that overall the longitudinal component of ${\bm p}$ is about 10\% of the radial component.

\begin{figure}[H]\begin{center}
     \includegraphics[scale=0.08]{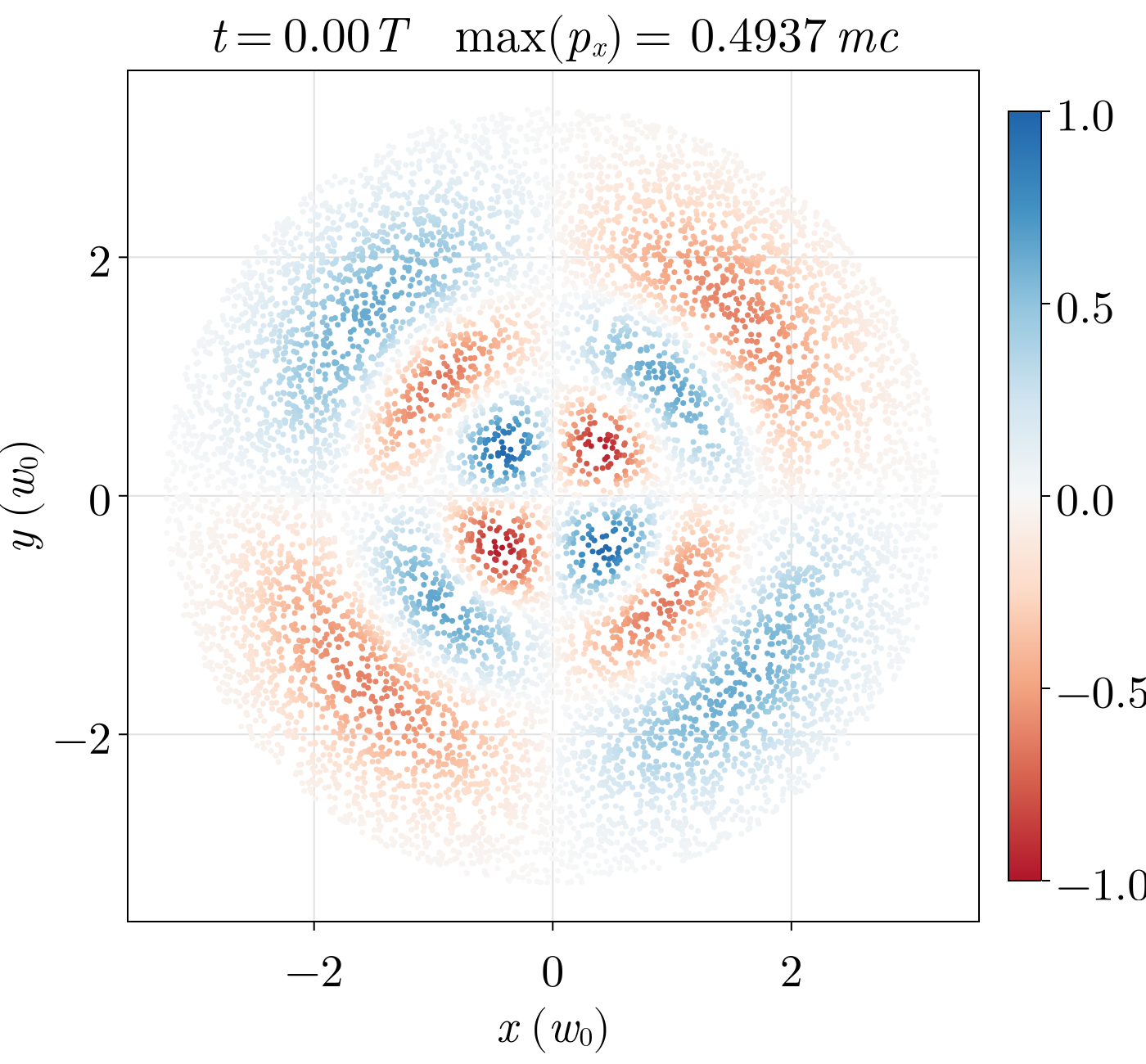}\includegraphics[scale=0.08]{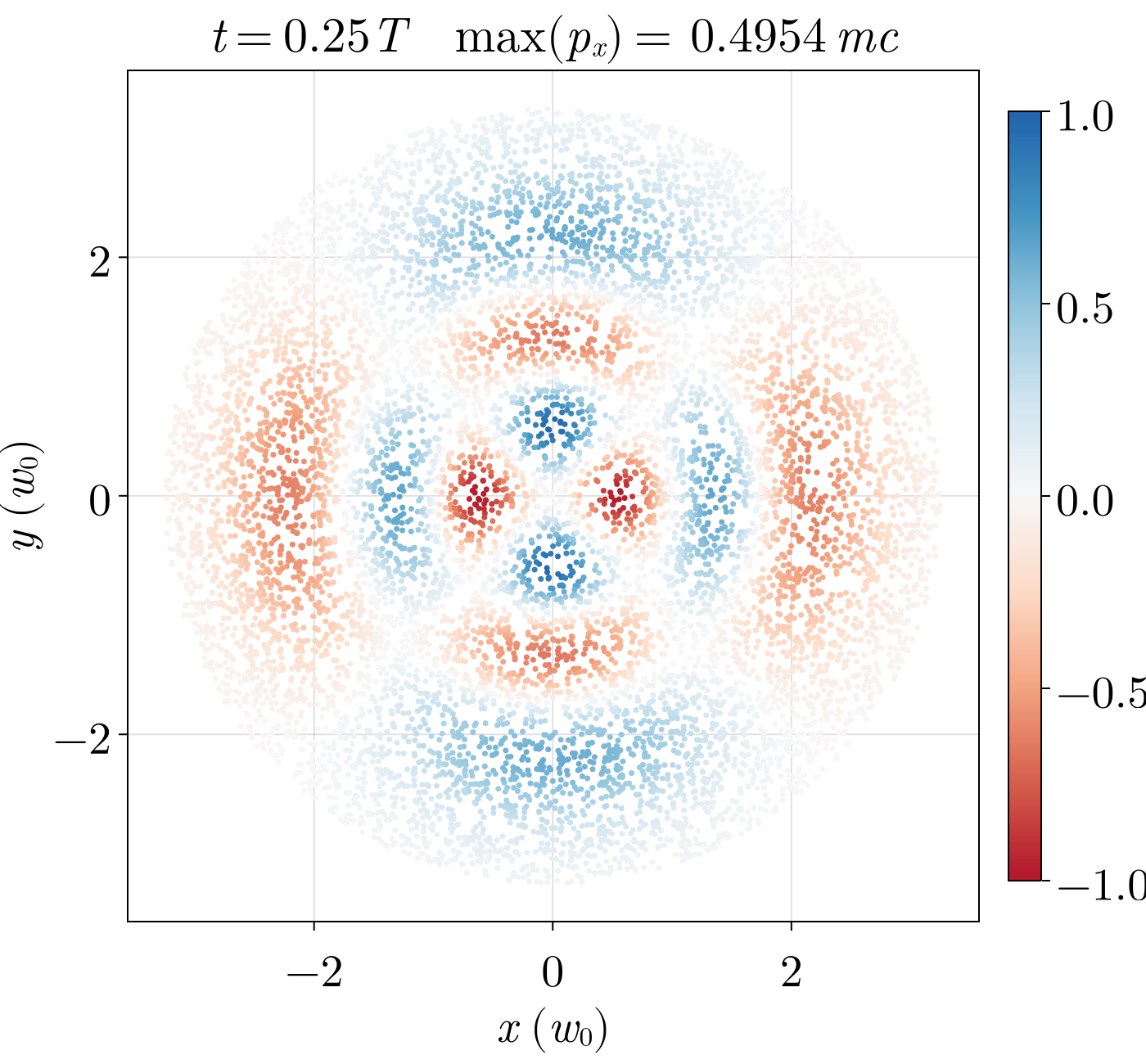}\includegraphics[scale=0.08]{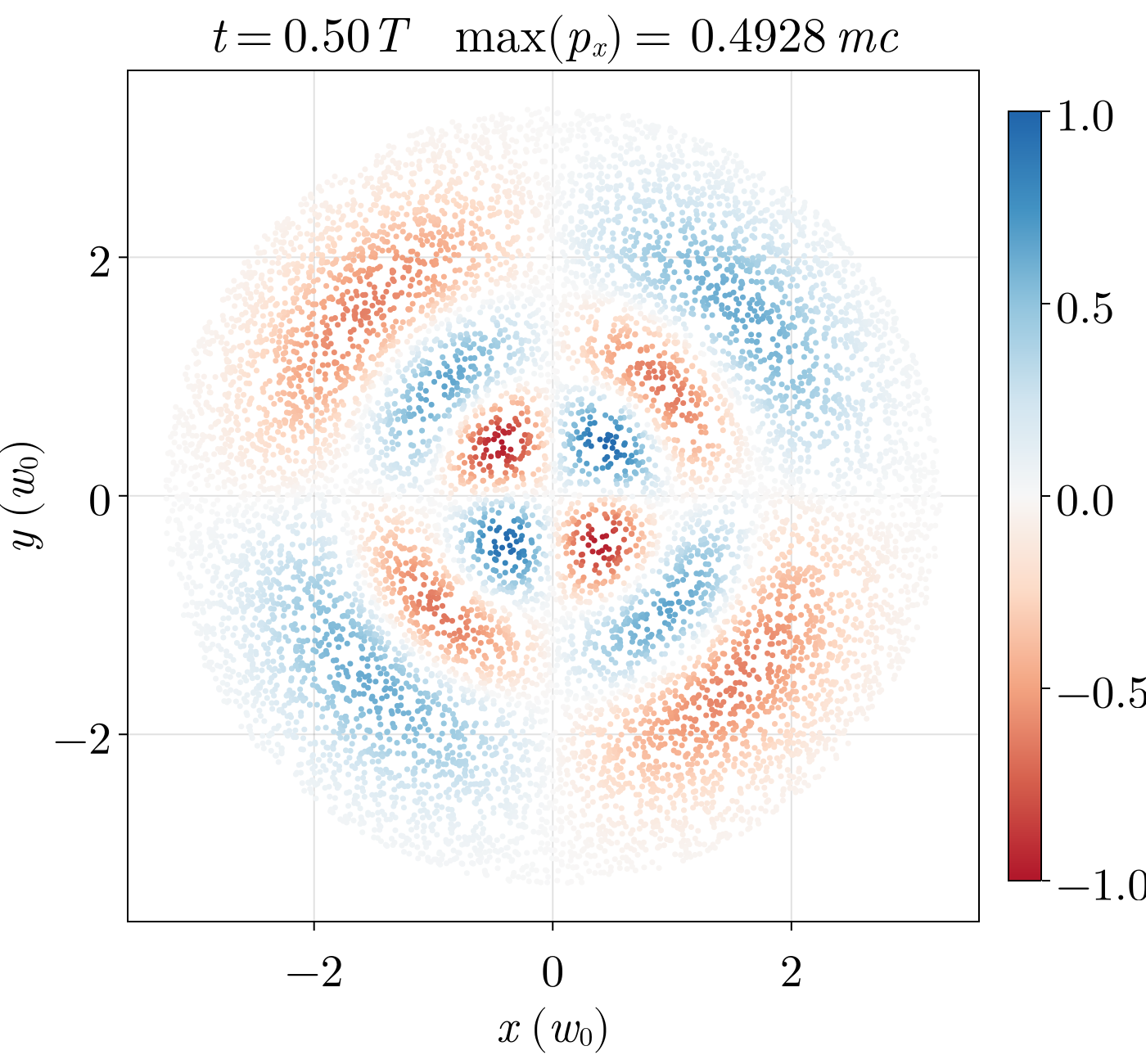}\includegraphics[scale=0.08]{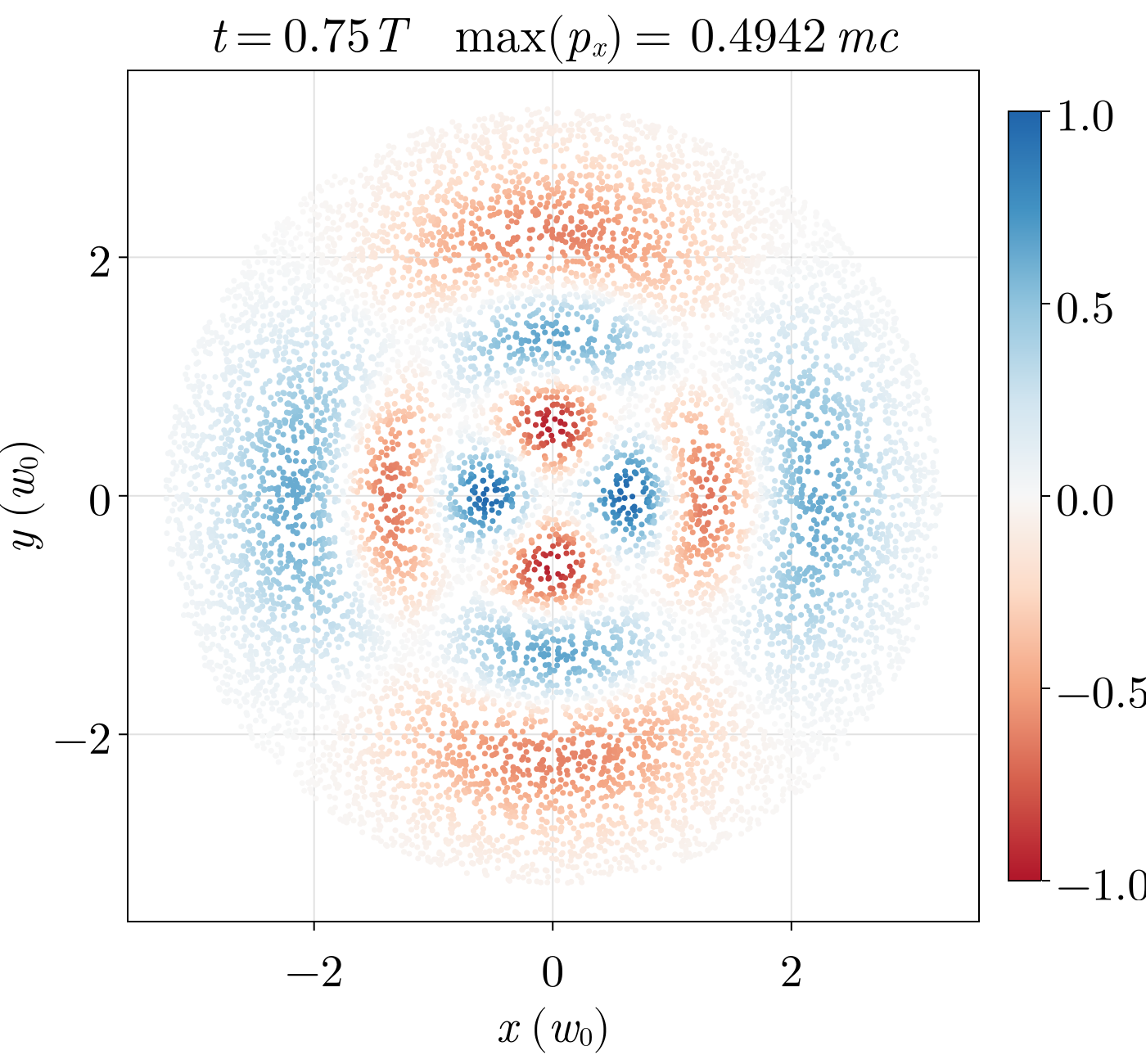}

     \includegraphics[scale=0.08]{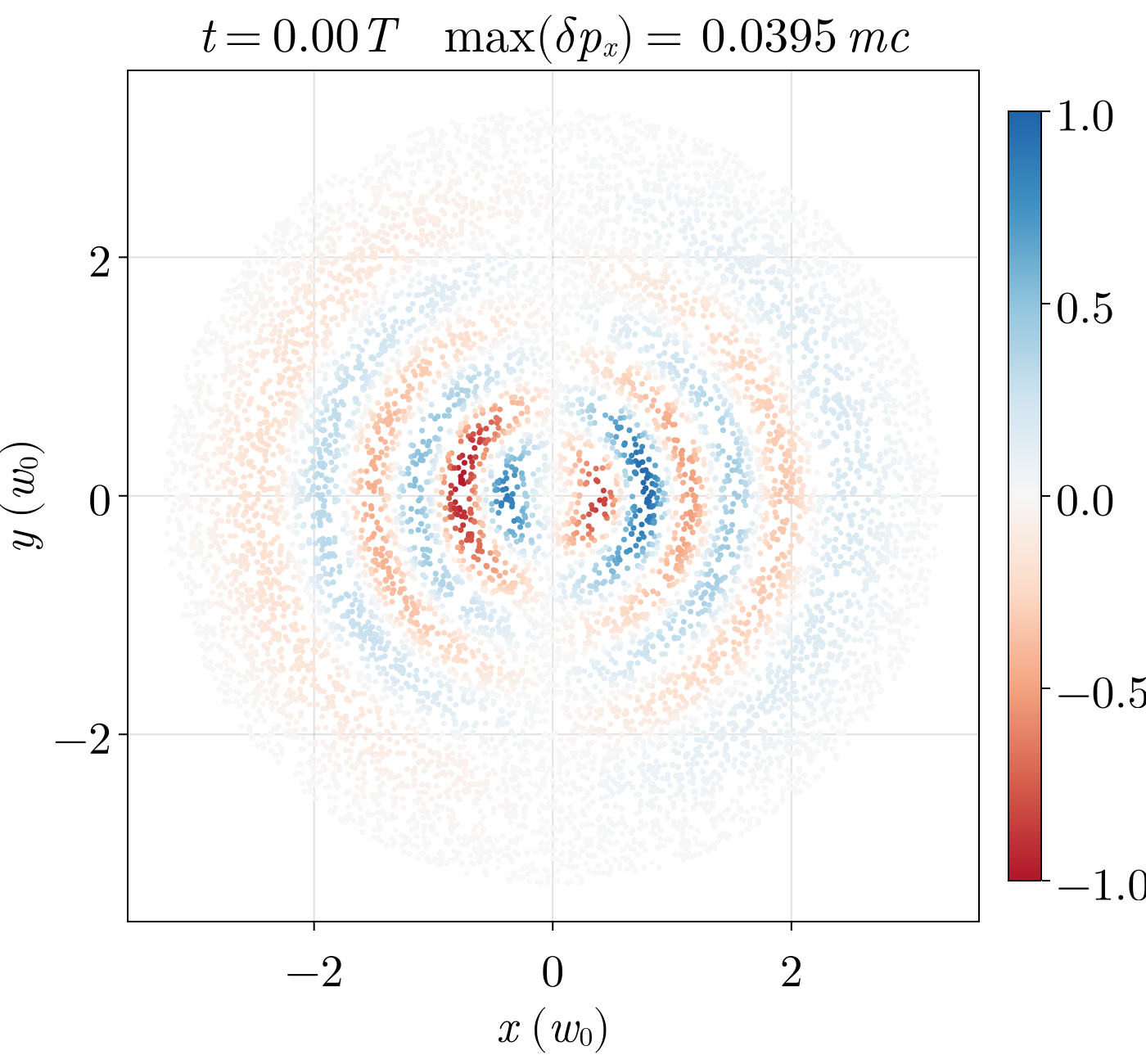}\includegraphics[scale=0.08]{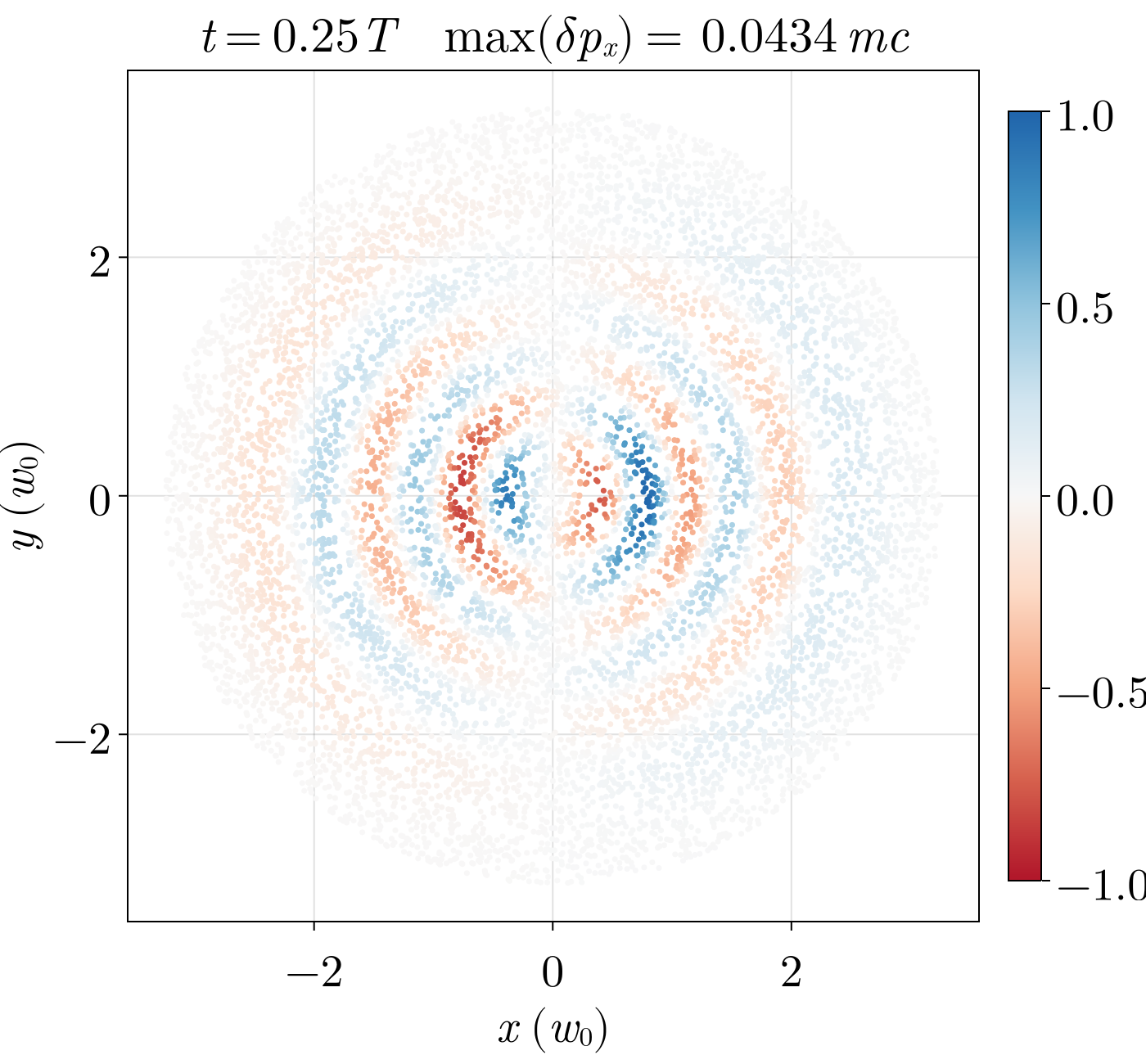}\includegraphics[scale=0.08]{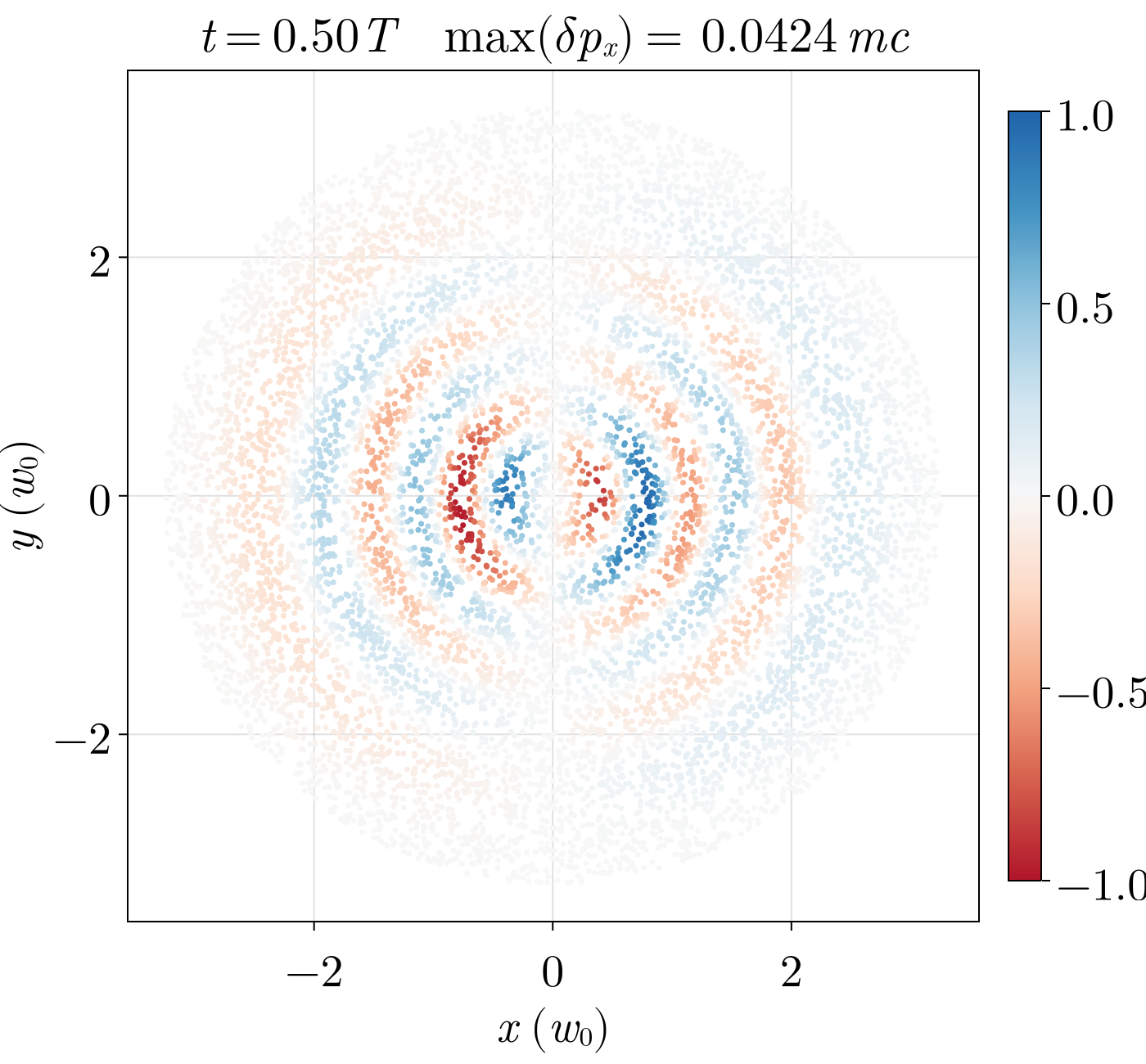}\includegraphics[scale=0.08]{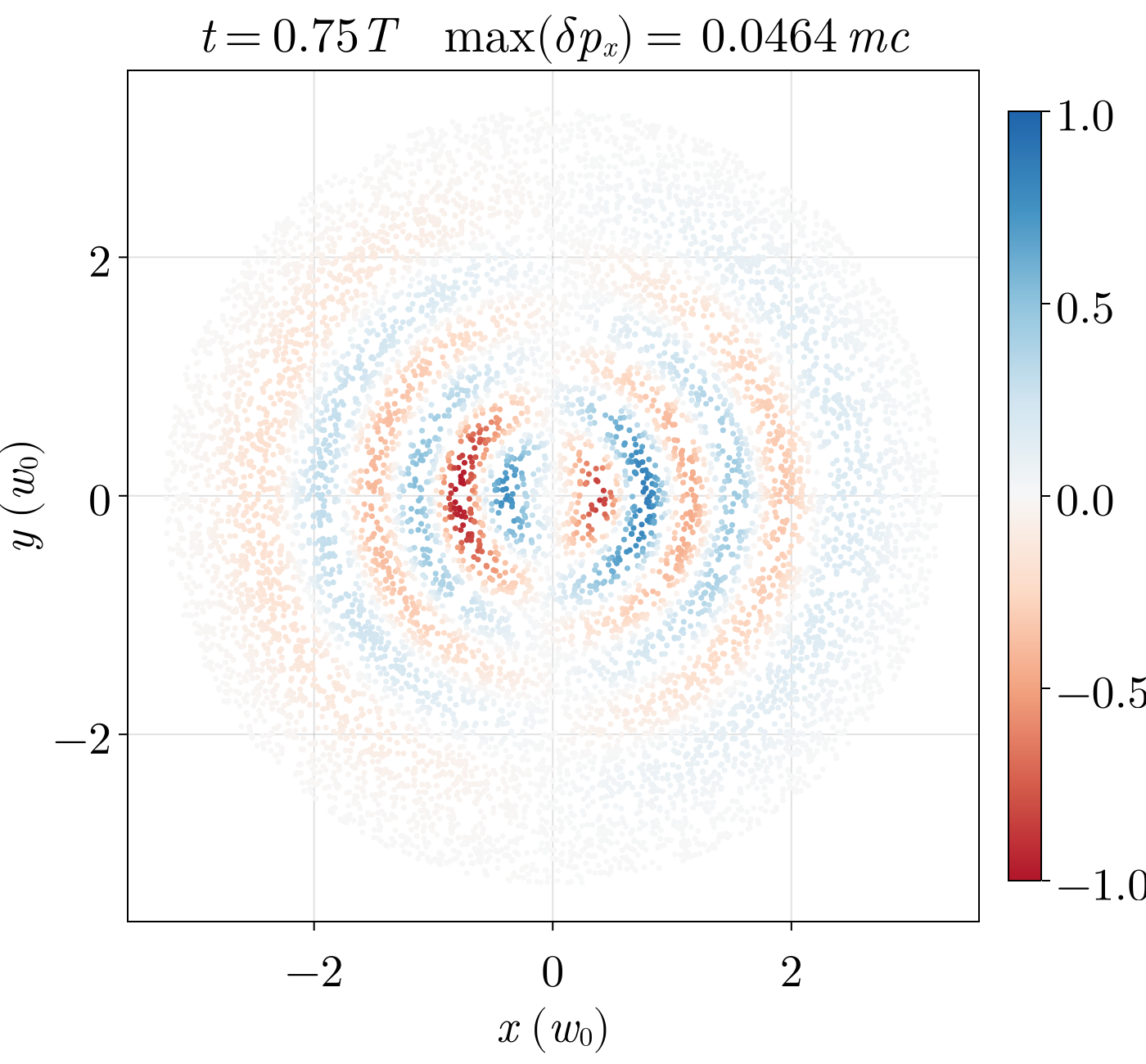}

     \vspace*{1cm}

     \includegraphics[scale=0.08]{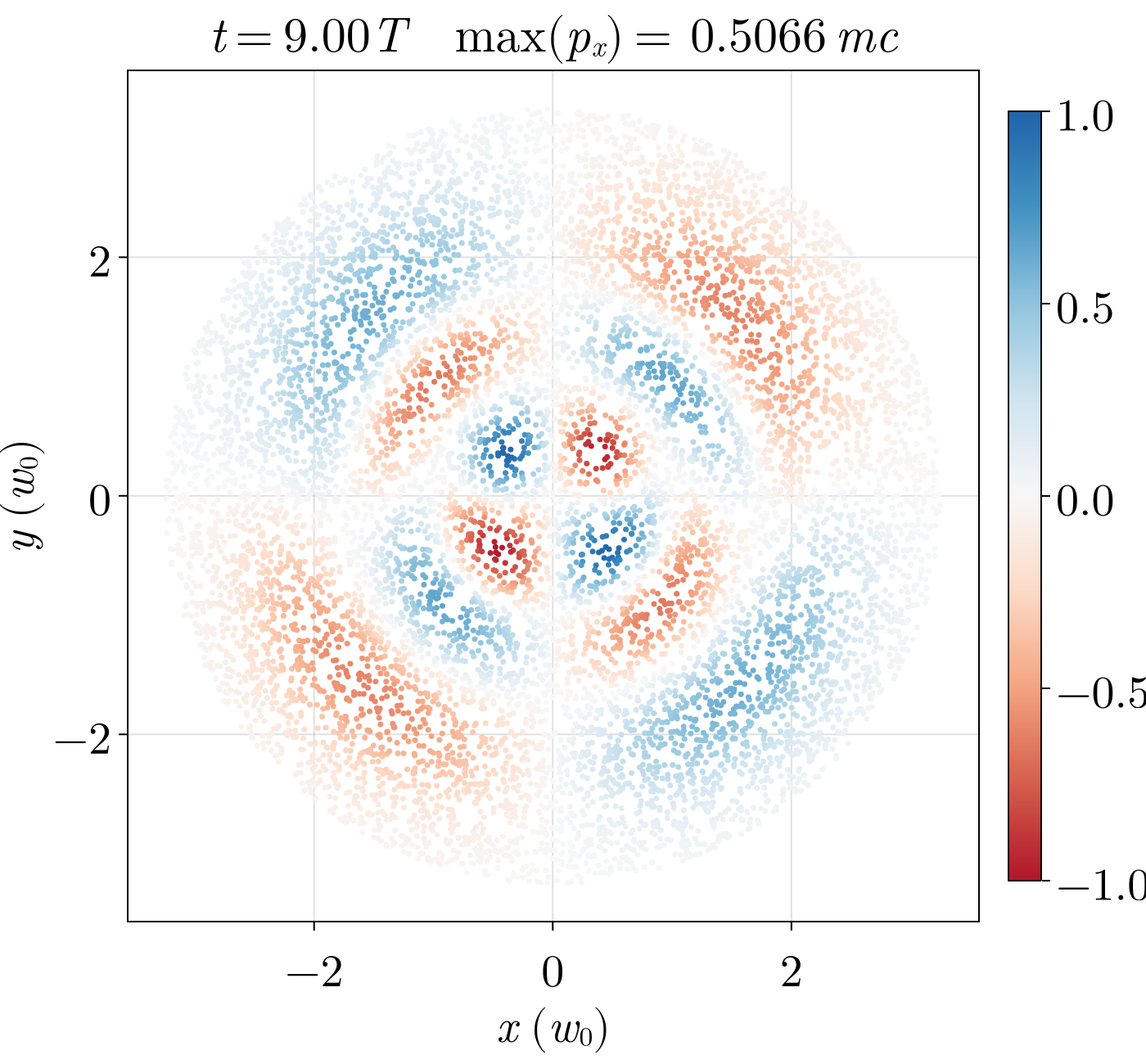}\includegraphics[scale=0.08]{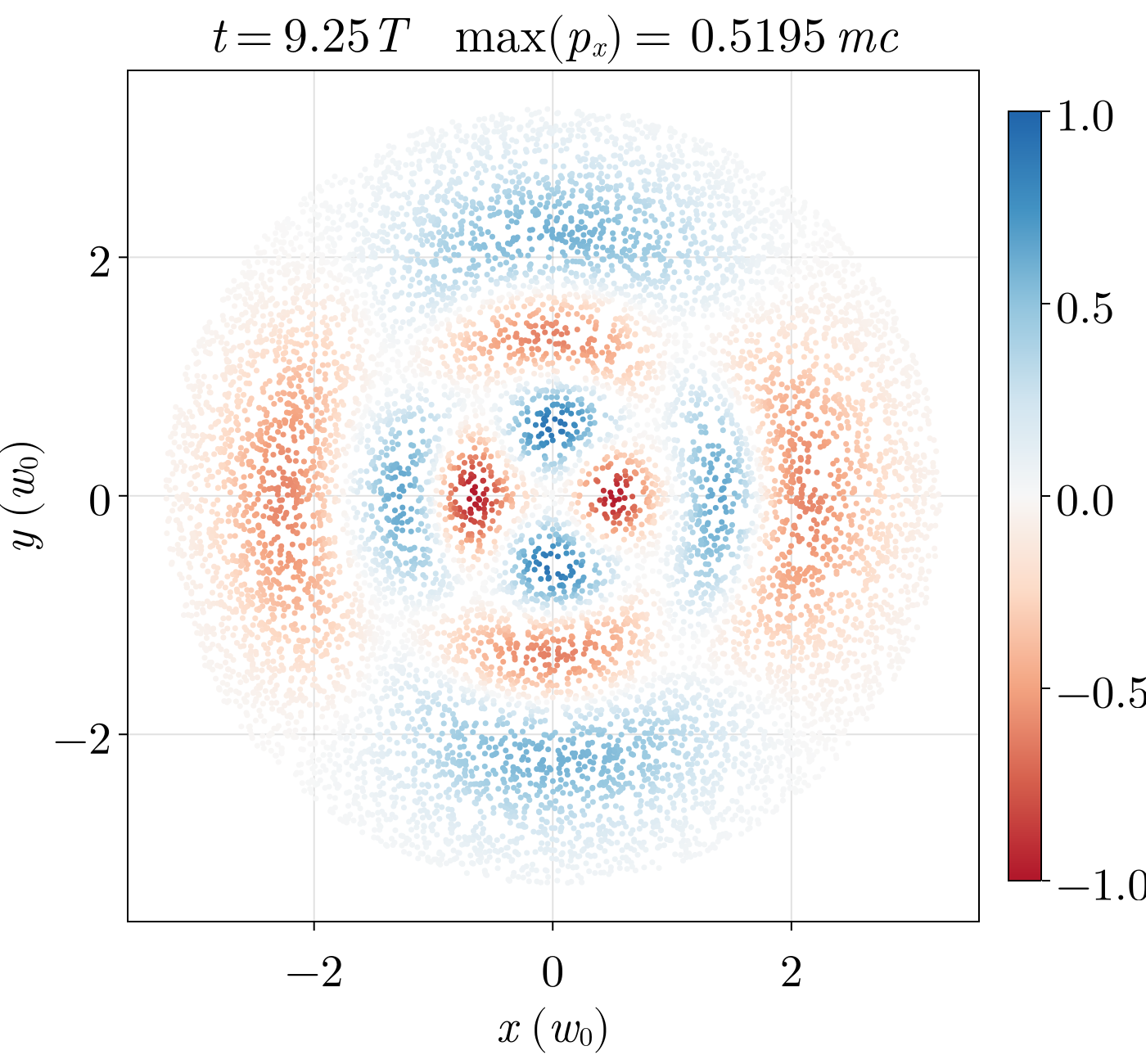}\includegraphics[scale=0.08]{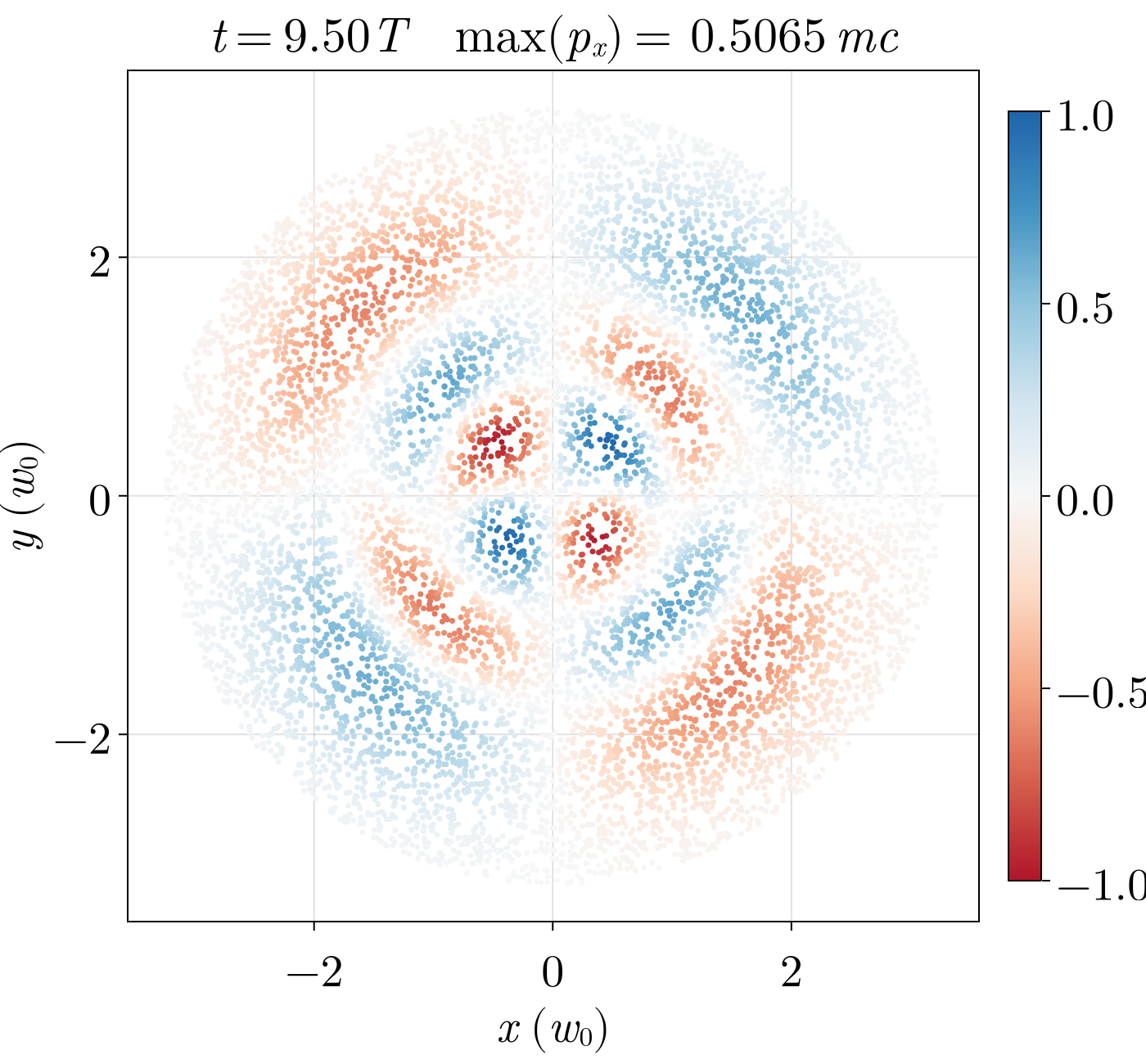}\includegraphics[scale=0.08]{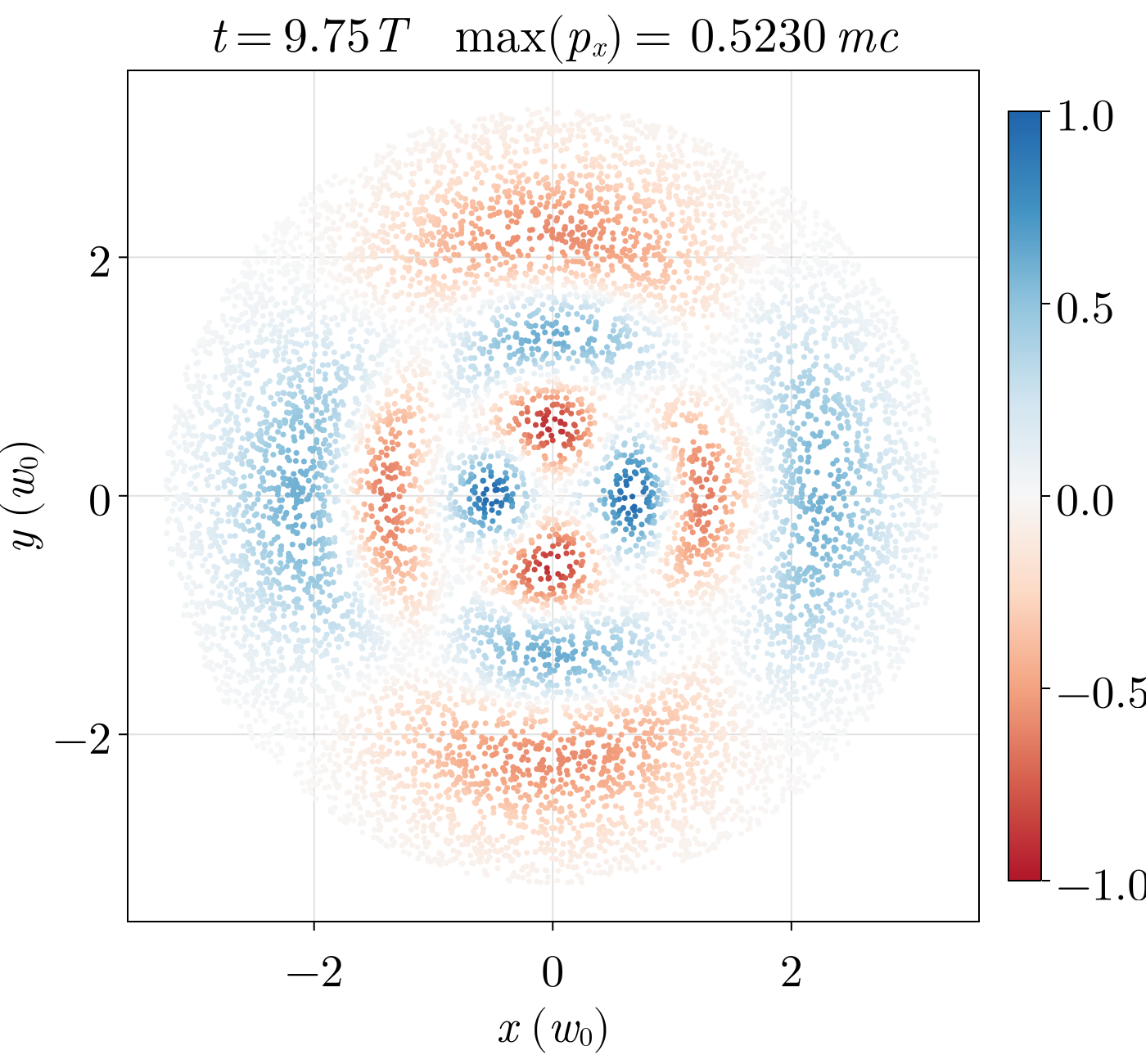}

     \includegraphics[scale=0.08]{supp-fig/f5/dpx/fig_0001.png}\includegraphics[scale=0.08]{supp-fig/f5/dpx/fig_0006.png}\includegraphics[scale=0.08]{supp-fig/f5/dpx/fig_0011.png}\includegraphics[scale=0.08]{supp-fig/f5/dpx/fig_0016.png}\end{center}

     \caption{The same as \ref{fig-px} but for $\xi_0 = 1.0$\label{fig-px2}}
\end{figure}

\begin{figure}[H]\begin{center}
     \includegraphics[scale=0.08]{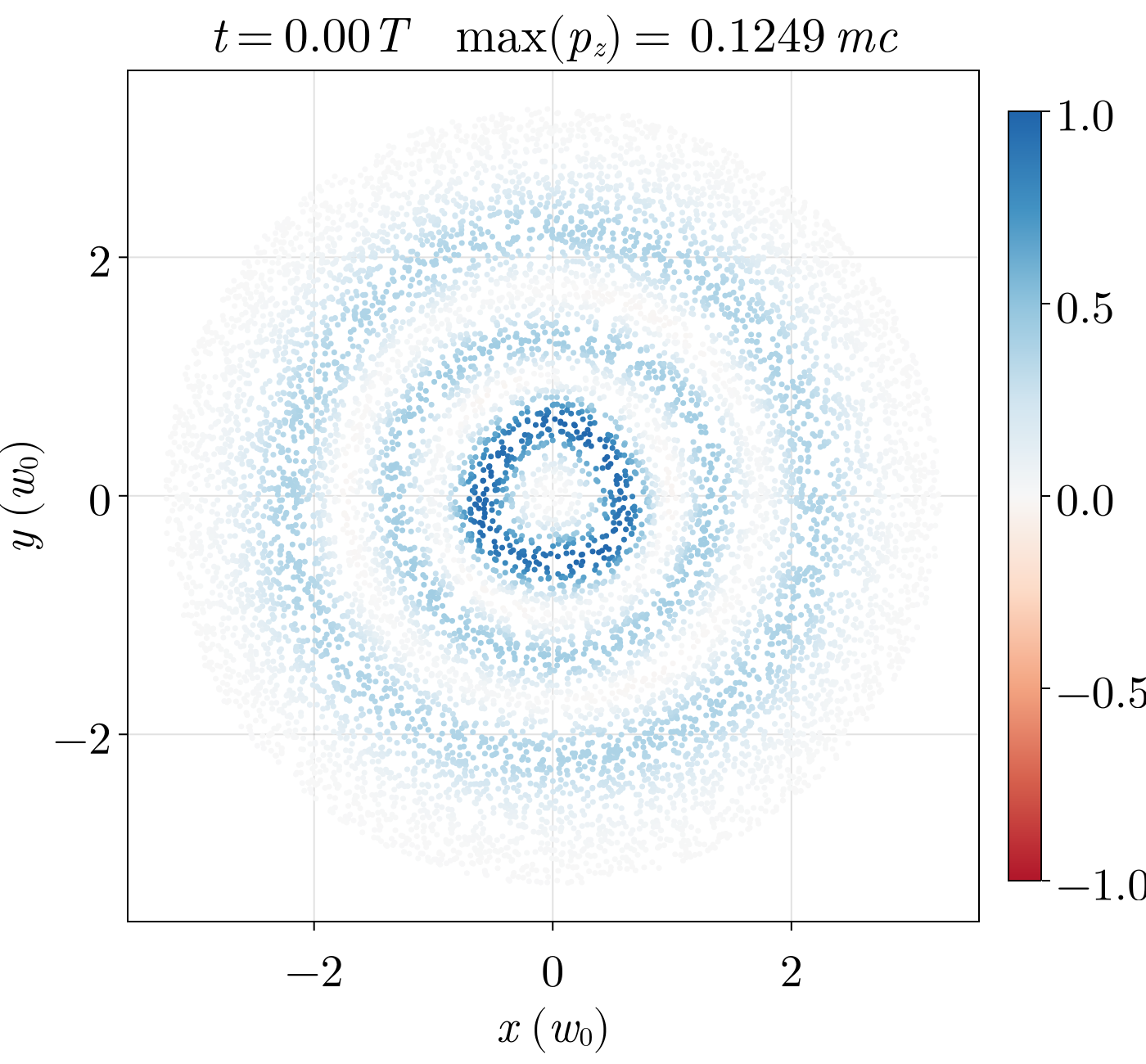}\includegraphics[scale=0.08]{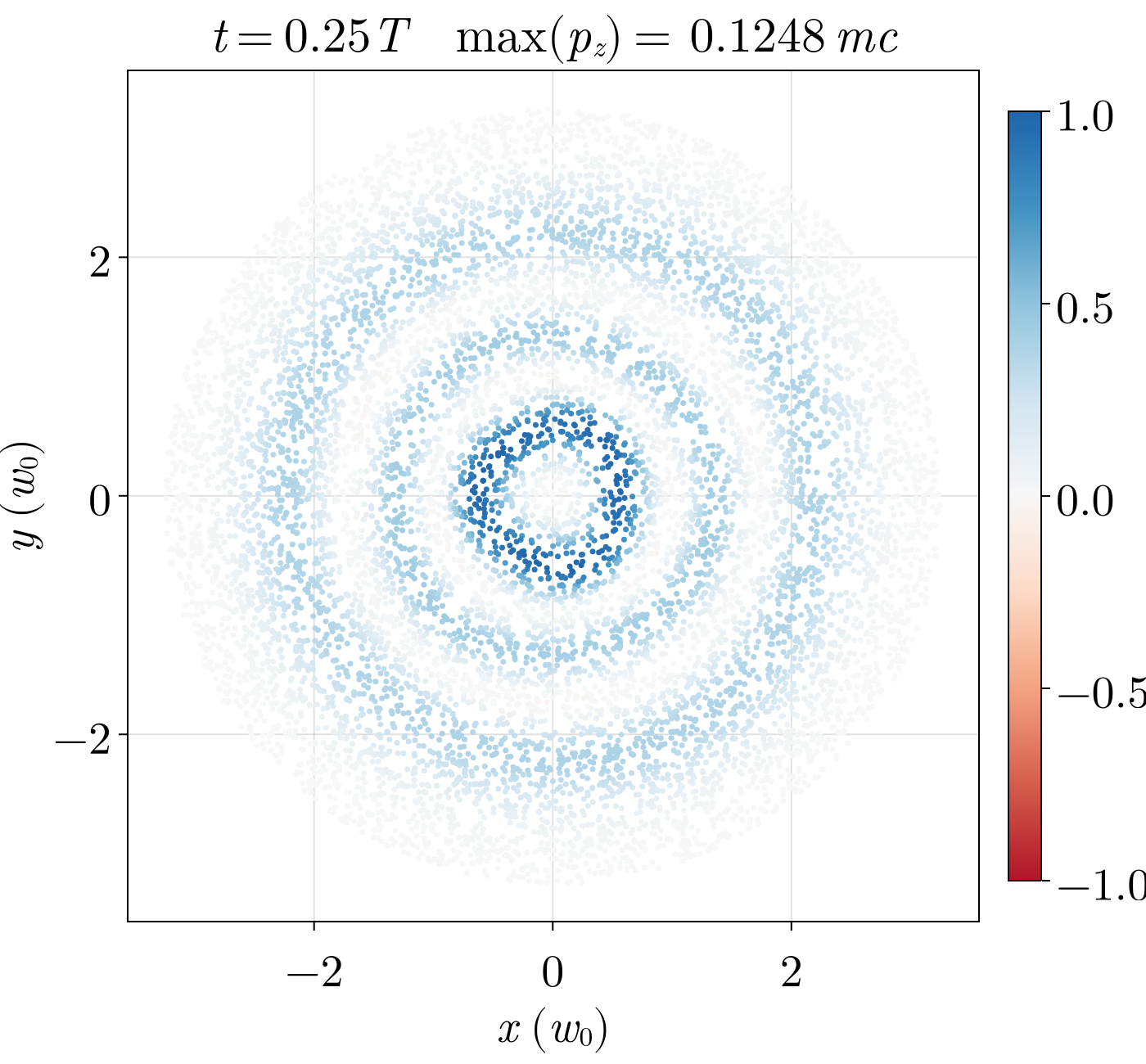}\includegraphics[scale=0.08]{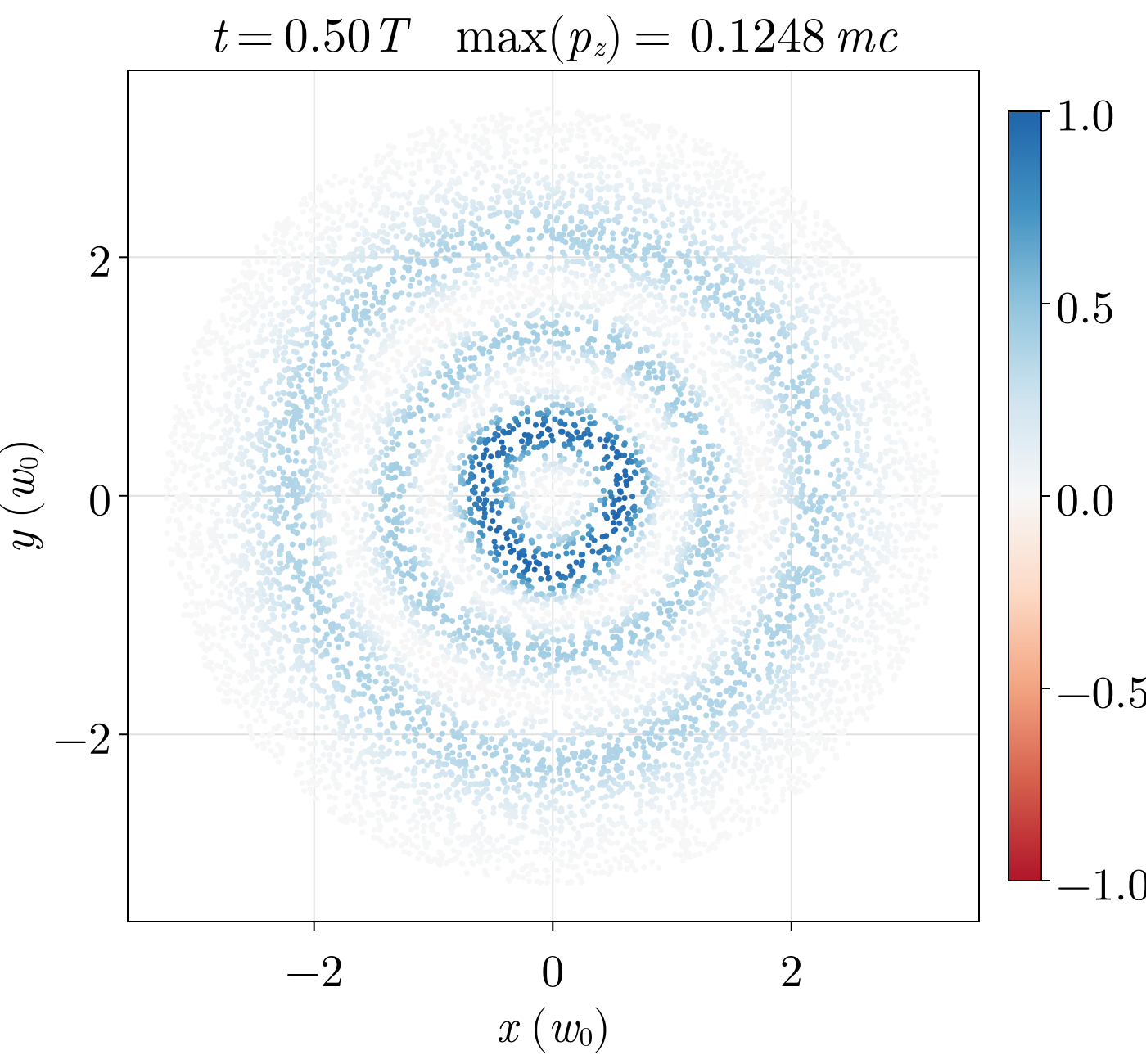}\includegraphics[scale=0.08]{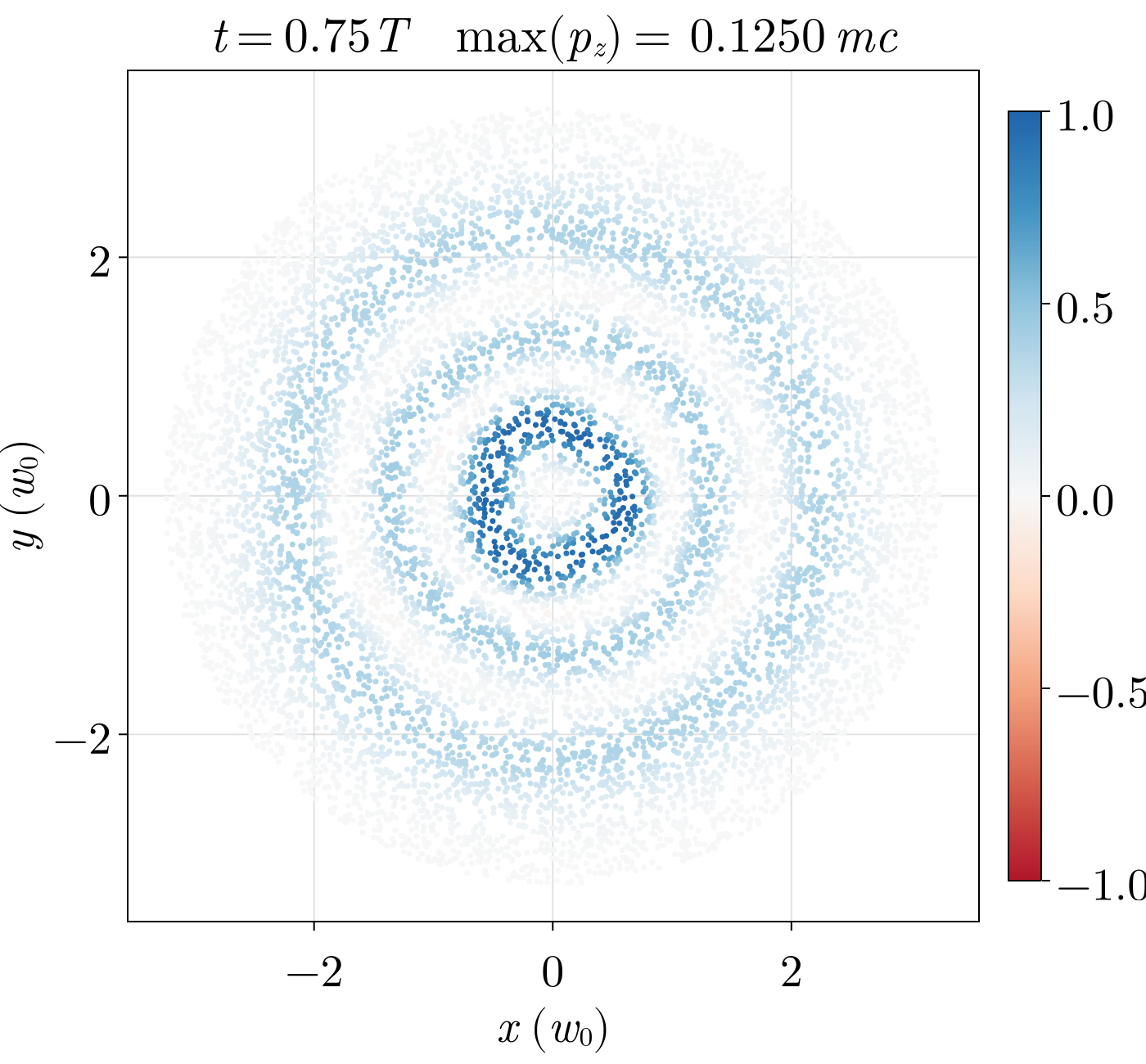}

     \includegraphics[scale=0.08]{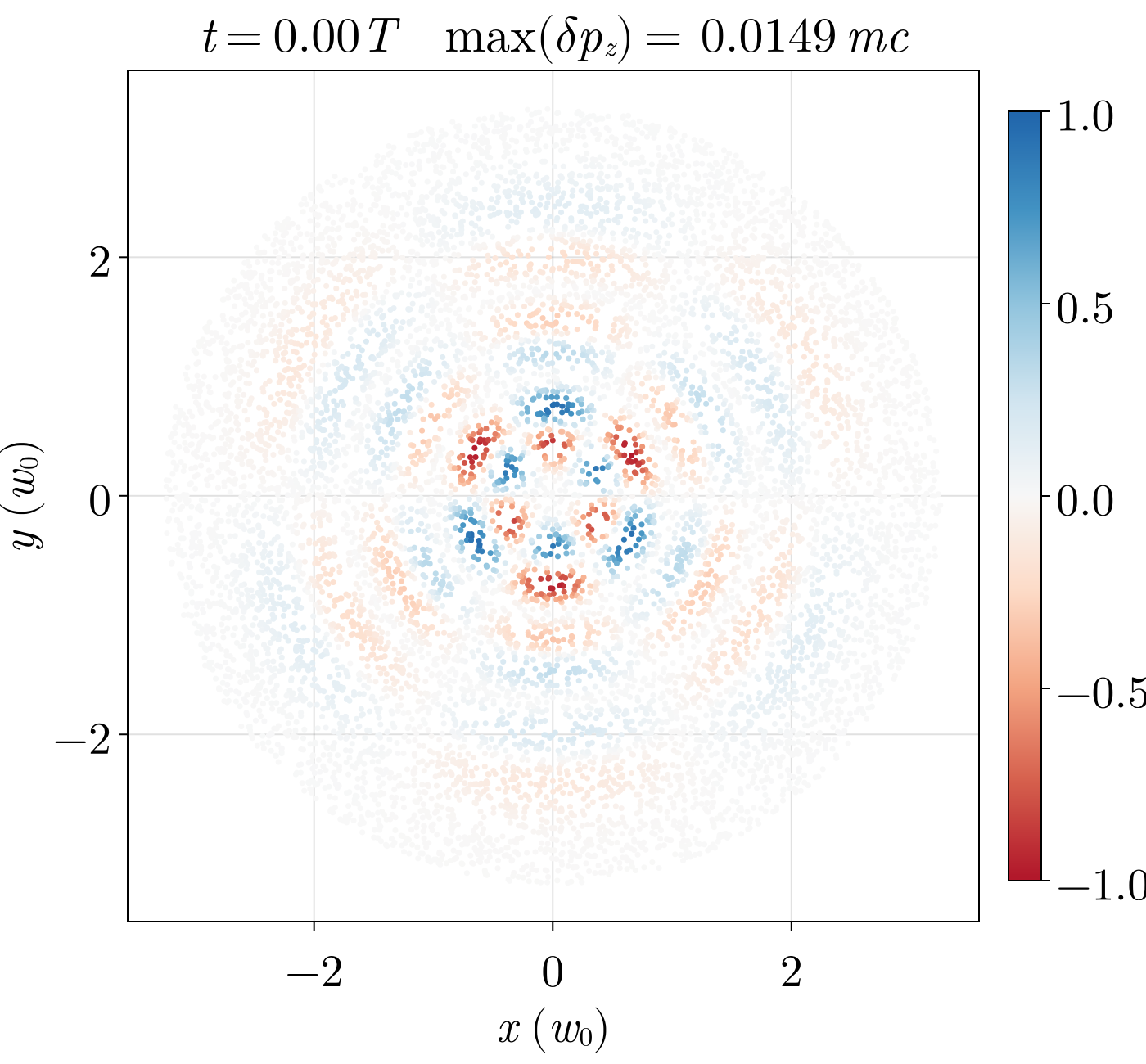}\includegraphics[scale=0.08]{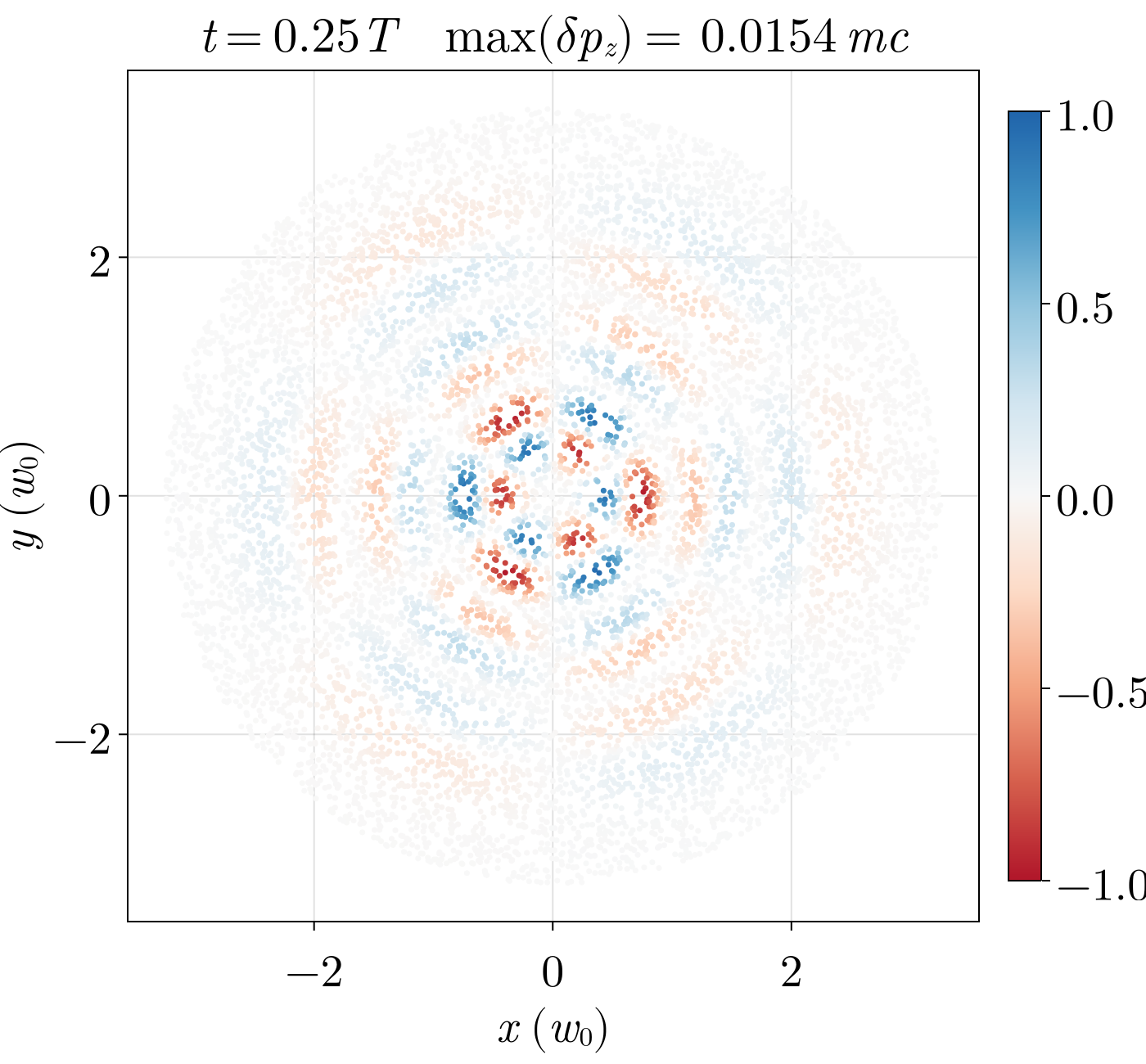}\includegraphics[scale=0.08]{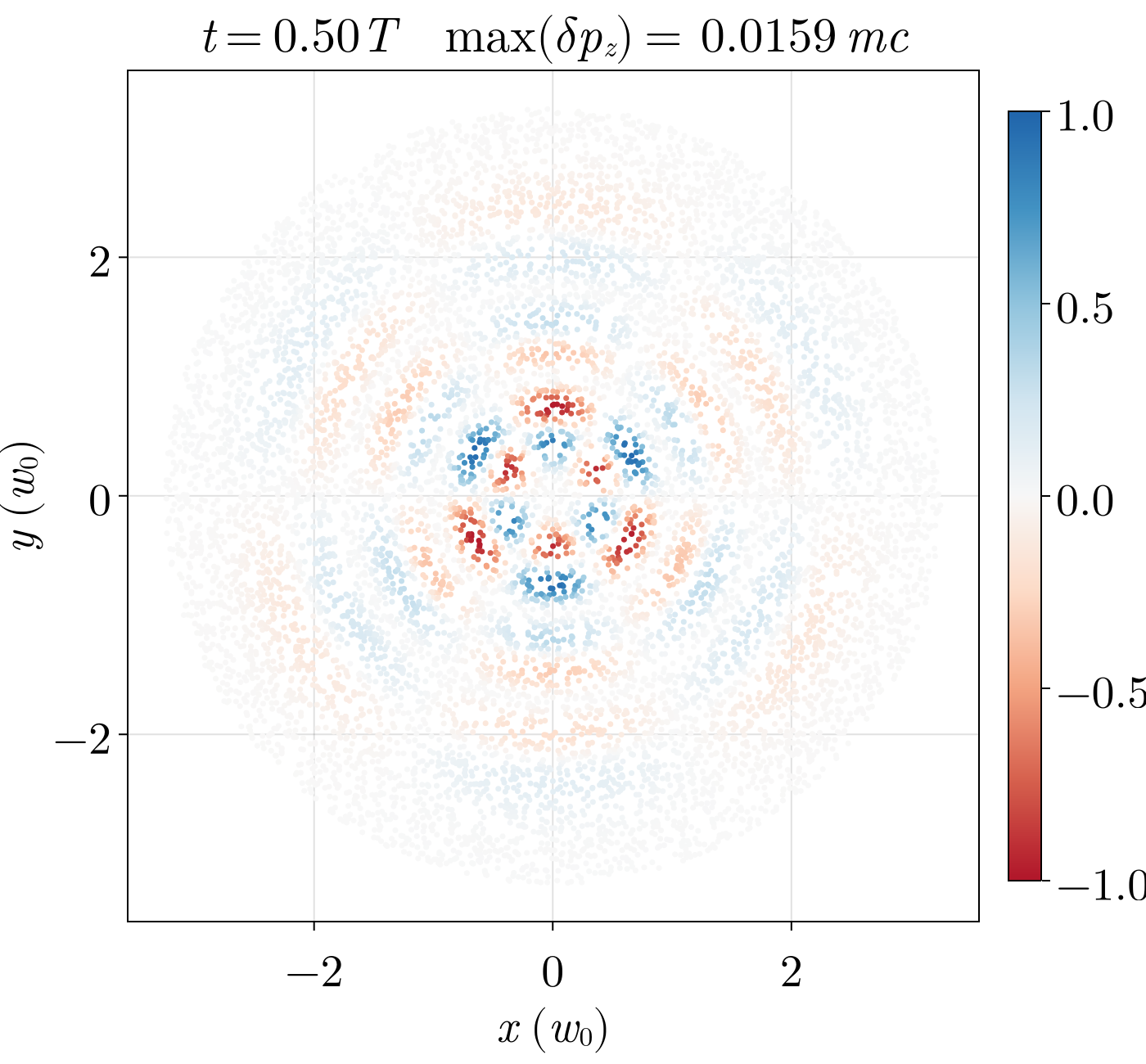}\includegraphics[scale=0.08]{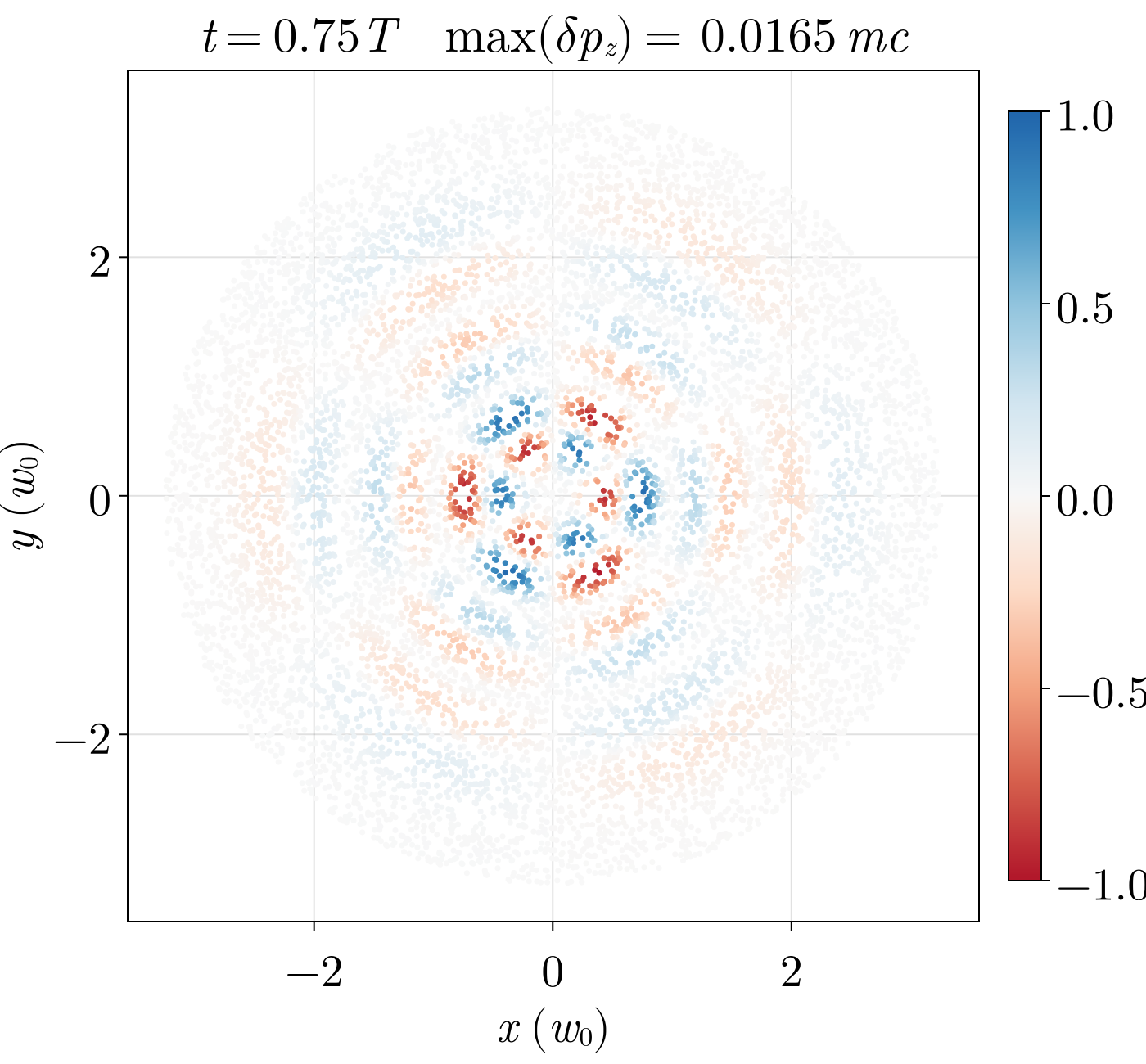}

     \vspace*{1cm}

     \includegraphics[scale=0.08]{supp-fig/f4/pz/fig_0181.png}\includegraphics[scale=0.08]{supp-fig/f4/pz/fig_0186.png}\includegraphics[scale=0.08]{supp-fig/f4/pz/fig_0191.png}\includegraphics[scale=0.08]{supp-fig/f4/pz/fig_0196.png}

     \includegraphics[scale=0.08]{supp-fig/f4/dpz/fig_0001.png}\includegraphics[scale=0.08]{supp-fig/f4/dpz/fig_0006.png}\includegraphics[scale=0.08]{supp-fig/f4/dpz/fig_0011.png}\includegraphics[scale=0.08]{supp-fig/f4/dpz/fig_0016.png}\end{center}

     \caption{The same as \ref{fig-pz} but for $\xi_0 = 1.0$\label{fig-pz2}}
\end{figure}

Next, we have investigated the behavior with respect to $\xi_0$; in Figs. \ref{fig-px2}, \ref{fig-pz2} we present the same comparison for $\xi_0=1$

Firstly, we notice that the orthogonal component of ${\bm p}$ increases proportional to $\xi_0$, and $p_z$ depends quadratically on $\xi_0$, as predicted by the analytical approximation (\ref{pan}). Also the overall errors are larger in this case: the relative error of $p_x$ changes from 8\% to 16\% in the time interval $t\in(0,10T)$ while for $p_z$ the error changes from 12\% to 25\%.

We have checked that properties presented here  for $p_L=2$, $m_L=2$ and circular polarization remain valid also for different indices $p$ and $m$ and for elliptic or linear polarization; also the behavior with the waist $w_0$ is that expected, i.e. the  accuracy of LPWA improves when $w_0$ increases.

The conclusion of the present analysis is that LPWA is sufficiently accurate for intensities $\xi_0 \lessapprox1$ and $w_0 \ge 75$, the values used in our numerical simulations.

\section{Thomson scattering of Laguerre Gauss beams}

\subsection{Fourier analysis of the emitted fields - large distance approximation}

Here we consider only the large distance  approximation of the electromagnetic fields emitted by a particle of charge $e_0$, moving along the trajectory ${\bm r}(t)$, in the vicinity of the point $ {\boldsymbol{\mathfrak{R}}}_0$ (here ${\boldsymbol{\mathfrak{R}}}_0$ is the initial position of the particle.)
\begin{align}
    {\bm E}({\bm x},t)={\frac{e_0}{4\pi\epsilon_0|{\bm x}_0|}\frac{{\bm n}_{0}\times[({\bm n}_{0}-{\boldsymbol\beta})\times {\bm a}]}{c^2(1-{\boldsymbol\beta}\cdot{\bm n}_{0})^3}\,\vline}_{\,t=t_r},\qquad {\bm B}({\bm x},t) = \frac1c{\bm n}_0\times{\bm E}({\bm x},t)
\end{align}
where ${\bm x}$ is the observation point,
\begin{align}
    {\bm x}_0={\bm x}-{\boldsymbol{\mathfrak{R}}}_0,\qquad {\bm n}_0=\frac{{\bm x}_0}{|{\bm x}_0|}\label{defn01}
\end{align}
$\boldsymbol{\beta}$ is the velocity measured in units of $c$ and ${\bm a}$ the particle acceleration.
We define the temporal Fourier transform
\begin{align}
    {\bm E}({\bm x},k) = \frac1{\sqrt{2\pi}}\int\limits_{-\infty}^{\infty} d (ct) {\bm E}({\bm x},t)e^{ik  (ct)}=\frac1{\sqrt{2\pi}}\int\limits_{-\infty}^{\infty} d (ct) {\frac{q}{4\pi\epsilon_0|{\bm x}_0|}\frac{{\bm n}_{0}\times[({\bm n}_{0}-{\boldsymbol\beta})\times {\bm a}]}{c^2(1-{\boldsymbol\beta}\cdot{\bm n}_{0})^3}\,\vline}_{\,t=t_r}e^{ik(ct)}
\end{align}
and we use the retarded time as a integration variable
\begin{align}
    t=t_r+\frac{|{\bm x}_0-{\bm r}_0|}c\approx t_r+\frac{|{\bm x}_0|}c-\frac{{\bm n}_0\cdot{\bm r}_0}c,\qquad dt=dt_r(1-{\boldsymbol{\beta}}\cdot{\bm n}_{R_0})\approx dt_r(1-{\boldsymbol{\beta}}\cdot{\bm n}_0);
\end{align}
with these we obtain
\begin{align}
    {\bm E}({\bm x},k) = \frac{q}{4\pi\epsilon_0|{\bm x}_0|}\frac1{\sqrt{2\pi}}\int\limits_{-\infty}^{\infty}d(ct_r) \frac{{\bm n}_{0}\times[({\bm n}_{0}-{\boldsymbol\beta})\times {\bm a}]}{c^2(1-{\boldsymbol\beta}\cdot{\bm n}_{0})^2}e^{ik((ct_r)+|{\bm x}_0|-{\bm n}_0\cdot{\bm r}_0)}
\end{align}

Next we use the identity
\begin{align}
    \frac{d}{dt}\frac{{\bm n}_0\times({\bm n}_0\times{\boldsymbol\beta})}{1-{\boldsymbol{\beta}\cdot{\bm n}_0}} & = \frac1{c({1-{\boldsymbol{\beta}\cdot{\bm n}_0}})^2}\left[{\bm n}_0({\bm n}_0\cdot{\bm a})-{\bm a}+{\bm a}({\boldsymbol{\beta}}\cdot{\bm n}_0)-{\boldsymbol{\beta}}({\bm n}_0\cdot{\bm a})\right]                                                 =\frac{{\bm n}_{0}\times[({\bm n}_{0}-{\boldsymbol\beta})\times {\bm a}]}{c(1-{\boldsymbol\beta}\cdot{\bm n}_{0})^2}
\end{align}
and we write an equivalent form of the Fourier transform
\begin{align}
    {\bm E}({\bm x},k) = \frac{e_0}{4\pi\epsilon_0|{\bm x}_0|}\frac1{\sqrt{2\pi}}\int\limits_{-\infty}^{\infty}d(ct_r)c\frac{d}{d(ct_r)}\left[\frac{{\bm n}_0\times({\bm n}_0\times{\boldsymbol\beta}(t_r))}{c(1-{\boldsymbol{\beta}(t_r)\cdot{\bm n}_0)}}\right]e^{i k(ct_r+|{\bm x}_0|-{\bm n}_0\cdot{\bm r}_0(t_r))}
\end{align}
After an integration by parts we obtain the well known formula \cite{jackson}
\begin{align}
     & {\bm E}({\bm x},k) =\frac{-i k e_0 e^{ik |{\bm x}_0|}}{4\pi\epsilon_0|{\bm x}_0|}\frac1{\sqrt{2\pi}}\int\limits_{-\infty}^{\infty}d(ct)\left[{{\bm n}_0\times({\bm n}_0\times{\boldsymbol\beta}(t))}\right]e^{ik(ct-{\bm n}_0\cdot{\bm r}_0(t))} \\
     & {\bm B}({\bm x},k)  = \frac{1}c{\bm n}_0\times{\bm E}({\bm x},k)
\end{align}

\subsection{The trajectory of a particle in a monochromatic field}

Consider a particle, initially in the position
\begin{align}
    {\bm r}(0) = {\boldsymbol{\mathfrak R}}_0
\end{align}
in the $Oxy$ plane. We denote by $(\rho_0, \varPhi_0)$ the cylindrical coordinates of the initial position of the particle and define
\begin{align}
    {\bm r}_0={\bm x}-{\boldsymbol{\mathfrak{R}}_0}
\end{align}

The particle interacts with an LG beam of not too large intensity, of indices $(p_L,m_L)$, the field is approximated as monochromatic, with a $\rho_0$ dependent amplitude, and an initial phase $m_L\varPhi_0$.
Consider a plane wave monochromatic field of frequency $\omega_L=ck_L$, propagating along the $Oz$ direction, described by the vector potential
\begin{align}
    {\bm A}(\phi)=A_0\left[ {\bm e}_x\zeta_x \cos(k_L\phi-m_L\varPhi_0)+{\bm e}_y\zeta_y\sin(k_L\phi-m_L\varPhi_0)\right],\qquad \zeta_x^2+\zeta_y^2=1,\qquad \phi=ct-z
\end{align}
Consider a charged particle, of charge $e_0$, in the presence of the field with the initial conditions
\begin{align}
    {\bm p}_i={\bm p}_0, \qquad{\bm A}({\bm r}_i,t_i)={\bm A}_0
\end{align}
i.e. the initial momentum is ${\bm p}_0$ and the value of the vector potential at the initial position of the particle is ${\bm A}_0$. With these, the particle trajectory ${\bm r}_0(t)$ can be written as an analytic function of the variable $\chi = ct-z_0$
\begin{align}
     & \bm{r}_{0\perp}(\chi)=\int_{c t_i-z_i}^\chi \mathrm{d} \phi \frac{\bm{p}_{0 \perp}-e \bm{A}(\phi)}{n_{\mathrm{L}} \cdot p_0}, \quad z_0(\chi)=\int_{c t_i-z_i}^\chi \mathrm{d} \phi F(\phi),\quad \chi = ct-z_0\nonumber                                    \\
     & F(\chi) =\frac{e^2 \mathcal{A}^2(\chi)}{2\left(n_{\mathrm{L}} \cdot p_0\right)^2}+\frac{p_{0 z}}{n_{\mathrm{L}} \cdot p_0}, \qquad e^2 \mathcal{A}^2(\chi)=e^2 \mathbf{A}^2(\chi)-2 e \mathbf{A}(\chi) \cdot \mathbf{p}_{0 \perp}, \qquad n_L=(1,{\bm e}_z)
\end{align}
We discuss the particular case:
\begin{align}
    t_i=0,\qquad z_i=0,\qquad {\bm p}_0 = 0
\end{align}
and use the notation
\begin{align}
    \xi_0 = \frac{-e_0 A_0}{m_ec}.
\end{align}
The trajectory can be written as
\begin{align}
     & {\bm r}_{0\perp}(\chi)=\frac{\xi_0}{k_L}\left[{\bm e}_x\zeta_x\sin(k_L\chi+m_L\varPhi_0)-{\bm e}_y\zeta_y\cos(k_L\chi+m_L\varPhi_0)\right]                                                           \\
     & z_0(\chi)=\frac{\xi_0^2}4\left[\chi+\frac{(\zeta_x^2-\zeta_y^2)\sin(2k_L\chi+2m_L\varPhi_0)}{2k_L}\right]-z_{in},\qquad z_{in} = \frac{\xi_0^2}4\frac{\zeta_x^2-\zeta_y^2}{2k_L}\sin(2m_L\varPhi_0).
\end{align}
We also write the velocity of the particle as a function of $\chi$
\begin{align}
     & {\boldsymbol\beta}=\frac{d{\bm r}(\chi)}{d\chi}\frac{d\chi}{d (ct)}=\frac{d{\bm r}(\chi)}{d\chi}(1-\beta_z) \\
     & \beta_z =\frac{\xi_0^2}4\left[1+(\zeta_x^2-\zeta_y^2)\cos(2k_L\phi-2m_L\varPhi_0)\right](1-\beta_z)
\end{align}
which can be put in an equivalent form
\begin{align}
     & \beta_z(\chi) =\frac{F(\chi)}{1+F(\chi)},\qquad F(\chi) = \frac{\xi_0^2}4\left[1+(\zeta_x^2-\zeta_y^2)\cos(2k_L\chi-2m_L\varPhi_0)\right]                  \\
     & {\boldsymbol{\beta}}_\perp(\chi)=\xi_0\left[{\bm e}_x\zeta_x\cos(k_L\chi-m_L\varPhi_0)+{\bm e}_y\zeta_y\sin(k_L\chi-m_L\varPhi_0)\right]\frac1{1+F(\chi)}.
\end{align}

\subsection{The field emitted by a particle in an LG beam}

Consider a particle, initially in the position
\begin{align}
    {\bm r}(0) = {\boldsymbol{\mathfrak R}}_0
\end{align}
in the $Oxy$ plane. We denote by $(\rho_0, \varPhi_0)$ the cylindrical coordinates of the initial position of the particle.

The particle interacts with an LG beam of not too large intensity, of indices $(p,m)$. Then the trajectory pf the particle can be approximated according to the formulas in the previous section,  with an intensity parameter $\xi$ dependent on the radial coordinate $\rho_0$.
\begin{align}
     & {\bm r}(t) -{\boldsymbol{\mathfrak R}}_0={\bm r}_0(t)\nonumber                                                                                                                                                      \\
     & {\bm r}_{0\perp}(\chi)=\frac{\xi(\varrho_0)}{k_L}\left[{\bm e}_x\zeta_x\sin(k_L\chi-m_L\varPhi_0)-{\bm e}_y\zeta_y\cos(k_L\chi-m_L\varPhi_0)\right]                                                                 \\
     & z_0(\chi)=\frac{\xi^2(\varrho_0)}4\left[\chi+\frac{(\zeta_x^2-\zeta_y^2)\sin(2k_L\chi-2m_L\varPhi_0)}{2k_L}\right] +z_{in}                                                                                          \\
     & \beta_z(\chi) =\frac{F(\chi)}{1+F(\phi)},\qquad  {\boldsymbol{\beta}}_\perp(\chi)=\xi(\varrho_0)\left[{\bm e}_x\zeta_x\cos(k_L\chi-m_L\varPhi_0)+{\bm e}_y\zeta_y\sin(k_L\chi-m_L\varPhi_0)\right]\frac1{1+F(\chi)} \\
     & F(\chi) = \frac{\xi^2(\varrho_0)}4\left[1+(\zeta_x^2-\zeta_y^2)\cos(2k_L\chi-2m_L\varPhi_0)\right]
\end{align}
We use the previous results in the expression of the emitted electric field
\begin{align}
    {\bm E}({\bm x},k) =\frac{-i k e_0 e^{ik |{\bm x}_0|}}{4\pi\epsilon_0|{\bm x}_0|}\frac1{\sqrt{2\pi}}\int\limits_{-\infty}^{\infty}d(ct)\left[{{\bm n}_0\times({\bm n}_0\times{\boldsymbol\beta}(t))}\right]e^{ik(ct-{\bm n}_0\cdot{\bm r}_0(t))}
\end{align}
In the previous integral we perform another change of variable from $t$ to $\chi=ct-z$, using
\begin{align}
    d\chi = c (1-\beta_z(\chi)) dt = c\frac1{1+F(\chi)} dt,\qquad \rightarrow \qquad dt=\frac1c d\chi(1+F(\chi))
\end{align}
is also convenient to redefine the velocity
\begin{align}
     & {\boldsymbol{\beta}} = \frac1{1+F(\chi)}\tilde{\boldsymbol{\beta}}(\chi)                                                                                  \\
     & \tilde{\boldsymbol{\beta}}_\perp(\chi) =\xi(\varrho_0)\left[{\bm e}_x\zeta_x\cos(k_L\chi-m_L\varPhi_0)+{\bm e}_y\zeta_y\sin(k_L\chi-m_L\varPhi_0)\right], \\
     & \tilde\beta_z(\chi) =F(\chi)=\frac{\xi^2(\varrho_0)}4\left[1+(\zeta_x^2-\zeta_y^2)\cos(2k_L\chi-2m_L\varPhi_0)\right]
\end{align}
and obtain the Fourier transform of the emitted electric field
\begin{align}
    {\bm E}({\bm x},k) =\frac{-i k e_0 e^{ik |{\bm x}_0|}}{4\pi\epsilon_0|{\bm x}_0|}\frac1{\sqrt{2\pi}}\int\limits_{-\infty}^{\infty}d\chi\left[{{\bm n}_0\times({\bm n}_0\times\tilde{\boldsymbol\beta}(\chi))}\right]e^{ik(\chi +z(\chi)-{{\bm n}_0\cdot{\bm r}_0(\chi)})}\label{exint}
\end{align}
and, denoting by $\theta_0$, $\phi_0$ the polar angles of ${\bm n}_0$ and using ${\bm n}_L={\bm e}_z$
we have
\begin{align}
    z(\chi)-{\bm n}_0\cdot{\bm r}_0(\chi)={\bm n}_L\cdot{\bm r}_0(\chi)-{\bm n}_0\cdot{\bm r}_0(\chi)=({\bm n}_L-{\bm n}_0)\cdot{\bm r}_0(\chi)
\end{align}

Next we write the explicit form of  $({\bm n}_L-{\bm n}_0)\cdot{\bm r}_0(\chi)$;
\begin{align}
    ({\bm n}_L-{\bm n}_0)\cdot{\bm r}_0(\chi)=
     & -\frac{\xi(\varrho_0)}{k_L}\sin\theta_0\sqrt{\cos^2\phi_0\zeta_x^2+\sin^2\phi_0\zeta_y^2}\nonumber                                                                                                                                     \\
     & \quad \times\left[\sin(k_L\chi-m_L\varPhi_0)\frac{\cos\phi_0\zeta_x}{\sqrt{\cos^2\phi_0\zeta_x^2+\sin^2\phi_0\zeta_y^2}}-\cos(k_L\chi-m_L\varPhi_0)\frac{\sin\phi_0\zeta_y}{\sqrt{\cos^2\phi_0\zeta_x^2+\sin^2\phi_0\zeta_y^2}}\right]
\end{align}
We introduce the notations:
\begin{align}
     & n_L=(1,{\bm n}_L)\equiv(1,{\bm e}_z),\qquad n_0=(1,{\bm n}_0),\qquad 1-\cos\theta_0=n_L\cdot n_0,\nonumber                                                                                                                                                                  \\
     & \gamma_0=\arctan\left(\frac{\sin\phi_0\zeta_y}{\cos\phi_0\zeta_x}\right),\qquad \Gamma_0=\frac{\xi(\varrho_0)}{k_L}\sin\theta_0\sqrt{\cos^2\phi_0\zeta_x^2+\sin^2\phi_0\zeta_y^2},\qquad \Delta_0 = n_L\cdot n_0\frac{\xi^2(\varrho_0)}{8k_L}(\zeta_x^2-\zeta_y^2)\nonumber \\
     & (1-\cos\theta_0)z_{in} = n_L\cdot n_0 \frac{\xi^2(\varrho_0}4\frac{\zeta_x^2-\zeta_y^2}{2k_L}\sin(2m_L\varPhi_0)=Z_0\label{param-gen}
\end{align}
and we can write
\begin{align}
    \chi+({\bm n}_L-{\bm n}_0)\cdot{\bm r}_0(\chi)= & \chi\left(1+n_L\cdot n_0\frac{\xi^2(\varrho_0)}4\right)-\Gamma_0\sin(k_L\chi-m_L\varPhi_0-\gamma_0)+\Delta_0\sin(2k_L\phi-2m_L\varPhi_0)+Z_0.
\end{align}
Using the Anger identity \cite{bessel}
\begin{align}
    e^{i z \sin \phi}=\sum_{n=-\infty}^{\infty} e^{i n \phi} J_n(z)
\end{align}
to expand the exponential in the expression (\ref{exint})  and we obtain
\begin{align}
    e^{ik(\phi +z(\phi)-{{\bm n}_0\cdot{\bm r}_0(\phi)})} & =e^{-ikZ_0}e^{ik\phi(1+n_L\cdot n_0\frac{\xi^2(\varrho_0)}4)}\sum\limits_{M=-\infty}^{\infty}e^{-iMk_L\phi}e^{iMm_L\varPhi_0}\sum\limits_{p=-\infty}^{\infty}e^{i(M+2p)\gamma_0}J_{M+2p}(k\Gamma_0)J_p(k\Delta_0)\label{expansion11}
\end{align}
We define a generalized Bessel function
\begin{align}
    \tilde J_M(k;\gamma_0, \Gamma_0, \Delta_0) = \sum\limits_{p=-\infty}^{\infty}e^{i(M+2p)\gamma_0}J_{M+2p}(k\Gamma_0)J_p(k\Delta_0)
\end{align}
which depends on the argument $k$ and the parameters $\gamma_0$, $\Gamma_0$, $\Delta_0$.
Then the expansion (\ref{expansion11}) can be further written as
\begin{align}
    e^{ik(\chi +z(\chi)-{{\bm n}_0\cdot{\bm r}_0(\chi)})} & =e^{ik\chi(1+n_L\cdot n_0\frac{\xi^2(\varrho_0)}4)}\sum\limits_{M=-\infty}^{\infty}e^{-iMk_L\chi}e^{iMm_L\varPhi_0+ikZ_0}\tilde J_M(k;\gamma_0,\Gamma_0,\Delta_0)
\end{align}
Also the integrand  in (\ref{exint}) can be written in the equivalent form
\begin{align}
    {\bm n}_0\times({\bm n}_0\times\tilde{\boldsymbol{\beta}})  = & \xi(\varrho_0)\left[{\boldsymbol{\nu}}_{0x}\frac12\zeta_x\left(e^{i(k_L\phi -m_L\varPhi_0)}+e^{-i(k_L\phi-m_L\varPhi_0)}\right)+{\boldsymbol{\nu}}_{0y}\frac1{2i}\zeta_y\left(e^{i(k_L\phi -m_L\varPhi_0)}-e^{-i(k_L\phi-m_L\varPhi_0)}\right)\right]\nonumber      \\
                                                                  & + \frac{\xi^2(\varrho_0)}4{\boldsymbol{\nu}_{0z}}\left[1+(\zeta_x^2-\zeta_y^2)\cos(2k_L\phi-2m_L\varPhi_0)\right]                                                                                                                                                   \\
    =                                                             & \frac{\xi(\varrho_0)}2{\boldsymbol{\nu}}_{-} e^{i(k_L\phi -m_L\varPhi_0)}+\frac{\xi(\varrho_0)}2{\boldsymbol{\nu}}_{+}e^{-(ik_L\phi -m_L\varPhi_0)}\nonumber                                                                                                        \\&+\frac{\xi^2(\varrho_0)}4{\boldsymbol{\nu}_{0z}}\left[1+\frac{(\zeta_x^2-\zeta_y^2)}2e^{2i(k_L\phi-m_L\varPhi_0)}+\frac{(\zeta_x^2-\zeta_y^2)}2e^{-2i(k_L\phi-m_L\varPhi_0)}\right],\nonumber\\
                                                                  & {\bm n}_0\times({\bm n}_0\times{\bm e}_{\{x,y,z\}})={\boldsymbol{\nu}}_{0\{x,y,z\}},\qquad {\boldsymbol{\nu}}_{\pm}=\zeta_x{\boldsymbol{\nu}}_{0x}\pm i\zeta_y{\boldsymbol{\nu}}_{0y},\qquad ({\boldsymbol{\nu}}_{\pm})^* = {\boldsymbol{\nu_{\mp}}}\label{defnupm}
\end{align}

With these the Fourier transform of the emitted electric field becomes
\begin{align}
    {\bm E}({\bm x},k)=\sum\limits_{N=-\infty}^{\infty}\tilde{\bm E}_N({\bm x})\delta(k-N \frac{k_L}{\alpha}),\label{Encirc}
\end{align}
with
\begin{align}
    \hspace*{-1.5cm}\tilde{\bm E}_N({\bm x};\varrho_0,\varPhi_0) = & \frac{-i k_N  e_0 e^{ik_N (|{\bm x}_0|+Z_0)}}{4\pi\epsilon_0|{\bm x}_0|\alpha}{\sqrt{2\pi}}e^{iNm_L\varPhi_0}\left[{\boldsymbol{\nu}}_{+}\frac{\xi(\varrho_0)}{2}\tilde J_{N-1}(k_N;\gamma_0,\Gamma_0,\Delta_0)+{\boldsymbol{\nu}}_{-}\frac{\xi(\varrho_0)}{2}\tilde J_{N+1}(k_N;\gamma_0,\Gamma_0,\Delta_0)+\right.\nonumber        \\
                                                                   & \hspace*{-1cm} \left.{\boldsymbol{\nu}}_{0z}\left(\frac{\xi^2(\varrho_0)}4\tilde J_N(k_N;\gamma_0,\Gamma_0,\Delta_0)+\frac{\xi^2(\varrho_0)}8(\zeta_x^2-\zeta_y^2)\tilde J_{N-2}(k_N;\gamma_0,\Gamma_0,\Delta_0)+\frac{\xi^2(\varrho_0)}8(\zeta_x^2-\zeta_y^2)\tilde J_{N+2}(k_N;\gamma_0,\Gamma_0,\Delta_0)\right)\right],\nonumber \\
                                                                   & \alpha = 1+\frac{\xi^2(\varrho_0)}4n_0\cdot n_L,\qquad k_N = N\frac{k_L}{\alpha};\label{defkn}
\end{align}
here we indicated explicitly the dependence on the cylindrical variables of the initial position of the electron $\varrho_0$, $\varphi_0$.
In the same approximation, the magnetic field  is
\begin{align}
    {\bm B}({\bm x},k) = \frac1c{\bm n}_0\times{\bm E}({\bm x},k)=\sum\limits_{N=-\infty}^{\infty}\tilde {\bm B}_N({\bm x})\delta(k-N\frac{k_L}{\alpha})\label{Bncirc}
\end{align}
with
\begin{align}
    \hspace*{-1cm} \tilde{\bm B}_N({\bm x};\varrho_0,\varPhi_0) = & \frac{-i k_N  e_0 e^{ik_N (|{\bm x}_0|+Z_0)}}{4\pi\epsilon_0|{\bm x}_0|c\alpha}{\sqrt{2\pi}}e^{iNm_L\varPhi_0}\left[{\boldsymbol{\mu}}_{+}\frac{\xi(\varrho_0)}{2}\tilde J_{N-1}(k_N;\gamma_0,\Gamma_0,\Delta_0)+{\boldsymbol{\mu}}_{-}\frac{\xi(\varrho_0)}{2}\tilde J_{N+1}(k_N;\gamma_0,\Gamma_0,\Delta_0)+\right.\nonumber        \\
                                                                  & \hspace*{-1.5cm}\left.{\boldsymbol{\mu}}_{0z}\left(\frac{\xi^2(\varrho_0)}4\tilde J_N(k_N;\gamma_0,\Gamma_0,\Delta_0)+\frac{\xi^2(\varrho_0)}8(\zeta_x^2-\zeta_y^2)\tilde J_{N-2}(k_N;\gamma_0,\Gamma_0,\Delta_0)+\frac{\xi^2(\varrho_0)}8(\zeta_x^2-\zeta_y^2)\tilde J_{N+2}(k_N;\gamma_0,\Gamma_0,\Delta_0)\right)\right],\nonumber \\
                                                                  & {\bm e}_{\pm} = \zeta_x{\bm e}_x\pm i \zeta_y {\bm e}_y                                                                                                                                                                                                                                                                               \\
                                                                  & {\boldsymbol{\nu}}_{\pm} = {\bm n}_0\times\left({\bm n}_0\times{\bm e}_{\pm}\right)=-{\bm{n}}_0\times{\boldsymbol{\mu}_{\pm}},\qquad {                                                                       \boldsymbol{\nu}}_z={\bm n}_0\times\left({\bm n}_0\times{\bm e}_{z}\right)=-{\bm n}_0\times{\boldsymbol{\mu}}_z\nonumber \\
                                                                  & {\boldsymbol{\mu}}_{\pm}={\bm n}_0\times{\boldsymbol{\nu}}_{\pm}=-{\bm n}_0\times{\bm e}_{\pm},\qquad {\boldsymbol{\mu}}_{0z}={\bm n}_0\times{\boldsymbol{\nu}}_{0z}=-{\bm n}_0\times{\bm e}_z                                                                                                                 \nonumber
\end{align}
From the previous results we  obtain the frequency spectrum of the emitted radiation, consisting in equidistant lines given be the well known formula of  non-linear Thomson scattering
\begin{align}
    \omega_N = ck_N = \frac{N\omega_L}{1+\frac{\xi^2(\varrho_0)}4n_0\cdot n_L}=\frac{N\omega_L}{1+\frac{\xi^2(\varrho_0)}4(1-\cos\theta_0)}\label{defomegan}
\end{align}
The previous result shows that if the retardation is included the emitted frequency depends on the direction through the polar angle $\theta_0$ of the direction of observation unity vector ${\bm n}_0$ (\ref{defn01}); however, we shall see that when the coherent sum of the field emitted by the entire electron distribution, at large distance from the focal plane  dependence is very weak.

Properties of the generalized Bessel functions
\begin{align}
    \tilde J_M(k;\gamma_0, \Gamma_0, \Delta_0)   & = \sum\limits_{p=-\infty}^{\infty}e^{i(M+2p)\gamma_0}J_{M+2p}(k\Gamma_0)J_p(k\Delta_0)                                                                                                       \\
    \tilde J_{-M}(-k;\gamma_0,\Gamma_0,\Delta_0) & = \sum\limits_{p=-\infty}^{\infty}e^{i(-M+2p)\gamma_0}J_{-M+2p}(-k\Gamma_0)J_p(-k\Delta_0)=\sum\limits_{p=-\infty}^{\infty}e^{i(-M+2p)\gamma_0}J_{M-2p}(k\Gamma_0)J_{-p}(k\Delta_0)\nonumber \\
                                                 & =\sum\limits_{p=-\infty}^{\infty}e^{-i(M+2p)\gamma_0}J_{M+2p}(k\Gamma_0)J_{p}(k\Delta_0)=[\tilde J_{M}(k;\gamma_0,\Gamma_0,\Delta_0)]^*
\end{align}
Then we can easily prove that
\begin{align}
    {\bm E}_{-N}({\bm x}) = (E_N({\bm x}))^*,\qquad {\bm B}_{-N}({\bm x}) = (B_N({\bm x}))^*
\end{align}

\subsubsection{The case of circular polarization}

In the case of circular polarization $\zeta_x=\epsilon_L\zeta_y$ with $\epsilon_L=\pm1$ the previous results have a much simpler form; we have
\begin{align}
     & \gamma_0\rightarrow \epsilon_L\phi_0,\qquad \Gamma_0\rightarrow\overline\Gamma_0=\frac{\xi(\varrho_0)}{\sqrt{2}k_L}\sin\theta_0,\qquad \Delta_0 = 0,\qquad Z_0=0\label{param-circ}\nonumber \\
     & \tilde J_M(k;\gamma_0, \Gamma_0, \Delta_0) \rightarrow e^{iM\epsilon_L\phi_0}J_{M}(k\tilde\Gamma_0)
\end{align}
If $\theta_0\ll1$, considering the behavior of Bessel functions at low arguments, we can also approximate the Fourier components of the electric and magnetic field as
\begin{align}
    \tilde{\bm E}_N({\bm x};\varrho_0,\varPhi_0) \rightarrow & \frac{-i k_N  e_0 e^{ik_N |{\bm x}_0|}}{4\pi\epsilon_0|{\bm x}_0|\alpha}{\sqrt{2\pi}}{\boldsymbol{\nu}}_{+}\frac{\xi(\varrho_0)}{2}e^{iNm_L\varPhi_0}e^{i(N-1)\epsilon_L\phi_0}J_{N-1}(k_N\frac{\xi(\varrho_0)}{\sqrt{2}k_L}\sin\theta_0)= \nonumber \\
                                                             & = e^{-iNm_L\varPhi_0}e^{i(N-1)\epsilon_L\phi_0}e^{ik_N |{\bm x}_0|}{\boldsymbol{\mathcal E}}_N({\bm x};\varrho_0,\varPhi_0)\nonumber                                                                                                                 \\
    \tilde{\bm B}_N({\bm x};\varrho_0,\varPhi_0) \rightarrow & \frac1{c} e^{iNm_L\varPhi_0}e^{i(N-1)\epsilon_L\phi_0}{\bm n}_0\times {\boldsymbol{\mathcal  E}}_N({\bm x};\varrho_0,\varPhi_0)\label{aproxenexact}
\end{align}
In the previous equation we introduced the  notation
\begin{align}
    {\boldsymbol{\mathcal  E}}_N({\bm x};\varrho_0,\varPhi_0)=\frac{-i k_N  e_0 }{4\pi\epsilon_0|{\bm x}_0|\alpha}{\sqrt{2\pi}}{\boldsymbol{\nu}}_{+}\frac{\xi(\varrho_0)}{2}J_{N-1}(k_N\frac{\xi(\varrho_0)}{\sqrt{2}k_L}\sin\theta_0) \label{approxexp}
\end{align}
that will be used in the next section.

\section{Integration over the initial electron distribution}

We consider the case of a charge distributed uniformly with the charge density $\sigma_0$ within a disk of radius $\rho_m$ in the plane $Oxy$, interacting with an $LG$ beam. We assume that the beam intensity is low, such that the amplitude of the oscillatory motion of the particle is low with respect to waist $w_0$ of the external beam. we also assume that the particles are initially at rest, in the focal plane $Oxy$.

The field emitted by this distribution will be calculated as the superposition of the fields emitted by the elementary charge $de_0=\sigma_0 da$ in each area element $da$ within the disk.

\begin{align}
    {\bm E}_{tot}({\bm x},t) = \int \limits_{\Sigma} da {\bm E}({\bm x},t)
\end{align}
and the Fourier transform becomes
\begin{align}
    {\bm E}_{tot}({\bm x},\omega) = \sum\limits_{N=-\infty}^{\infty} \tilde{\bm E}_{N,tot}({\bm x},\omega),\qquad
    \tilde {\bm E}_{N,tot}({\bm x})=\int\limits_0^{\rho_m}d\varrho_0\varrho_0\int\limits_0^{2\pi}d\varPhi_0 \tilde{\bm E}_N({\bm x};\varrho_0,\varPhi_0)\delta(\omega-\omega_N)\label{int-En2}
\end{align}
and  $\omega_N$ is given by Eq. (\ref{defomegan}).
Similarly, the Fourier transform of the total magnetic field radiated is
\begin{align}
    {\bm B}_{tot}({\bm x},\omega) = \sum\limits_{N=-\infty}^{\infty} \tilde{\bm B}_{N,tot}({\bm x}),\qquad    \tilde {\bm B}_{N,tot}({\bm x})=\int\limits_0^{\rho_m}d\rho_0\rho_0\int\limits_0^{2\pi}d\varPhi_0 \tilde{\bm B}_N({\bm x};\varrho_0,\varPhi_0)\delta(\omega-\omega_N)
\end{align}
In the equations above $\tilde{\bm E}_N$, $\tilde {\bm B}_N$ are defined in  Eqs. (\ref{Encirc},\ref{Bncirc}), where the electron charge $e_0$ must be replaced by the surface charge density $\sigma_0$ and $\omega_N$ is a function of $\varrho_0$, $\varPhi_0$ trough the angle $\theta_0$ (see Eq. (\ref{defomegan}). If we consider only the scattering at very small angles ($\theta_0\ll1$) and $\xi_0\le1$ then $\omega_N$ can be written as
\begin{align}
    \omega_N\approx N \omega+\epsilon(\varrho_0,\varPhi_0),\qquad \epsilon(\varrho_0,\varPhi_0)\ll1
\end{align}
Then we define the $ N$ th Fourier component of the total electric or magnetic field  field as an integral over a small interval $ \Delta \omega \ll \omega_L$
\begin{align}
     & \boldsymbol{E}_{N, t o t}(\boldsymbol{x}) \equiv \int_{N \omega_L-\Delta \omega/ 2}^{N \omega_L+\Delta \omega/ 2} d \omega \boldsymbol{E}_{N, t o t}(\boldsymbol{x};\varrho_0,\varPhi_0) =\int_0^{\varrho_m} d \varrho_0 \varrho_0 \int_0^{2 \pi} d \varPhi_0 \tilde{\boldsymbol{E}}_N(\boldsymbol{x};\varrho_0,\varPhi_0)\nonumber \\
     & \boldsymbol{B}_{N, t o t}(\boldsymbol{x}) \equiv \frac{1}{c} \int_0^{\varrho_m} d \varrho_0 \varrho_0 \int_0^{2 \pi} d \varPhi_0\left(\boldsymbol{n}_0 \times \tilde{\boldsymbol{E}}_N(\boldsymbol{x};\varrho_0,\varPhi_0)\right)\label{ebint}
\end{align}
In the expression of $\tilde{\bm E}_N$ in the previous equations $k$ must be replaced by $k_N$ calculated according  to the formula (\ref{defkn}).
\subsection{Analytical estimation of the phase of the emitted field in the circular polarization case}

We consider the expression of the total emitted electric field of given harmonic index $N$ (\ref{ebint})
\begin{align}
    \tilde {\bm E}_{N,tot}({\bm x})=\int\limits_0^{\rho_m}d\rho_0\rho_0\int\limits_0^{2\pi}d\varPhi_0 \tilde{\bm E}_N({\bm x};\varrho_0,\varPhi_0)
\end{align}
and use here the approximation of $\tilde{\bm E}_N$ (\ref{aproxenexact})
\begin{align}
    \hspace*{-1cm}\tilde {\bm E}_{N,tot}({\bm x}) & = \int\limits_0^{\varrho_m}d\varrho_0\varrho_0 \int\limits_0^{2\pi}d\varPhi_0 e^{iNm_L\varPhi_0}e^{i(N-1)\epsilon_L\phi_0}e^{ik_N |{\bm x}_0|}{\boldsymbol{\mathcal E}}_N({\bm x};\varrho_0,\varPhi_0)
\end{align}
Next we analyze  the integral over $\varPhi_0$; let us denote the cylindrical coordinates of the observation point by $(\varrho,\varPhi,Z)$
\begin{align}
    {\bm x}=\varrho\cos\varPhi{\bm e}_x+\varrho\cos\varPhi{\bm e}_y+Z{\bm e}_z
\end{align}
In a typical calculation $Z$ is fixed and $\varrho$, $\varPhi$ changes (we record the radiation on a screen perpendicular on the beam propagation direction at the distance $Z$ from the focal plane).
The distance $|x_0|$  is a function of $\varrho_0$, $\varPhi_0$
\begin{align}
    |{\bm x}_0|(\varrho_0,\varPhi_0)=\sqrt{(\varrho\cos\varPhi-\rho_0\cos\varPhi_0)^2+(\varrho\sin\varPhi-\rho_0\sin\varPhi_0)^2+Z^2}
\end{align}
and $e^{iNk_L|{\bm x}_0|}$ is a very fast oscillating function;  we use the stationary phase method for estimating the integral over $\varPhi_0$.
We use the general formula \cite{oscillatory}
\begin{align}
    {\cal I}_{\varPhi_0}= \int d\varPhi_0 e^{if(\varPhi_0)} g(\varPhi_0)\approx\sum\limits_{y_0}\sqrt\frac{2\pi}{|f''(y_0)|}g(y_0)e^{if(y_0)+i\sigma\frac{\pi}4}
\end{align}
where $y_0$ {are} the stationary points of $f(\varPhi_0)$
\begin{align}
    f'(y_0)=0
\end{align}
and $\sigma$ is the sign of the second derivative $f''(y_0)$

In our case
\begin{align}
    f(\varPhi_0)=k_N|{\bm x}|_0(\varrho_0,\varPhi_0)
\end{align}
and its derivative is
\begin{align}
    f'(\varPhi_0)=-\frac{k_N \varrho \varrho_0 \sin(\varPhi-\varPhi_0)}{\sqrt{\varrho^2+Z^2+\varrho_0^2-2 \varrho \varrho_0 \cos(\varPhi-\varPhi_0)}}
\end{align}
whose zeroes are
\begin{align}
    \Phi_0^{(1)} = \varPhi,\qquad \Phi^{(2)}_0=\varPhi+\pi
\end{align}
and the integral over $\varPhi_0$ becomes
\begin{align}
     & {\cal I}_{\varPhi_0}=e^{i(m_LN\varPhi+\epsilon_L(N-1)\varPhi)}{\bm h}({\bm x};\rho_0),\nonumber                                                                                                                           \\
     & h({\bm x};\varrho_0)=\sqrt{2\pi i}\left[e^{iNk_L\sqrt{Z^2+(\varrho-\varrho_0)^2}}{\boldsymbol{\mathcal E}}({\bm x};\varrho_0,\varPhi)\sqrt{\frac{\sqrt{Z^2+(\varrho-\varrho_0)^2}}{Nk_L\varrho\varrho_0}}\right.\nonumber \\
     & \qquad\qquad\left.-i(-1)^{Nm_L}e^{iNk_L\sqrt{Z^2+(\varrho+\varrho_0)^2}}{\boldsymbol{\mathcal E}}({\bm x};\varrho_0,\varPhi+\pi)\sqrt{\frac{\sqrt{Z^2+(\varrho+\varrho_0)^2}}{Nk_L\varrho\varrho_0}}\right]\label{htot}
\end{align}

With these the expression of the total  electric field measured in ${\bm x}$ is
\begin{align}
    \tilde {\bm E}_{N,tot}({\bm x}) & =e^{i(m_LN\varPhi+\epsilon_L(N-1)\varPhi)}\int\limits_0^{\rho_m}d\varrho_0\varrho_0{\bm h}({\bm x};\varrho_0)
\end{align}
whose dependence on $\varPhi$ is of the form
\begin{align}
    {\bm E}_{N,tot}({\bm x})\sim e^{iq\varPhi},\qquad q = N m_L +(N-1)\epsilon_L
\end{align}

Finally, we can estimate the behavior of the remaining integral over $\varrho_0$; we write the function $h(\varrho_0)$ as
\begin{align}
    h({\bm x},\varrho_0)=e^{iNk_L\sqrt{Z^2+(\varrho-\varrho_0)^2}}{\bm h}_1({\bm x};\varrho_0)+e^{iNk_L\sqrt{Z^2+(\varrho+\varrho_0)^2}}{\bm h}_2({\bm x},\varrho_0)
\end{align}
where ${\bm h}_i({\varrho;\varrho_0})$ can be identified from the comparison with \ref{htot}.
In the integral over $\varrho_0$ the two exponential functions $e^{iNk_L\sqrt{Z^2+(\varrho\pm\varrho_0)^2}}$ are, again, fast oscillating; from them, only $e^{iNk_L\sqrt{Z^2+(\varrho-\varrho_0)^2}}$ has a stationary point at $\varrho_0=\varrho$, if $\varrho\le\varrho_m$, with $\varrho_m$ the upper limit of integral. As a consequence, if $\rho\lessapprox {\cal O}(\varrho_m)$ the integral over $\varrho_0$ depends slowly on $\varrho$, while for $\varrho\gtrapprox{\cal O}(\varrho_m)$ the result is a fast oscillating function of $\varrho$.

\bibliography{supp}